\documentclass[sigconf,x11names]{acmart}

\usepackage{popets,arabtex,CJKutf8,amsthm,enumitem,amsmath,hyperref,siunitx,subcaption,mathtools,xspace, framed,quoting}
\usepackage[utf8]{inputenc}

\usepackage{amssymb}
\hypersetup{colorlinks=true,allcolors=black}
\usepackage[capitalize]{cleveref}
\renewcommand{\ref}[1]{\cref{#1}}
\setcode{utf8}
\newcommand{\correct}[1]{}
\newcommand{\hide}[1]{}

\newenvironment{shadedquotation}
 {\colorlet{shadecolor}{LavenderBlush2}
 \begin{shaded*}
  \quoting[leftmargin=0pt, vskip=0pt]
 }
 {\endquoting
 \end{shaded*}
}

\newenvironment{shadedquotationn}
     {\colorlet{shadecolor}{Azure2}
     \begin{shaded*}
          \quoting[leftmargin=0pt, vskip=0pt]
         }
         {\endquoting
     \end{shaded*}
}

\newenvironment{shadedanswer}
 {\colorlet{shadecolor}{Ivory1}
     \begin{shaded*}
      \quoting[leftmargin=0pt, vskip=0pt]
     }
     {\endquoting
     \end{shaded*}
}

\newcommand{\dataName}{CryptoQA\xspace}

\setcopyright{popets}
\copyrightyear{2025}

\acmYear{}
\acmVolume{}
\acmNumber{}
\acmDOI{}
\acmISBN{}
\acmConference{}
\begin{document}

\title[CryptoQA]{CryptoQA: A Large-scale Question-answering Dataset for AI-assisted
Cryptography}



\author{Mayar Elfares, Pascal Reisert, Tilman Dietz, Manpa Barman, Ahmed Zaki, Ralf Küsters, Andreas Bulling}
\affiliation{%
  \institution{University of Stuttgart}
  \country{Germany}}


\renewcommand{\shortauthors}{Elfares et al.}
\begin{abstract}
    Large language models (LLMs) excel at many general-purpose natural language processing tasks. 
    However, their ability to perform deep reasoning and mathematical analysis, particularly for complex tasks as required in cryptography, remains poorly understood, largely due to the lack of suitable data for evaluation and training.
    To address this gap, we present \dataName, the first large-scale question-answering (QA) dataset specifically designed for cryptography.
    \dataName~contains over two million QA pairs drawn from curated academic sources, along with contextual metadata that can be used to test the cryptographic capabilities of LLMs and to train new LLMs on cryptographic tasks.
    We benchmark 15 state‑of‑the‑art LLMs on \dataName, evaluating their factual accuracy, mathematical reasoning, consistency, referencing, backward reasoning, and robustness to adversarial samples.
    In addition to quantitative metrics, we provide expert reviews that qualitatively assess model outputs and establish a gold‑standard baseline.
    Our results reveal significant performance deficits of LLMs, particularly on tasks that require formal reasoning and precise mathematical knowledge.
    This shows the urgent need for LLM assistants tailored to cryptography research and development.
    We demonstrate that, by using \dataName, LLMs can be fine-tuned to exhibit better performance on cryptographic tasks. 
\end{abstract}


\maketitle

\begin{figure*}
    \centering
    \includegraphics[width=\linewidth]{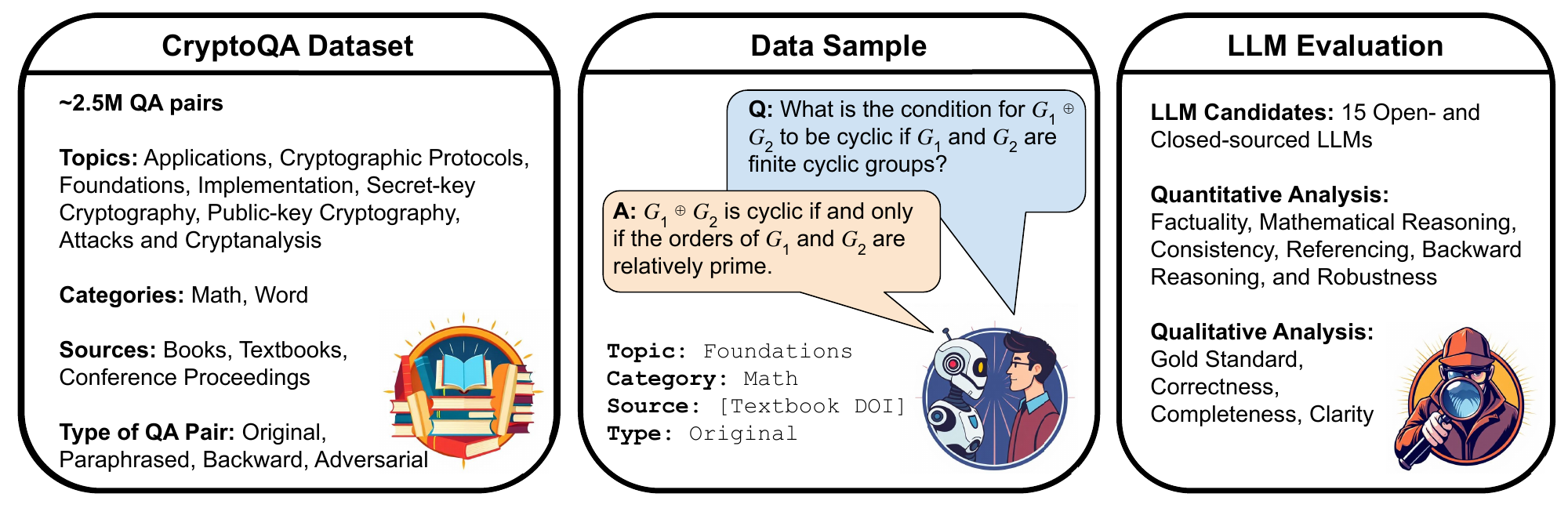}
    \caption{Our new dataset \dataName~is the first large-scale cryptography dataset with over 2.5M QA pairs with rich metadata, i.e., different topics, categories, sources, and types of QA pairs (cf. \cref{sec:dataset}). We use \dataName~to quantitatively and qualitatively evaluate existing state-of-the-art LLMs for their cryptographic capabilities, such as factuality and mathematical reasoning, w.r.t. different metrics (cf. \cref{sec:evaluation,sec:results}).
    (Icons generated by deepai.org.)}
    \label{fig:teaser}
\end{figure*}

\section{Introduction}\label{sec:intro}
Artificial intelligent (AI) assistive tools, particularly those based on large language models (LLMs), are fundamentally reshaping how we interact with, design, and apply scientific methods \cite{kumar2023mycrunchgpt, zhang2025exploring}.
In recent years, LLMs have achieved remarkable progress on a broad spectrum of natural-language-processing (NLP) tasks, including text comprehension, generation, and reasoning \cite{jiang2024mixtral, qwen2025technicalreport, deepseekr12025, meta2024llama3.1, openai2025gpt5, xai2025grok4}. 
One of the reasons for this success is the ever-increasing availability of web-crawled language datasets that can be used for training and evaluation.
In contrast, training LLMs or assessing their abilities to solve tasks from specialised fields, such as cryptography, remains a challenge \cite{xu2025llmcyber, li2025cipherbank, noever2023llmsforciphers, jing2025secbench} because domain-specific datasets are often not available.
Nevertheless, in practice, LLMs are used for domain-specific tasks in nearly every field.
For example, in our survey, an overwhelming number of cryptographers reported using LLMs daily for their work (cf. \cref{sec:results}).
Our survey aligns well with recent research that discusses the growing adoption of LLMs in cryptography research and practice \cite{al2023chatgpt, mohammed2023ai}.
Naturally, using LLMs for domain-specific tasks without a good understanding of how correct the answers are poses a significant risk.
The danger of wrong answers remaining unnoticed is particularly high for LLMs, given their impressive performance on NLP tasks and the resulting trust in LLMs that many of us have developed in recent years \cite{ding2025citations}.
For cryptographic tasks, these errors in model outputs can have substantial consequences, potentially leading to flawed analyses, incorrect proofs, or vulnerabilities in cryptographic protocols \cite{coffey2017ameliorating, el2021overview}.
Beyond technical risks, such errors carry broader social implications, as cryptography underpins secure communication, financial transactions, and digital privacy. Inaccurate reasoning by LLMs could compromise data security, undermine public trust in digital systems, or facilitate malicious exploitation \cite{rosenzweig2013alarming, chung2017critical}. 
This highlights the critical need for evaluating LLM capabilities and limitations in cryptography for the following key reasons: 

\noindent\textbf{Assessing reliability.}
    The reliability of LLM predictions is crucial for both expert and non-expert users in cryptography. Non-experts—such as software developers or engineers who implement or modify cryptographic protocols—may fail to detect incorrect model outputs, potentially introducing security vulnerabilities. Although experts, such as cryptography researchers, are more likely to identify erroneous predictions by consulting the relevant literature, inaccurate responses may still mislead them initially, reducing research efficiency and slowing the broader adoption of LLMs in cryptographic practice.
    
    \noindent\textbf{Guiding model design.} Identifying the failure modes of current models allows the development of specialised architectures, pre‑training strategies, and domain‑specific corpora that improve the performance of LLMs on cryptographic tasks.
    
     \noindent\textbf{Benchmarking progress.} Standardised evaluations on recognised benchmarks are required for cryptographers to track improvements over time and to make informed decisions regarding the suitability of models for different use cases.

A foundational step towards LLM-assistance in cryptography is creating large-scale, domain-specific datasets that capture the nuanced knowledge and reasoning patterns essential for cryptographic tasks, and that can then be used for the evaluation and training of AI models.
We therefore introduce \dataName\footnote{We publicly release the dataset and source code to ensure reproducibility and facilitate future research at \cite{code}. We provide our implementation as a self-contained Python notebook designed to run seamlessly on Google Colab, requiring no prior environment configuration. This design choice ensures full reproducibility while lowering the technical barrier for practitioners. In particular, the notebook is tailored for cryptographers who may not have prior experience with machine learning or AI toolchains.}, the first large‑scale cryptography dataset comprising more than two million question‑answer (QA) pairs that span all major sub‑fields of cryptography (see \cref{fig:teaser}). 
While existing datasets rely primarily on deterministic or expert-validated approaches to ensure reliability, they are often limited in scale and domain coverage. In contrast, we ensure high-quality data by only drawing content from peer-reviewed articles, such as journal articles, textbooks, or conference proceedings. 
To cover a wide variety of journals and conferences, and to consistently add metadata to the articles, we started with articles from the International Association for Cryptologic Research's (IACR) ePrint Archive \cite{iacrEprint}.
The eprints were then matched with peer-reviewed articles, and the peer-reviewed version was used for our dataset, along with the corresponding metadata from \cite{iacrEprint}, e.g., topics and sources(cf. \cref{sec:dataset} for more details). 
This allowed for a scalable and consistent annotation without incurring the cost of manual labelling (cf. \cref{sec:dataset} for more details).
In total, we collected \num{1388} documents and corresponding metadata, spanning textbooks \num{82}, books \num{216}, and conference proceedings \num{1088}. 
Since generating QA pairs for this large collection of cryptographic data cannot feasibly be achieved through manual annotation, we use an LLM, namely \texttt{DeepSeek-V3} \cite{liu2024deepseek}, to generate the QA pairs automatically (cf. \cref{sub:QA_generation} for more details on the QA pair generation). 
We ultimately received \num{1067690} pairs of questions and answers from cryptography, which we are included in our dataset \dataName~under \texttt{type = original}. By applying transformations such as paraphrasing, backward reasoning, and adversarial modifications to these pairs (cf. \cref{sec:results}), \dataName~\cite{code} includes in total $\approx 2.5M$ QA pairs.

We then use our dataset \dataName~to evaluate state‑of‑the‑art LLMs, both quantitatively and qualitatively, on the  capabilities\footnote{LLM \textit{ability} is the capacity to perform a task, like mathematical reasoning, while \textit{capability} is the effective and efficient application of that ability, often in a specific context, such as solving a set of math problems in a dataset.
The analyses and results presented in this paper reflect observed capabilities of LLMs rather than their underlying abilities. 
In particular, our results provide valuable insight into how models perform in applied settings, but cannot fully capture the intrinsic reasoning or generalisation capacities of the models. Establishing reliable methods to disentangle and measure fundamental abilities remains an open research challenge for LLMs in general.} most pertinent to cryptography: factuality, mathematical reasoning, consistency, referencing, backward reasoning and robustness. 
Our experiments reveal significant performance gaps, highlighting the need for continued research and specialised training to enable effective LLM integration in cryptography-related research and development.
For instance, our top-performing LLM, Qwen2.5-72B-Instruct \cite{qwen2025technicalreport}, only achieved approximately 70\% accuracy on general cryptographic questions across the entire dataset, while its performance on mathematically oriented questions was lower, around 60\%.
To complement our quantitative results, we report a qualitative evaluation ($N = 50$ participants) involving domain experts in cryptography with different expertise levels (BSc, MSc, and PhD).
This expert review sets a gold standard\footnote{In NLP, a gold standard is a dataset or annotation considered authoritative and of the highest quality, typically verified through human evaluation by experts, and used as the definitive reference for model assessment.} for human performance, assessing LLM response quality along dimensions such as correctness, completeness, and clarity. 
Our evaluation shows that state-of-the-art LLMs perform on par with trained M.Sc. students in cryptography, but lag behind domain experts, such as senior researchers, on tasks that require deep cryptographic knowledge, mathematical reasoning, and complex cross-domain inference. 
LLMs excel at conceptual and textbook-style questions but struggle with more advanced topics.

Overall, our evaluation enables us to compare and rank the current state-of-the-art LLMs in terms of their capabilities to correctly and completely solve cryptographic tasks. 
We expect our results to provide a good indication for researchers and non-experts on which LLMs to use for their cryptography-related questions.
Additionally, we expect that future LLMs can be tested against \dataName~using our implementation, included in \cite{code}. 
Moreover, \dataName~cannot only be used to test LLMs, but also to improve future LLMs by integrating \dataName~into the training set or by fine-tuning already existing models. 
For our proof concept, we follow the latter approach and fine-tune the model that performed best in our evaluation, namely, \texttt{Qwen2.5-72B-Instruct} \cite{qwen2025technicalreport}.
Our results show that the fine-tuned model outperforms its generic version by 7–13\% on different evaluation metrics.

\noindent\textbf{Our Contributions.} 
\begin{itemize}[leftmargin = *]
    \item We present \dataName, the first large-scale, domain-specific\linebreak question-answer dataset tailored to cryptography. The dataset comprises over two million high-quality QA pairs extracted and refined from cryptographic literature, along with their metadata, to facilitate nuanced analyses and the training of AI models across cryptographic subdomains.
    \item We perform a systematic evaluation of 15 state-of-the-art LLMs with respect to their capabilities in cryptographic reasoning.
    Across our experimental evaluation, the Qwen2.5-72B-Instruct model consistently achieves the highest performance across the majority of cryptographic tasks.
    Additionally, future LLMs can be tested using \dataName, providing a standardised framework for benchmarking and facilitating progress toward AI-assisted cryptography.
    \item We demonstrate that training (or fine-tuning existing) LLMs on \dataName~yields a significantly improved performance in cryptographic tasks compared to standard LLMs.
    For instance, fine-tuning Qwen2.5-72B-Instruct yields an improvement of 7–13\% across the evaluated metrics (cf. \cref{sec:results}).
\end{itemize}

\section{Related Work}\label{sec:relatedwork}
\subsection{Assistive Tools for Cryptography}

A variety of assistive tools have long underpinned cryptographic research and development. 
Formal verification frameworks such as ProVerif \cite{blanchet2018proverif}, Tamarin \cite{meier2013tamarin}, DY* \cite{dy21} and CryptoVerif \cite{blanchet2007cryptoverif} enable symbolic analysis of cryptographic protocols, while proof assistants like Coq \cite{chlipala2013certified} and Isabelle/HOL \cite{nipkow2002isabelle} support the construction and verification of cryptographic proofs.
Mathematical software such as SageMath \cite{denny2013sage} is also widely employed for number‑theoretic and algebraic computations.
Although these tools are powerful, they are highly specialised and narrowly scoped,
often demanding deep expertise to operate effectively on these tasks rather than for general‑purpose cryptography.
Recently, researchers in the cryptography community have begun to explore LLMs as assistive tools for a broader spectrum of tasks \cite{li2025constructing,ren2025towards}.
LLM assistants promise general-purpose support via natural-language interfaces, lowering the barrier to entry for non-experts and accelerating workflows for seasoned practitioners \cite{ren2025towards}. 
However, despite this potential, due to the lack of datasets, current LLMs have not yet been rigorously evaluated in cryptography, where correctness, precision, and contextual understanding are paramount. 
As such, their utility remains largely anecdotal, underscoring the need for dedicated evaluation benchmarks and metrics.

\subsection{Cryptography-related Datasets}

As the capabilities of LLMs have advanced rapidly, the demand for diverse and challenging QA benchmarks to train and evaluate them has grown \cite{rogers2023qa}.
Despite increasing interest in applying LLMs to cryptography, the field still lacks datasets that assess general cryptographic reasoning.
Existing resources focus on specialised tasks with limited samples rather than broad QA evaluation.
For example, the CipherDataset \cite{noever2023llmsforciphers} comprises 654 test cases of classical ciphers (e.g., Caesar, Vigenère) and evaluates LLM decryption capabilities.
Similarly, CipherBank \cite{li2025cipherbank}, containing 2,358 problems based on nine symmetric ciphers.
Maskey et al. \cite{maskey2025benchmarkinglargelanguagemodels} expanded this line of work with a cryptanalysis benchmark that includes both classical and modern schemes (AES, RSA) across 4,509 samples, demonstrating that LLMs struggle with any cipher requiring key inference or non‑trivial algebraic manipulation.
Other datasets -- CipherSpectrum \cite{Wickramasinghe_2025}, SecBench \cite{jing2025secbench}, USTCTFC2016 \cite{wang2017malware}, CIC‑IDS2018 \cite{sharafaldin2018toward}, CIC‑IoT2022 \cite{dadkhah2022towards}, Bot‑IoT \cite{koroniotis2019towards}, ISCX‑VPN2016 \cite{draper2016characterization}, and ISC‑Tor2016 \cite{lashkari2017characterization} -- focus on network‑security topics such as malware communications, intrusion detection, botnet traffic, and VPN/Tor traffic. 
Further datasets -  such as SecQA \cite{liu2023secqa} AICrypto \cite{wang2025aicrypto}, NYU CTF \cite{shao2025nyuctf}, CVE-Bench \cite{zhu2025cvebench}, PentestGPT \cite{deng2024pentest} and CyberGym \cite{wang2025cybergym}- benchmark some LLM capabilities but remain limited in size ($<1$K manual samples).

While cryptography‑specific QA resources remain scarce, the broader mathematical‑QA domain has seen substantial development. 
Early work by Kushman et al. introduced 514 algebra word problems sourced from student queries on algebra.com \cite{kushman2014learning}. 
Roy et al. contributed the IL dataset with 1,404 elementary‑level arithmetic problems and the CommonCore dataset \cite{roy2016solvinggeneralarithmeticword} containing 600 multi‑step problems.
The large‑scale AQuA‑RAT dataset offers over 100,000 multiple‑choice questions with rationales, partially derived from standardised tests (GMAT, GRE) and expanded via crowdsourcing \cite{ling2017program}.
To address data‑quality concerns, MathQA \cite{amini2019mathqa} refined 37,000 math problems with formal operation annotations.
Finally, GSM8K \cite{cobbe2021training} provides 8,500 middle‑school‑level math word problems accompanied by natural‑language solutions.

Although all of these datasets represent important advances, they are highly specialised, typically concentrating on particular cipher families or narrowly defined tasks. 
Moreover, most operate at purely syntactic, symbolic or mathematical levels, without addressing the broader dimensions of modern cryptography.
Consequently, a substantial gap remains for a comprehensive, large‑scale cryptography QA dataset that enables both knowledge‑based and computation‑based evaluation across a wide spectrum of cryptographic concepts.

\section{The CryptoQA Dataset}\label{sec:dataset}

The collection of high-quality data corpora is essential for effective training and evaluation of AI models. 
Prior research \cite{gunasekar2023textbooks,li2025scilitllmadaptllmsscientific, abdin2025phi, zhao2023beyond}
emphasised that corpora with the characteristics of an exemplary textbook for human learners tend to yield superior performance compared to unstructured, heterogeneous, or domain-agnostic text corpora.
Over the years, the academic community has produced an extensive collection of scientific literature and textbooks that serve as authoritative resources for human learning (cf. \cref{sec:relatedwork}). 
Motivated by this, we have constructed a large-scale, domain-specific textual corpus centred on cryptography.
The curation process involves the targeted collection of content from trusted scholarly sources, followed by rigorous preprocessing to ensure that the resulting dataset is suitable for downstream tasks such as LLMs' evaluation and training.

\subsection{Dataset Collection}

To ensure the scientific quality and consistency of our data, we sourced articles from the \emph{IACR ePrint archive} \cite{iacrEprint}, which provides access to a wide range of cryptography publications.
We then identified their corresponding peer-reviewed versions (if available) and included only the peer-reviewed versions in our collection.
To ensure linguistic consistency and high data quality, we limited the dataset to English-language publications, excluding non-English works and specific publication types, such as university theses and commemorative publications. 
The data acquisition consisted of two main steps: 
(1) Download of the full document (e.g. pdf), and (2) retrieval of metadata for each publication.
This resulted in a total of \num{1388} documents and corresponding metadata, spanning \texttt{textbooks} (\num{82}), \texttt{books} (\num{216}), and \texttt{conference proceedings} (\num{1088}).
In total, we downloaded \num{34998} individual chapters with the corresponding metadata (\num{1200} textbook chapters, \num{2855} book chapters, and \num{30943} proceeding chapters).
This corpus forms the backbone of our dataset construction pipeline and enables the generation of cryptography-specific question-answer pairs at scale.

\subsection{QA Generation} \label{sub:QA_generation}
The generation of QA pairs at the scale required for this work cannot feasibly be achieved through manual annotation. 
In addition, the use of LLMs for dataset creation has recently gained momentum due to their scalability and increasing fluency in domain-specific content generation \cite{alismail2025survey}.
Consequently, we adopt a fully automated pipeline leveraging LLMs for QA data generation.

\noindent\textbf{Model Selection.} To produce high-quality and contextually relevant QA pairs, we utilised \texttt{DeepSeek-V3} \cite{liu2024deepseek} as our base model.
We selected \texttt{DeepSeek-V3} as the generative model due to its favourable balance of performance, cost, and openness. \texttt{DeepSeek-V3} is a \num{671}-billion-parameter sparse Mixture-of-Experts (MoE) model that employs a transformer-based architecture with Multi-Head Latent Attention. It achieves competitive results on English benchmarks, outperforming or matching proprietary models in key areas like code generation and mathematical reasoning, and surpassing them in other tasks \cite{liu2024deepseek}.
In addition, \texttt{DeepSeek-V3} supports a maximum context length of 64k tokens (input + output), with an output token cap of 8k (cf. \cref{tab:llm_comparison} for a comparison with other LLMs). This high-context window allows us to process complete chapters with minimal truncation, preserving semantic coherence during generation.
Most importantly, we selected \texttt{DeepSeek-V3} as the ground-truth answer generator in our evaluation pipeline since it consistently produces concise and precise responses \cite{liu2024deepseek},
which are particularly suitable for use as reference answers in automatic evaluation frameworks; when applying metrics such as BLEU and ROUGE (cf. \cref{sec:evaluation}), the evaluation process typically relies on the overlap of specific n-grams between the generated candidate response and the reference. If the ground-truth answers are unnecessarily verbose or contain redundant information, they may introduce noise into these n-gram–based comparisons, allowing evaluation metrics to more accurately capture the quality of candidate models’ outputs. However, in cases where answers inherently require extended explanations—such as formal proofs—\texttt{DeepSeek-V3} prioritises content completeness over strict adherence to formatting constraints, through its 'scoring policy' and 'soft constraints', producing longer outputs to preserve correctness and semantic fidelity \cite{liu2024deepseek}\footnote{Reliance on a single model for ground-truth generation introduces a potential bias toward the linguistic and stylistic patterns of \texttt{DeepSeek-V3}, which should be taken into account when interpreting evaluation outcomes. Hence, in \cref{sec:results}, we report this source-model bias and further discuss this limitation.}.

\noindent\textbf{Querying Strategy.}
In-context learning (ICL) \cite{garg2022can} is a prompting-based paradigm wherein examples of a desired task are embedded directly into the prompt to guide the output generation of an LLM without updating its internal parameters. This technique narrows the solution space by demonstrating a target behaviour, effectively enabling task conditioning within a single forward pass of the model \cite{brown2020language}. ICL typically takes the form of N-shot prompting, where $N$ examples are provided to steer the model's behaviour toward a specific output format or style.
In our work, we adopt zero-shot prompting to generate structured QA pairs from the backbone dataset (i.e. the full chapters). 
Zero-shot prompting is the minimal case of ICL, where the 'context' consists only of natural language task instructions, e.g. a format schema.
In addition, our generation approach is loosely inspired by \emph{retrieval-augmented generation (RAG)} \cite{lewis2020retrieval} techniques. In classical RAG, a retrieval system first selects a small number of relevant documents based on a query, and these are then provided as context to a generative model. This grounding improves factual accuracy and mitigates hallucination by anchoring generation to source material.
In our work, we adapt this idea by processing chapters as grounding sources for the generation of QA pairs, 
making our approach deterministic, as opposed to retrieval-selective. It also ensures that QA pairs are generated uniformly across the corpus, promoting balanced topic representation and enhancing the dataset’s utility for fine-grained analysis and evaluation of cryptographic reasoning. 
In addition, the prompt requested at least 25 question–answer pairs per chapter, thereby mitigating under-generation caused by vague or poorly specified instructions.

\noindent \texttt{DeepSeek-V3} responses are returned in this \emph{Markdown} format with \LaTeX~syntax (cf. \cref{apx:examples} and \cref{apx:sample_responses} for more examples): 
\begin{shadedquotation}\noindent\textbf{QA example:}\\
    \text{**Q**}: Show that if one-way functions exist, then \verb|\(\mathsf{P} \neq \mathsf{NP}\)|.\\
    \text{**A**}: If \verb|$\mathsf{P}=\mathsf{NP}$|, then for any polynomial-time function \verb|$f$|, the language \verb|$L_f=\{y \mid \exists x:\, f(x)=y\}$| lies in \verb|$\mathsf{P}$|. By standard search-to-decision reductions, a polynomial-time decision algorithm for \verb|$L_f$| implies a polynomial-time algorithm that recovers a preimage \verb|$x$| for any \verb|$y$|, making \verb|$f$| efficiently invertible. Hence, no polynomial-time function can be one-way under \verb|$\mathsf{P}=\mathsf{NP}$|, and thus the existence of one-way functions implies \verb|$\mathsf{P}\neq\mathsf{NP}$|.
\end{shadedquotation}

Despite prompting for a consistent format, minor deviations were present in the outputs. To ensure comprehensive coverage and account for \texttt{DeepSeek-V3}'s inconsistency, we therefore assert the number of generated QA pairs per chapter and re-queried the LLM when fewer pairs were produced. The same process was applied for format inconsistencies to avoid ill-structured responses by employing regular expressions.

\subsection{Dataset Statistics}

This generation yielded 1,067,690 QA pairs after removing duplicates and near-duplicates (e.g., arising from LLM-generated variations of similar content).
Each QA pair is augmented with the following metadata (cf. \cref{apx:stats} for further statistics):
\begin{itemize}[leftmargin = *]
    \item \textbf{Topics:} We used the IACR ePrint Archive Metadata Harvesting \cite{iacrEprint} feature to set the different author-specified topics. The predefined topics are: \texttt{Applications}, \texttt{Cryptographic Protocols}, \texttt{Foundations}, \texttt{Implementation},  \texttt{Secret-key Cryptography},\\
    \texttt{Public-key Cryptography}, and \texttt{Attacks and Cryptanalysis}. A few QA pairs were assigned \texttt{Other} when no topic was available.

    \item \textbf{Categories:} We then specified categories - \texttt{Word} (assigned to 863,074 QA pairs) and \texttt{Math} (assigned to 204,616 QA pairs) - 
    to describe the nature of the question based on the presence of mathematical calculations (specified by \LaTeX~symbols).

    \item \textbf{Sources:} We report the source's DOI. Each file was further sorted into one of three categories: \texttt{Textbooks}, \texttt{Books}, and \texttt{Conference Proceedings}. Although textbooks and books contributed significantly to the backbone corpus, their representation in the final dataset is somewhat lower. This under-representation is primarily a consequence of the filtering phase, during which duplicate and near-duplicate QA pairs were removed. Since textbooks and books tend to cover foundational and overlapping material (e.g., definitions of standard cryptographic primitives or the explanation of classical algorithms), they naturally produced more redundant questions compared to more diverse and novel content found in conference proceedings. This distribution reflects the dominant role of conference proceedings in contemporary cryptographic research, both in terms of volume and diversity of content.

    \item \textbf{Types:} We further indicate whether the QA pair is \texttt{Original} (i.e. directly extracted from the source, $\approx$ 1.2M samples) or modified, e.g. \texttt{Paraphrased} ($\approx$ 1.2M samples) or injected with \texttt{Adversarial} ($\approx$ 1K samples) or \texttt{Backward} examples ($\approx$ 1K samples) (cf. \cref{sec:results} for further details).
\end{itemize}

This annotation allows for targeted training and evaluation of models on different question modalities.
Accordingly, we partitioned the \dataName dataset into training and testing subsets using an 80:20 ratio. During the split, particular care was taken to preserve the metadata distribution across the two subsets (cf. \cref{apx:stats}).

\section{Evaluation of Existing LLMs}\label{sec:evaluation}
In this section, we describe the experimental framework employed to assess the cryptographic reasoning capabilities of state-of-the-art (SOTA) LLMs. Specifically, we detail how models were evaluated using the \dataName dataset, including the candidate models and the metrics used to quantify performance, enabling insight into the models' specific strengths and limitations in cryptography.

\subsection{LLM Candidates} 
We evaluate $15$ LLM candidates for benchmarking, with open-source and closed-source LLMs. The choice follows CipherBank's \cite{li2025cipherbank} candidate LLMs with certain models substituted by recent versions and complemented by the inclusion of more recent state-of-the-art LLMs (i.e. the most queried on OpenRouter \cite{openrouter}, cf. \cref{apx:llm-comparison} for more details):

\begin{itemize}[leftmargin = *]
    \item \textbf{Open-Source LLMs:}  Mistral AI’s \texttt{Mixtral-8x22B Instruct} \cite{jiang2024mixtral}, 
    Alibaba’s \texttt{Qwen2.5-72B-Instruct} \cite{qwen2025technicalreport}, \texttt{DeepSeek-V3} \cite{liu2024deepseek}, and \texttt{DeepSeek-R1} \cite{deepseekr12025}.
  \item \textbf{Closed-Source LLMs} (API access):  Meta’s \texttt{Llama-3.1-70B-}\linebreak \texttt{Instruct} and \texttt{Llama-3.3-70B-Instruct} \cite{meta2024llama3.1}, OpenAI’s \texttt{GPT-4o} and \texttt{4o-mini} \cite{hurst2024gpt}, OpenAI's \texttt{o1} and \texttt{o1-mini} \cite{openai2024openaio1card}, OpenAI's \texttt{GPT-5} \cite{openai2025gpt5}, and X-AI's \texttt{grok-4} \cite{xai2025grok4},
    DeepMind’s \texttt{Gemini-}\linebreak \texttt{2.5-Pro} and \texttt{Gemini-2.5-Flash} \cite{comanici2025gemini25pushingfrontier}, and Anthropic’s\linebreak \texttt{Claude-Sonnet3.5} \cite{anthropic2024claude35sonnet}.
\end{itemize}

All models were evaluated with a fixed temperature of 0.2 and top-p sampling set to 1.0 to encourage deterministic behaviour (further information can be found in \cref{tab:llm_comparison}). The evaluation was conducted on the test split of \dataName and the LLMs provided by OpenAI's OpenRouter API \cite{openrouter}.

\subsection{Evaluation Metrics} 
For quantitative assessment, we report the following evaluation metrics:

\begin{itemize}[leftmargin = *]
    \item \textbf{BLEU} evaluates how closely a generated text matches one or more reference texts \cite{papineni2002machine}. 
    BLEU measures n-gram precision, i.e., how many contiguous sequences of n words in the candidate output also appear in the reference text (i.e. the \dataName answer).  
    BLEU does not consider semantic meaning and penalises valid paraphrases that use different wording.
    In this work, we utilise the default NLTK configuration \cite{bird-loper-2004-nltk}.

    \item \textbf{ROUGE} compares overlaps between the generated and reference texts \cite{lin2004rouge}. 
    It focuses on recall rather than precision, though some variants consider both:
    ROUGE-N measures the n-gram recall e.g., ROUGE-1 for unigrams, ROUGE-2 for bigrams.
    ROUGE-L is based on the longest common subsequence between the candidate and reference text. 
    ROUGE is good for measuring coverage of reference content. 
    However, like BLEU, it is based on surface form overlaps and does not handle paraphrasing or semantics well. In this work, we calculated ROUGE-1, ROUGE-2, ROUGE-L and ROUGELsum (i.e. a summarisation-oriented variant of ROUGE-L that treats the whole answer as one sequence rather than sentence by sentence). 

    \item \textbf{Perplexity} \cite{jelinek1977perplexity} is a standard measure of how well a language model predicts a given sequence of text. It evaluates the model’s uncertainty when generating tokens.

    \item \textbf{METEORscore} \cite{banerjee2005meteor} is a text evaluation metric that goes beyond exact word matches by considering stems, synonyms, and order, giving a more human-aligned quality score.

    \item \textbf{BERTscore} measures semantic similarities between texts using deep contextual embeddings from BERT \cite{devlin2019bert}. 
    Then, the cosine similarity is calculated between tokens, and the best matches are aggregated to produce a final score (precision, recall, and F1 variants).

    \item \textbf{LLM-as-judge} \cite{dietz2025llm} uses an LLM to compare outputs or score them against rubrics (e.g. a 10-point scale). This approach offers a scalable and flexible approach to evaluating generated text. They provide consistent, semantically informed judgments and reduce reliance on costly human evaluation \cite{raina2024llm}. 
    We used our base model DeepSeek-V3 as well as other models as judges (cf. \cref{sec:results} and \cref{apx:other-judges}). 
    
    \item \textbf{Human evaluation} refers to the process of systematically assessing model outputs by domain experts to establish a reliable benchmark for performance \cite{wang2023aligning}. In our study, cryptographers, spanning different expertise levels, manually evaluated randomly sampled questions drawn from the dataset distribution. 
\end{itemize}

Finally, while each LLM evaluation metric is effective for targeted comparisons, they should be viewed as complementary rather than definitive. To ensure a balanced assessment, we report all the above metrics to assess lexical overlap (BLEU, ROUGE1 ROUGE2, ROUGEL, ROUGESum, and METEOR), semantic embedding similarity (BERT-precision, BERT-recall, and BERT-F1), fluency and uncertainty (perplexity), as well as holistic human and human-like (LLM-as-judge) evaluations in \cref{apx:comprehensive-results}. In \cref{sec:results}, we mainly focus on four key metrics (i.e. ROUGEL, BERT-F1, LLM-as-judge, and human evaluation) to maintain clarity, highlighting the most informative aspects of model performance.

\section{Results}\label{sec:results}

In this section, we present and analyse the results of our evaluation of the LLM candidates on the \dataName dataset with our evaluation metrics, providing a multifaceted view of the models’ capabilities.

Firstly, we report our evaluation metrics for our candidate LLMs under different setups, by comparing the LLMs generated responses to our \dataName's answers to: (1) the entire test set, (2) the different stratifications of the test set accross the different metadata, (3) the paraphrased questions, (4) the test set given the source for referencing, (5) the backward reasoning subset, and (6) the adversarial subset.
Secondly, we report the results of our user (expert) survey for: (i) a gold standard (i.e. human baseline) by comparing the human answers vs the LLM answers on our subset with different expertise levels, (ii) a human evaluation of the dataset quality, and (iii) a human evaluation for the LLM responses.
All metrics range from 0 to 1, where higher values are better (except for perplexity, where values are normalised, the lower the better). See \cref{apx:sample_responses} for sample examples.

\subsection{Quantitative Results}
Leveraging the dataset's rich metadata, which enables detailed structural and semantic analysis, we evaluated the candidate LLMs across a set of capabilities identified as critical for cryptography.
These evaluations were performed by partitioning the dataset according to its metadata attributes, allowing fine-grained assessment of model performance across distinct tasks. For example, isolating math questions from word questions allows us to assess an LLM's mathematical reasoning capabilities independently, rather than conflating them with its broader cryptographic performance. Accordingly, we partitioned the dataset using all our metadata, enabling systematic evaluation across the full range of annotated dimensions.

We assess all capabilities on the \dataName~test split (with \textit{type 'original'}), except for paraphrasing consistency, backward reasoning, and robustness.
For the latter three, we crafted some data samples by modifying some of our dataset samples and labelling them with a different \textit{type} attribute.
For paraphrasing consistency (i.e. \textit{type 'paraphrased'}), we changed our entire dataset's questions using varied phrasings with the help of our LLM base model \texttt{DeepSeek-V3}  (cf. \cref{apx: paraphrasing} for more details).
For backward reasoning (i.e. \textit{type 'backward'}) and robustness (i.e. \textit{type 'adversarial'}), we generated a subset of \num{2000} handcrafted samples by modifying some of our dataset samples.
As is common in prior work \cite{farzand2024out, hinkin1998brief, boateng2018best}, the samples for the latter two subsets were chosen and independently refined by different experts.

In the following, we briefly explain the different capabilities and the quantitative results achieved by the LLMs. All capabilities below were tested on the \dataName~test split (with \textit{type 'original'}) unless otherwise stated.


\begin{figure}[h]
    \centering
    \begin{subfigure}[b]{0.49\textwidth}
        \centering
        \includegraphics[width=\textwidth]{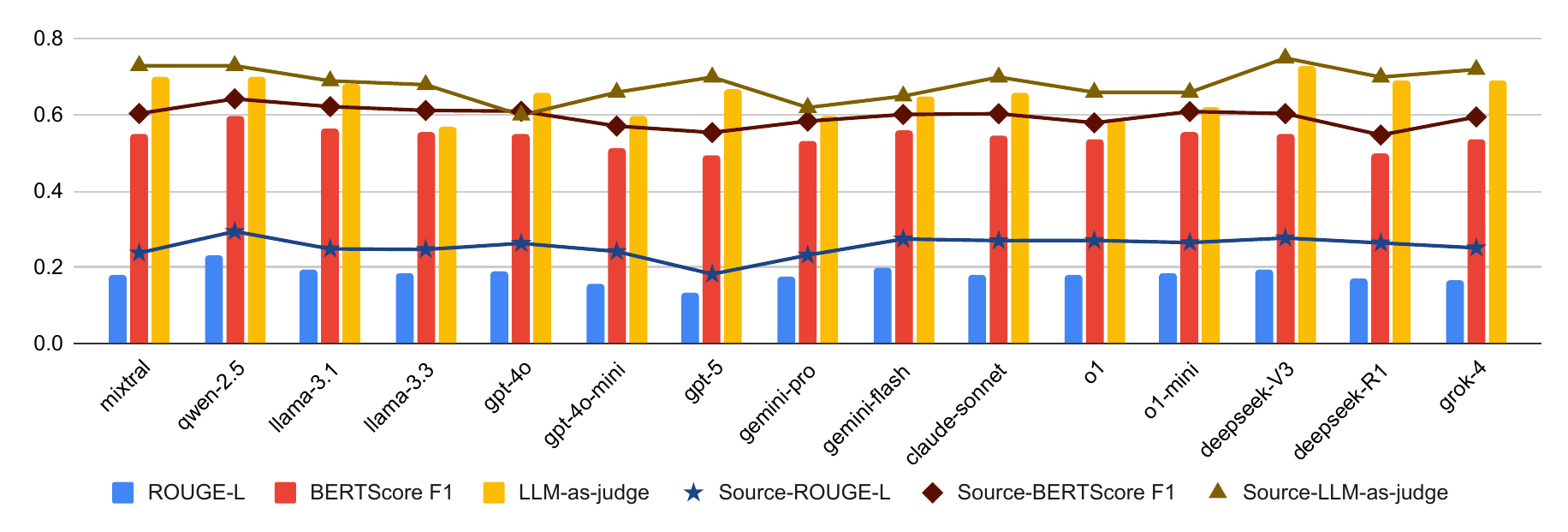}
        \caption{Original dataset (without DOIs: bars, with DOIs: lines)}
        \label{fig:original}
        \label{fig:sourced}
    \end{subfigure}
    \begin{subfigure}[b]{0.49\textwidth}
        \centering
        \includegraphics[width=\textwidth]{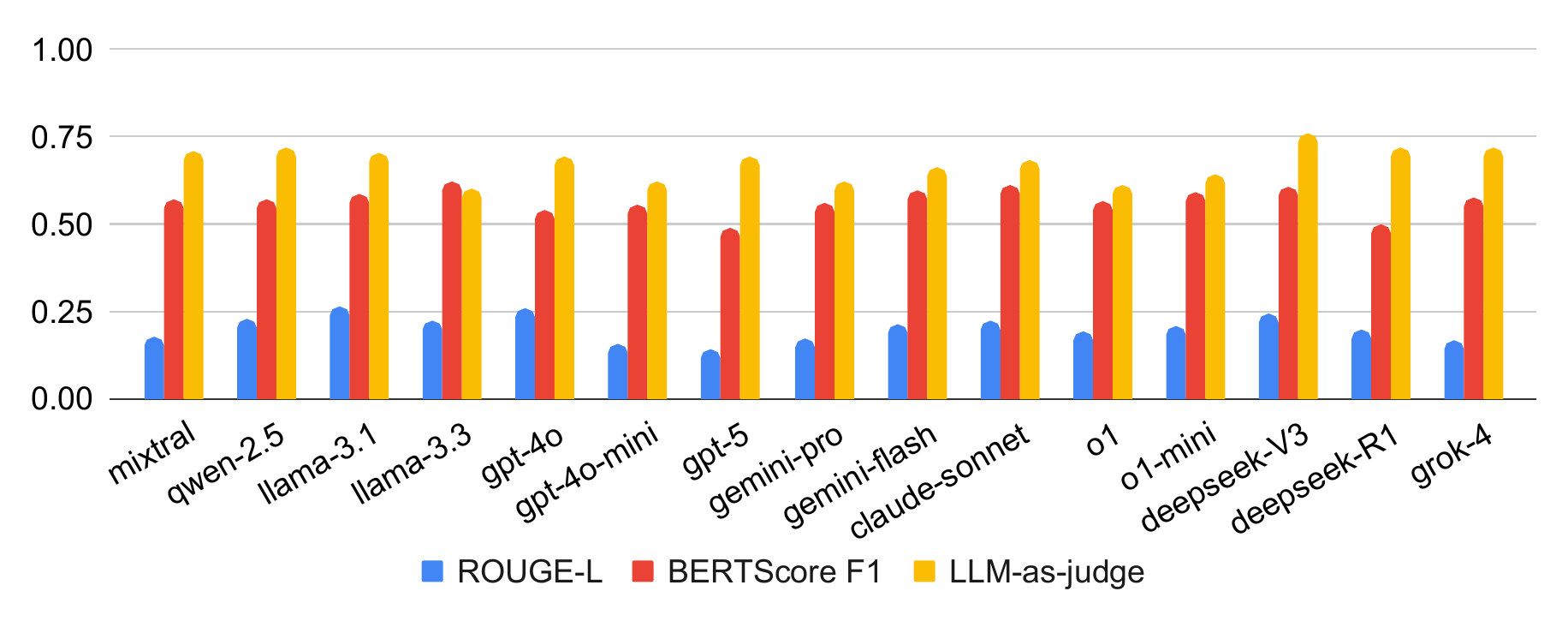}
        \caption{Paraphrased dataset}
        \label{fig:paraphrased}
    \end{subfigure}

    \caption{The main evaluation metrics when prompting our candidate LLMs with the original dataset versus the source (as DOI) along the original dataset (\Cref{fig:original}), and the paraphrased dataset (\cref{fig:paraphrased}). More comprehensive figures showing other evaluation metrics can be found in \cref{apx:comprehensive-results}.}
    \label{fig:main-res}
\end{figure}


\noindent\textbf{Factuality} refers to the model’s ability to produce outputs that are factually correct, verifiable, and consistent with established domain knowledge or specified sources \cite{wang2024factuality}. This dimension of evaluation is particularly critical in domains such as cryptography, where accuracy is essential and errors can have significant implications. \\
We evaluate the LLM candidates' responses against our \dataName answers. On the test set, as shown in \cref{fig:original}, all evaluated LLMs remain far from achieving comprehensive coverage or accuracy across all questions, indicating substantial gaps in their overall cryptographic reasoning.
However, \texttt{Qwen2.5-72B-}\texttt{Instruct} consistently (but slightly) outperforms other LLMs (significance tested with a Friedman test across all 15 models ($\chi^2(14) = 67.3$, $p < 0.001$), followed by a Nemenyi post-hoc pairwise comparisons to identify specific differences between models ($p < 0.05$ for Qwen)), followed by \texttt{Deepseek-v3} and \texttt{GPT-5} on average.
    \texttt{Qwen2.5-72B-Instruct} is explicitly trained on a mixture of high-quality web data, textbooks, scientific publications, and multilingual sources, which likely includes substantial coverage of mathematical and cryptographic content. Its instruction-tuning process further enhances domain alignment by optimising the model for structured, step-by-step reasoning, which is a skill crucial for cryptographic reasoning. Additionally, its robust handling of non-English material (e.g. Chinese and Arabic textbooks that integrate English cryptographic terminology) broadens its exposure to diverse cryptographic explanations, strengthening factual recall and conceptual generalisation compared to models more narrowly tuned on general-purpose English corpora\footnote{This cross-linguistic behaviour can be both beneficial and problematic. While retaining native or domain-specific terms can enhance precision, unintended language switching may impede comprehension. For instance, the \texttt{Qwen-2.5-72b-instruct} model occasionally mixes languages in responses: \textit{'The ancient Egyptians occasionally used \<تبديل الرموز الهيروغليفية> (hieroglyphic substitution of symbols) to obscure meaning [...].'} where it is good to have the original Arabic term,
    and
    \textit{'This design ensures that the internal state remains unpredictable and \begin{CJK*}{UTF8}{gbsn}难以被逆向工程，从而提供强大的安全保障。\end{CJK*}'} (translating to: “and difficult to reverse-engineer, thus providing strong security guarantees”) where it is hard to understand the meaning without translation.
    This behaviour likely stems from the model’s multilingual training data and token prediction mechanisms, compounded by the use of English technical terms in many Chinese cryptography textbooks (not included in our dataset). These observations underscore the importance of careful prompt design and explicit language specification, especially in domain-specific tasks such as cryptography, where unexpected language shifts can reduce clarity and complicate evaluation.}.\\
    We also observe a clear source-model bias: Our empirical results indicate that \texttt{DeepSeek-V3} remains in the top-3 LLMs in our LLM-as-judge evaluation setting. This performance disparity can be attributed, in part, to the fact that \texttt{DeepSeek-V3} was used as the judge, introducing a potential source-model bias. This phenomenon aligns with findings by Panickssery et al. \cite{panickssery2024llm}, 
    who demonstrate that LLMs exhibit a notable preference for their own outputs during self-evaluation tasks (e.g. LLM-as-judge). Such self-preferential behaviour raises important concerns regarding the objectivity and fairness of LLM benchmarking when the model used for evaluation is also the source of the reference data. We further test with different LLMs as judges and found the same behaviour (same rank of LLMs) in \cref{apx:other-judges}, hence supporting our results.\\
    Finally, it is important to consider data-source alignment effects. Since the dataset was generated by \texttt{DeepSeek-V3}, differences in token distribution, phrasing, or solution patterns may unintentionally favour models that share similar training corpora characteristics. Therefore, while Qwen’s superior scores suggest stronger domain-specific reasoning, the relative performance gaps may partly reflect compatibility with the dataset’s style rather than purely intrinsic cryptographic ability.\\
    Nonetheless, our results remain consistent with prior independent research (cf. \cref{apx:llm-comparison}). In particular, contemporary benchmark studies identify \texttt{Qwen2.5-72B-Instruct} and \texttt{DeepSeek-V3} models as the current state-of-the-art on multitask language understanding and mathematical tasks that most closely approximate the logical, analytical, and symbolic-reasoning demands inherent to cryptography (cf. \cref{apx:llm-comparison}). Thus, the superior performance of these models in our experiments aligns with their established strengths in domains requiring advanced reasoning capabilities.
    In parallel, recent studies reveal that a growing proportion of research publications are being written with LLM assistance \cite{liang2024mapping, liang2024monitoring}, which naturally appears in our dataset. Although the adoption of LLMs for manuscript generation in cryptography remains relatively limited \cite{liang2024mapping}, transparent reporting is particularly crucial when evaluating models on data that may be partially or fully synthesised. Nonetheless, as LLMs continue to advance in mathematical reasoning, such trends may shift, underscoring the need for clearer standards of disclosure in future research.\\
    Furthermore, our evaluation shows that several LLM families and their reduced-scale counterparts, such as o1 vs. o1-mini, GPT-4o vs. GPT-4o-mini, and Gemini-pro vs Gemini-flash, exhibit on-par performance. This observation suggests that model size alone is not a decisive factor for domain-specific performance once instruction tuning and optimisation are applied (cf. \cref{tab:llm_comparison} for comparison). Similar observations have been reported in other domains \cite{wang25slm} where the smaller variants present clear advantages: they are computationally more efficient, require significantly fewer resources for inference, and offer faster response times while maintaining comparable accuracy.

\begin{figure}[h]
    \centering
    
    \includegraphics[width=0.49\textwidth]{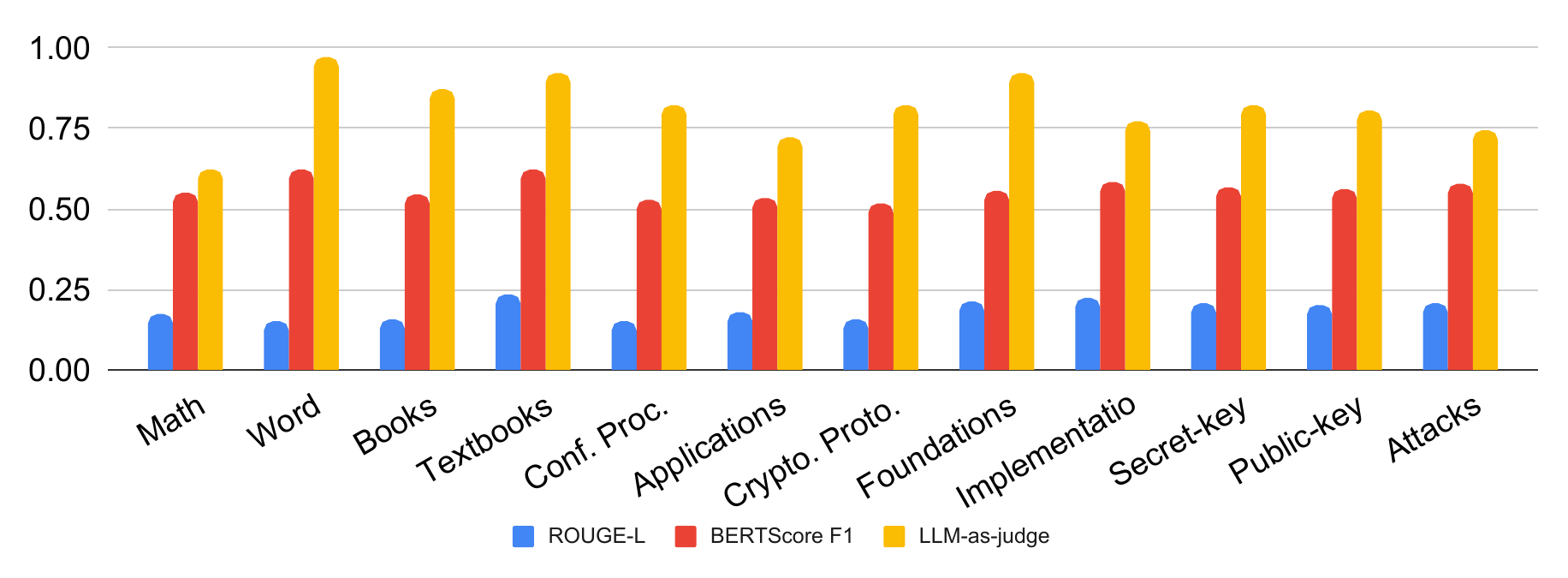}
    \caption{The averaged scores over all LLMs for the different metadata attributes. Further metrics can be found in \cref{apx:comprehensive-results}.}
    \label{fig:stratification}
    
    \end{figure}

\noindent\textbf{Mathematical Reasoning} refers to the ability of an LLM to understand, manipulate, and derive conclusions using mathematical concepts \cite{ahn2024large}, encompassing both \emph{numerical calculations} \cite{roy2016illinois, roy2016solvinggeneralarithmeticword} and \emph{symbolic representations} \cite{patel2021nlp}. This capability ensures that LLMs can handle both \emph{quantitative} and \emph{formal reasoning} tasks, which are critical for cryptography \cite{meier2013tamarin, dy21}.\\
Our results in \cref{fig:stratification} (on the same test set as before) indicate that LLMs exhibit lower performance on mathematical questions compared to word-based problems, which aligns with prior observations reported in the literature \cite{amini2019mathqa, zhang2024mathverse, zhang-xiong-2025-backmath}. A more fine-grained analysis using different evaluation metrics reveals nuanced behaviour: lexical similarity metrics tend to yield higher scores on mathematical questions, likely due to the restricted and highly structured vocabulary inherent to mathematical expressions. In contrast, semantic similarity metrics  provide superior evaluation on word problems, reflecting the greater variability in language and expression in these questions, where superficial lexical overlap is less indicative of correct reasoning.\\
    We further evaluated the performance of LLMs across different dataset splits stratified by source and topic to investigate potential variations in model behaviour. The results indicate that questions originating from \texttt{textbook} sources and those belonging to \texttt{foundations} topics consistently yield the highest performance. This trend is likely attributable to the greater exposure of the LLMs to similar content during pretraining, as textbooks and foundational materials are more prevalent and standardised. Consequently, the models are better able to recognise and reproduce patterns associated with these sources and topics, leading to superior performance relative to less common or more specialised domains. Here, we report the average performance scores across all evaluated LLMs, as the observed differences between individual models were not statistically significant (tested with a Friedman test ($\chi^2$(14) = 12.7, $p$ = 0.56)).

\noindent\textbf{Consistency} refers to the extent to which a language model produces stable and logically coherent outputs when presented with semantically equivalent inputs or when asked to repeat a task under similar conditions \cite{wu2025estimating}. This is particularly important for cryptography, where reasoning often depends on precise logical structures and deterministic procedures. We assess two types of consistencies. Firstly, \emph{paraphrasing consistency} refers to the ability of a language model to provide logically aligned answers when the same question is posed using different phrasings \cite{yu2025training}. High paraphrasing consistency indicates that the model captures the underlying meaning of the input rather than being influenced by linguistic differences. 
    Hence, we presented the same question or prompt in multiple phrasings and examined whether the LLM returned consistent answers (cf. \cref{apx: paraphrasing} for examples). The paraphrased questions we changed were generated from our original data set with the help of our LLM base model \texttt{DeepSeek-V3}  (cf. \cref{apx: paraphrasing} for more details).    
    Secondly, \emph{self-consistency} refers to the model's ability to generate stable and non-contradictory outputs when queried multiple times with the same prompts \cite{ahmed2023better}. We tested self-consistency by sampling multiple completions from the same prompt (of the non-paraphrased split of the dataset) and compared the results.\\
    We observe that all SOTA LLMs demonstrate high self-consistency, meaning that they generate highly similar outputs when presented with identical inputs across repeated trials (e.g. two trials in our case, given our large-scale dataset), as well as high paraphrasing consistency, indicating that their responses remain stable when semantically equivalent questions are rephrased. 
    This behaviour suggests that these models possess strong language understanding and semantic generalisation capabilities.\\
    As shown in \cref{fig:original,fig:paraphrased}, the results for the original dataset and the paraphrased dataset are on-par. Quantitatively, we conducted pairwise statistical comparisons of the mean evaluation metrics between (1) answers to the original vs the paraphrased dataset for paraphrasing consistency and (2) answers of two trials of the original dataset (for self-consistency). Since each sample in the first dataset corresponded directly to a sample in the second, we applied paired statistical tests to evaluate differences in average metric scores. Specifically, we used standard paired t-tests under the assumption of normally distributed score differences and Wilcoxon signed-rank tests as a non-parametric alternative when normality could not be assumed. For all metrics, the resulting p-values were above the 0.05 significance threshold, indicating that the differences in mean scores between the two datasets were not statistically significant. These results suggest that the datasets exhibit comparable overall text quality and evaluation behaviour across metrics.
    This high degree of stability across repeated and paraphrased inputs not only reflects the robustness of the learned semantic space but also indicates that SOTA LLMs exhibit a form of contextual invariance—the ability to preserve reasoning and output fidelity despite superficial linguistic variation.\smallskip

\noindent\textbf{Confidence} refers to the model's internal estimate of how likely its generated output is correct, appropriate, or aligned with the given prompt \cite{liu2025uncertainty}. When a language model generates a response token-by-token, it assigns a probability distribution over the vocabulary at each step. The log-probability of the selected token represents the model's internal belief in the correctness or appropriateness of that token given the current context. Aggregating these log-probabilities across an answer allows us to approximate the model's overall confidence in a given output (cf. \cref{apx:confidence} for an example).
\\
    While log-probability provides a theoretically grounded measure of a language model’s uncertainty/confidence, it is important to recognise that many LLMs are poorly calibrated: they can assign high confidence to incorrect outputs or low confidence to correct ones. Prior studies \cite{kadavath2022language, jiang2021can} have documented this tendency, noting that LLMs frequently exhibit overconfidence, particularly in tasks that require multi-step reasoning or retrieval of factual knowledge outside the distribution of their pretraining data. \\
    For evaluation,  we compute perplexity and confidence for each LLM wherever token-level log-probabilities are available, either through open-source implementations or when explicitly provided by the API provider (cf. \cref{apx:confidence} and \cref{apx:comprehensive-results}). Perplexity is calculated as the exponentiation of the mean per-token cross-entropy, which is defined as the negative log-likelihood of each token in the reference sequence given the preceding context.
    Confidence is derived from these log-probabilities by transforming the negative cross-entropy into a normalised score in the range $[0,1]$.\\
    For models where token-level probabilities are unavailable - as is common with closed-source APIs - we approximate perplexity by a local, open-source autoregressive model (i.e. our base model \texttt{DeepSeek-V3}). The generated text from the target model is input into the local model to compute per-token log-probabilities and cross-entropy, which are then used to estimate both perplexity and a surrogate confidence score. While these approximations may not exactly match the original model’s internal likelihoods, they provide robust relative estimates for assessing output quality, ranking responses, and comparing uncertainty across prompts. \\
    Furthermore, we prompted the models to self-report their confidence for each generated output. However, we observed that nearly all models consistently reported confidence values exceeding 90\%, regardless of the correctness of their responses. This overconfidence highlights the necessity of evaluating confidence in combination with other measures, such as factuality or consistency.

\noindent\textbf{Referencing} is the ability of language models to generate or utilise intra-textual and inter-document references \cite{byun2024reference}. Intra-textual references refer to the model’s ability to correctly reference earlier parts of a document, such as linking a claim to a previously stated theorem or algorithm or resolving expressions like 'as mentioned above' or 'as previously stated'. Inter-document references involve citing external scholarly works, standards, or protocols, which are essential for supporting claims and situating contributions within the broader body of cryptographic literature.
    These abilities remain challenging to achieve for LLMs, especially in underexplored areas \cite{fuad2024llm}. 
    Unlike prior work, which often actively filters such references out \cite{li2025scilitllmadaptllmsscientific}, we include them, e.g., 'What is the key idea behind Proposition 1?' or 'What simulation tool was used to evaluate the proposed methods?' or '[...] as seen in Table 2'.
    We then evaluate these questions via retrieval-augmented prompts, following \cite{zhang2025reference}, by adding the source's DOI to the question.
    \\
    As shown in \cref{fig:sourced}, including the source DOI in the prompts significantly enhanced the performance of the LLMs, particularly in cases with explicit references, such as 'as shown in Theorem 1' or 'according to X et al.'. By providing the model with a direct pointer to the original source, it can better contextualise the content, retrieve relevant factual information, and disambiguate numerical or tabular data, thereby improving alignment with the primary literature and reducing hallucinations. In contrast, without DOI prompting, some of our candidate models frequently responded with statements such as ‘I do not know’ or ‘please provide context’, indicating an inability to anchor their reasoning. For example, \texttt{GPT-5} has a tendency to seek context when a DOI is not given: \textit{'I am not familiar with the specific acronyms "TLP-3" and "SIVCS" [...]. If you can provide more details or clarify the field, I might be able to offer a more accurate response.'}
    Although this avoids hallucinations and acknowledges gaps, it still does not answer the question for the user.
    Additionally, with DOIs, a subset of LLMs were unable to retrieve or utilise the DOI information effectively, e.g Mixtral-8x22B Instruct:  \textit{'$10.1007/978-3-030-00464-4\_5$ does not provide a specific reference'},
    highlighting variability in source integration capabilities across models and suggesting that explicit reference inclusion is a beneficial but not universally exploitable strategy for enhancing domain-specific reasoning \cite{byun2024reference}.\smallskip
    
\begin{figure}[h]
        \centering
        \includegraphics[width=0.49\textwidth]{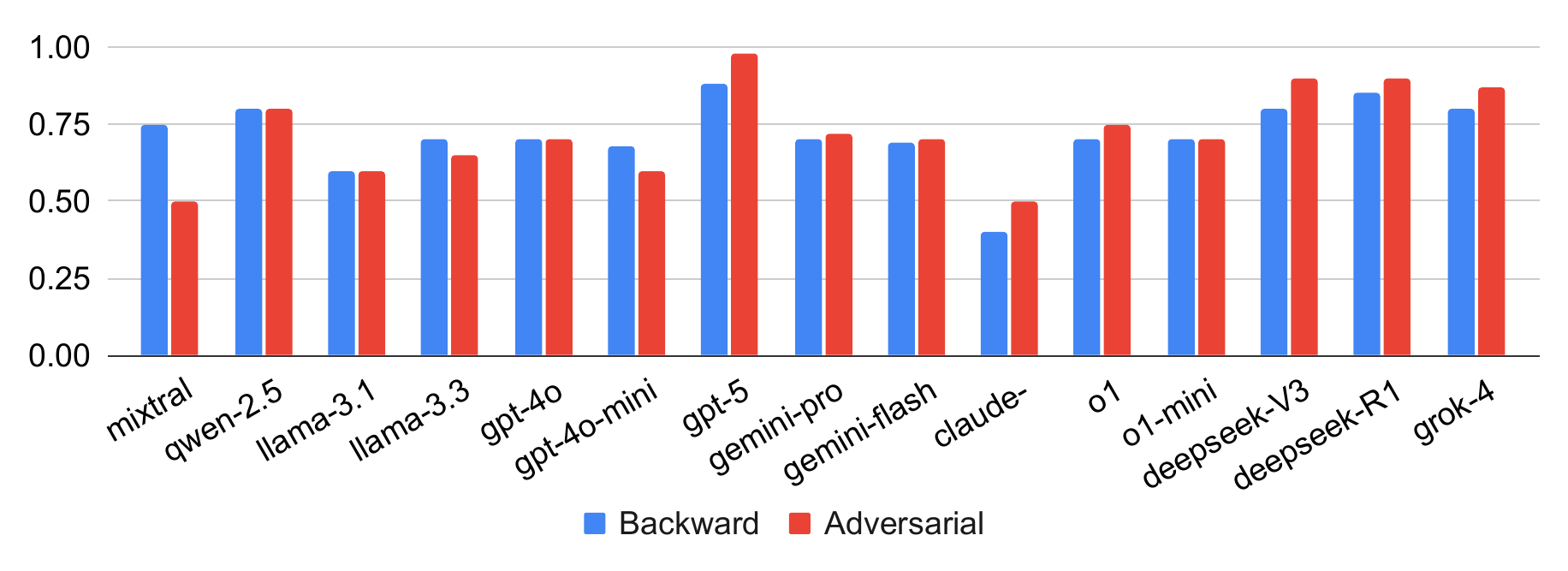} 
        \caption{The LLMs' performance on our backward and adversarial subsets.}
        \label{fig:backward-adversarial}
        
    \end{figure}

\noindent\textbf{Backward Reasoning} (a.k.a. the reversal curse \cite{berglund2023reversal}) is a concept that relates to how LLMs handle inverting or reversing logical, causal, or mathematical relationships \cite{zhang-xiong-2025-backmath, deb2023fill}. 
    Backward reasoning proceeds from conclusion to cause, or from effect to underlying rule. It is the inverse of forward reasoning (cause to effect).
    \begin{shadedquotation}
            \textbf{Backward reasoning example:} \\
            \textbf{Forward:} \textit{Given plaintext 'HELLO' and key 3, what is the ciphertext using a Caesar cipher?} [Expected Answer: KHOOR]\\
            \textbf{Backward} \textit{The plaintext 'HELLO' has a ciphertext 'KHOOR' in Caesar cipher. What key was used?}
    \end{shadedquotation}
    Evaluating and improving backward reasoning is critical for goal-directed problem solving in cryptography. For instance, decrypting a ciphertext, recovering a key, or proving a security property often involves reasoning from the desired result (e.g., plaintext, key validity, or protocol correctness) backwards.
    For backward reasoning (i.e. \textit{type 'backward'}), we generated a subset of \num{1000} handcrafted samples by modifying some of our dataset samples.
    As is common in prior work \cite{farzand2024out, hinkin1998brief, boateng2018best}, the data samples were chosen and independently refined by different experts.
    \\
    Although backward reasoning remains a well-documented challenge in large language models (often referred to as the reversal curse) \cite{chen2025reverse, berglund2024reversalcurse, jiang-etal-2024-forward, zhang-xiong-2025-backmath}, our evaluation (\cref{fig:backward-adversarial}) indicates that, as exceptions, \texttt{DeepSeek-R1}, \texttt{DeepSeek-V3}, \texttt{Grok-4}, and \texttt{GPT-5} achieved comparatively strong performance on the backward-reasoning dataset. A likely explanation for this outcome is that the dataset primarily consisted of relatively simple instances, typically involving numbers of up to two digits, which may not sufficiently stress-test the limits of backward inference. These findings suggest that while certain state-of-the-art models appear capable of overcoming aspects of the reversal curse under constrained conditions, further investigation with more complex and computationally demanding examples is required to assess their robustness in realistic cryptographic scenarios.
    Future work should therefore focus on designing datasets with higher complexity to better evaluate the depth and generalisability of backward reasoning in LLMs.

\noindent\textbf{Robustness} refers to the ability of a language model to maintain reliable, accurate, and coherent performance when exposed to variations or perturbations in input, such as adversarial prompts \cite{jiang2025survey}. 
    Adversarial prompts deliberately introduce ambiguity, misleading cues, or logically inconsistent elements to stress-test a model’s reasoning abilities \cite{madrueno25advancing, wang2023generating}. 
    These prompts are particularly effective for evaluating the robustness of LLMs in high-stakes domains such as cryptography, where correctness and precision are paramount. 
    \\
    To systematically construct adversarial prompts, we manually applied controlled perturbation techniques to existing dataset samples, generating variants that target known failure modes in cryptographic reasoning. 
    These transformations include: (i) definition-confusion prompts, which deliberately conflate or misuse terminology to test whether the model correctly distinguishes closely related concepts—for example: \textit{'Q: Is AES a form of public-key encryption used in secure email protocols?'}, (ii) implausible primitive-combination prompts, which pair cryptographic mechanisms that cannot logically interact, thereby evaluating the model’s ability to reject invalid operations, such as: \textit{'Q: Can we use SHA-256 to decrypt a message encrypted with RSA?'}, (iii) semantic-reversal prompts, which invert core assumptions or protocol properties to assess the model’s ability to detect false logical implications '\textit{Q: If Alice shares her private key publicly, what does this improve in RSA?}', and (iv) hallucination-probe prompts, designed to expose whether the model fabricates non-existent algorithms or mechanisms when confronted with contradictory or unfamiliar terminology, such as: '\textit{Q: Explain how the Schnorr–Bellman–Gentry algorithm enhances SHA-based ECC.}'. 
    By adversarially modifying authentic dataset questions using these principled techniques, we obtain a robust adversarial evaluation set that challenges LLMs under misleading, contradictory, or structurally invalid conditions.
  \\
    Similarly to the previous capabilites, we observed heterogeneous model behaviour in response to our adversarial examples (\cref{fig:backward-adversarial}). Some LLMs generated hallucinated answers, indicating susceptibility to conceptual conflation or logical traps. In contrast, other models explicitly responded with warnings such as \textit{'the \\ Schnorr–Bellman–Gentry algorithm does not exist in the cryptographic literature'} or otherwise refused to provide an answer, demonstrating awareness of the inconsistency or impossibility embedded in the prompt. This divergence highlights differences in model calibration, factual grounding, and reasoning robustness, and underscores the value of adversarial evaluation for assessing model reliability and safety in domain-specific tasks like cryptography.

\subsection{Qualitative Results}
To establish a human baseline for comparison and evaluation, each participant was asked to perform two tasks for a subset of cryptographic questions: first, to provide their own answer to the question, and second, to evaluate three model-generated answers corresponding to three distinct LLMs, including the base model (\texttt{DeepSeek-V3}, our ground truth), to assess the dataset quality.
Participants were given the option to respond with 'I do not know' when uncertain about an answer.

Each participant answered $20$ questions in total, drawn from a larger set of \num{1000} questions (following the same dataset distribution), ensuring that responses covered a representative portion of the dataset while distributing the workload across participants.
The evaluation involved rating each response according to predefined criteria:
\begin{itemize}[leftmargin = *]
    \item \textbf{Factual correctness:} This criterion measures whether the response contains accurate and verifiable information. An answer is factually correct if it aligns with established cryptographic knowledge, formal definitions, and known properties of algorithms, protocols, or theorems. Any instance of incorrect terminology, misidentified algorithms, or false claims (e.g., stating RSA is symmetric) would indicate a failure in factual correctness.

    \item \textbf{Conceptual completeness:} Conceptual completeness refers to whether the answer addresses all essential components of the question with sufficient depth. It evaluates if the response logically follows from the premises, covers the required steps or arguments, and does not omit important details. For example, when asked to explain a digital signature scheme, the answer should mention key generation, signing, and verification—not just a partial subset.

    \item \textbf{Clarity of explanation:} Clarity assesses the linguistic and structural quality of the response. A clear answer is concise, coherent, and unambiguous, using appropriate technical language without unnecessary jargon. It should be understandable by a graduate-level reader with a background in cryptography, facilitating comprehension without sacrificing precision.
\end{itemize}

Our user survey received $50$ responses with (10 non-experts, 8 BSc, 22 MSc, and 10 PhD). Of all collected responses, only one non-expert was excluded due to failure on the attention check, which consisted of a simple two-digit addition task.
Most participants (94\%) reported that they use LLM assistants daily. Other demographic variables (e.g., age, gender, nationality) were found to be statistically insignificant for our analysis, except for age, which correlates with expertise. Hence, we only report expertise levels in the following. All responses were collected anonymously and used exclusively for the purposes of this study (cf. \cref{apx:human_evaluation} for response examples).

This form of evaluation, although limited in scale, provides insights beyond traditional computational metrics by capturing nuanced reasoning errors and highlighting model limitations in understanding and articulating complex cryptographic concepts\footnote{Recruiting qualified cryptography experts willing to invest time in the study was challenging, resulting in a limited number of expert evaluations. Nonetheless, these assessments provide valuable contextual insight that complements our quantitative results, offering a deeper understanding of LLM performance and reasoning within cryptographic tasks.}.

    \begin{figure}[h]
        \centering
        \includegraphics[width=0.49\textwidth]{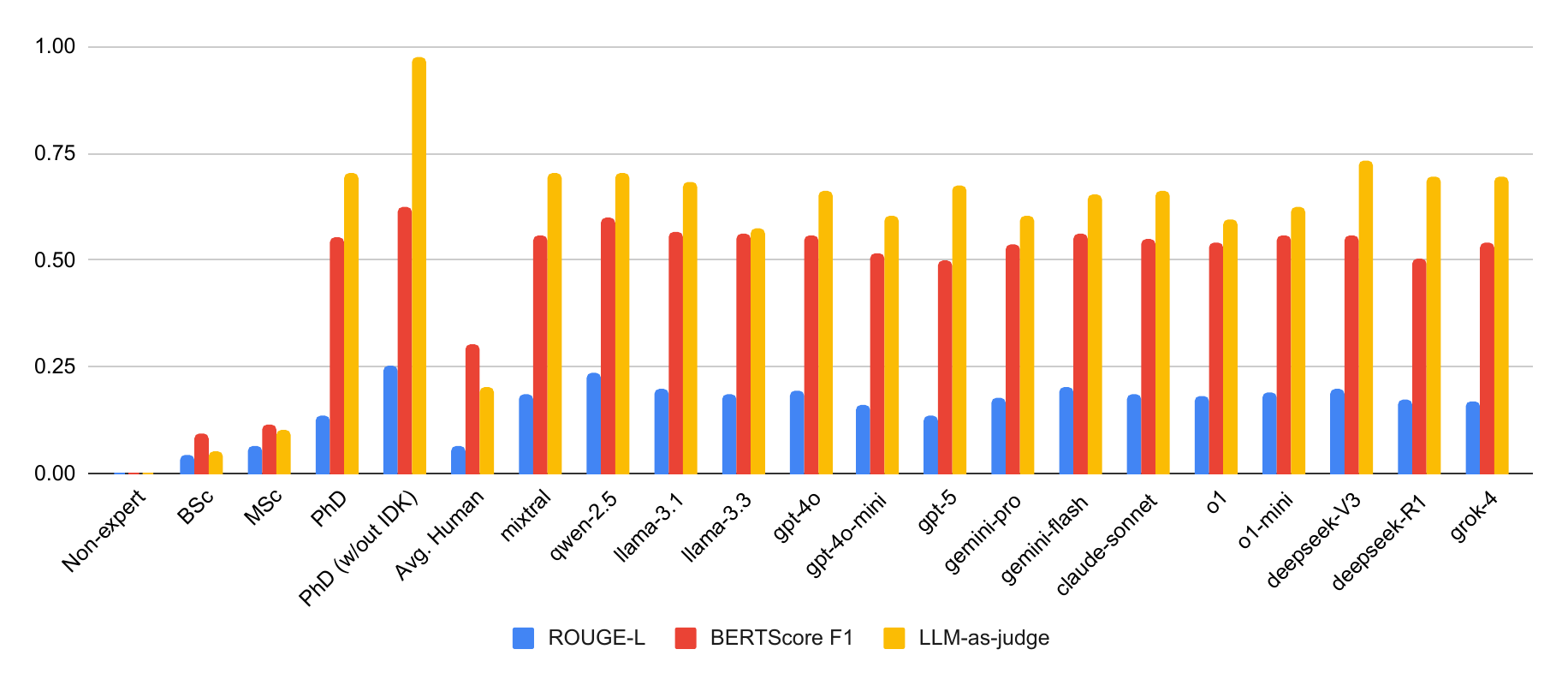}
        \caption{Gold Standard: The human's and the LLM's performance on our qualitative subset by only evaluating the answers. The first items represent each expert level, the 'PhD (w/out IDK)' represents the PhD holders responses when they were certain (excluding the ' I do not know' responses), and the 'Avg. Human' represent the average performance across all expertise levels.
        More comprehensive results can be found in \cref{apx:comprehensive-results}
        }
        \label{fig:gold}
               
    \end{figure}

    \begin{figure}[h]
        \centering
        \includegraphics[width=0.49\textwidth]{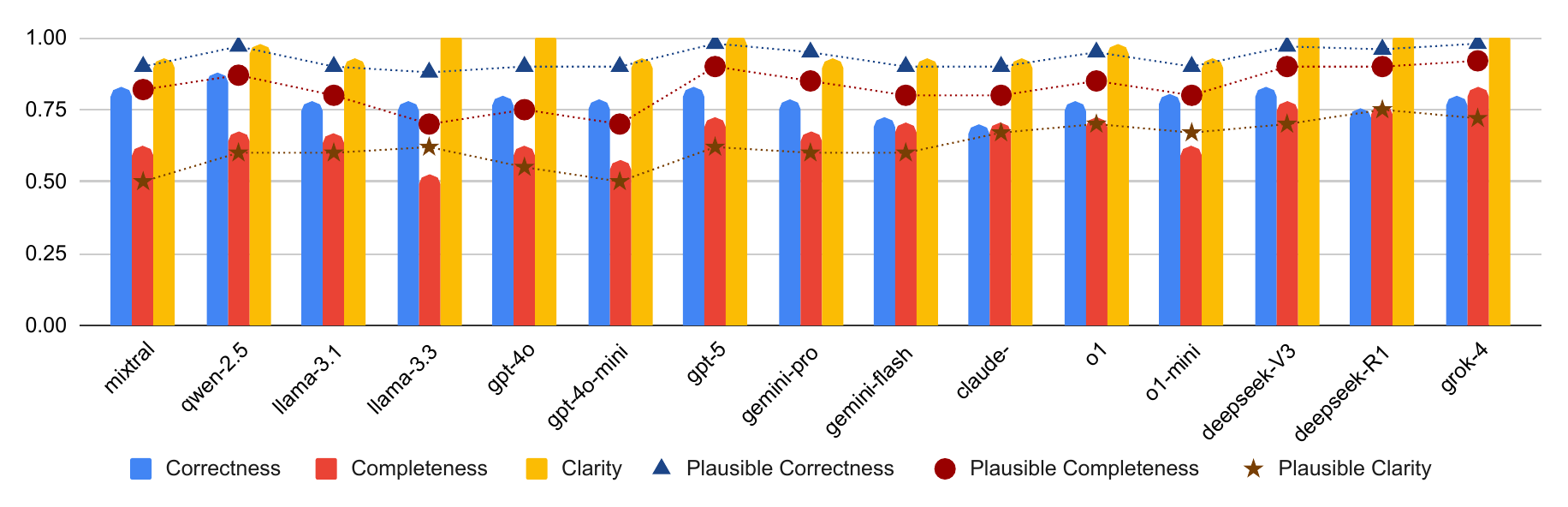} 
        \caption{The human evaluation of the LLMs as given by the human scores when their own answer was correct represented by columns, and when their own answer was incorrect or uncertain (the 'I do not know' option) represented by lines. The latter case is typically higher, indicating misleadingness when the user does not know the answer to the question and accepts the plausibly-sounding answer given by the LLM.}
        \label{fig:human}
    \end{figure}

\noindent{\textbf{Gold Standard (i.e. human baseline)}.} We observed that, as shown in \cref{fig:gold} (see LLM-as-judge), the performance of most LLMs significantly outperforms that of BSc and MSc holders in cryptography (tested with an independent samples t-test between avg. LLM and the avg. of each expertise level, $p < 0.05$). This aligns with expectations, as typical bachelor's and master’s programs cover a broad range of subfields within cryptography, providing students with foundational knowledge without deep specialisation in every area. Consequently, while BSc and MSc holders can competently answer many questions, they are not domain experts, and LLMs achieve a slightly higher performance across general cryptography tasks.
\\
To assess expert-level performance, we also evaluated responses from doctoral participants (10 PhD holders, cf. \cref{fig:gold}). We computed the same metrics and additionally recalculated these metrics while excluding responses labelled as 'I do not know'. This dual analysis revealed that domain-specific PhD experts consistently outperform LLMs in their areas of specialisation, demonstrating deeper conceptual understanding and reasoning capability. At the same time, even expert humans are not able to answer all questions spanning the full spectrum of cryptography subfields, highlighting the challenge of achieving comprehensive coverage. These findings underscore both the strengths and limitations of LLMs relative to human expertise: while LLMs perform at the level of well-trained students across broad domains, specialised human experts retain an advantage in specific areas, particularly when questions demand in-depth reasoning or familiarity with niche cryptographic constructs.\smallskip

\noindent{\textbf{Misleadingness}.}
In our study, human participants evaluated LLM-generated answers along three dimensions: correctness, completeness, and clarity, while also providing their own responses to each question. To meaningfully aggregate these evaluations, we adopted a conditional scoring approach, as shown in \cref{fig:human}. Specifically, for questions where a participant’s own answer was correct, we directly aggregated their evaluation scores for the corresponding LLM response. Conversely, for questions where the participant’s answer was incorrect or uncertain,
we computed a plausibility-based metric, referred to as the 'plausible correctness/completeness/clarity', which captures the extent to which LLM outputs appear convincing despite being factually incorrect.
\\
Analysis of these conditional metrics revealed a clear trend: participants’ domain expertise (i.e. non-experts, BSc, MSc, and PhD holders) is inversely proportional to misleadingness. That is, experts were less likely to overrate misleading or incorrect LLM outputs, whereas less knowledgeable participants were more susceptible to perceiving plausible-sounding but incorrect answers as credible.
Hence, the utility of LLMs is dependent on the user’s expertise: individuals with strong backgrounds in cryptography are better positioned to critically evaluate the model’s outputs. In contrast, less experienced users may over-rely on LLM-generated content, increasing the risk of accepting incorrect or misleading information.

Finally, our overall evaluation indicates that while LLMs can serve as helpful assistants in cryptography, providing guidance, explanations, and general problem-solving support, they exhibit notable limitations in certain domains. These observations underscore the importance of exercising caution and critical judgment when integrating LLMs into cryptographic workflows, an emerging trend within the field \cite{li2025constructing, gligoroski2025llm}, and highlight the need for developing more specialised, domain-adapted LLMs.\smallskip

\noindent{\textbf{Dataset Generation Quality}.} 
In addition to curating data from reliable sources, we assess the quality of the generated dataset 
by conducting human evaluation on a representative subset.
The generated dataset answers (generated by \texttt{DeepSeek-V3}) were rated relatively high across evaluation metrics, in \cref{fig:gold,fig:human}. These results provide empirical support for our selection of \texttt{DeepSeek-V3} as the base model for dataset generation, indicating that its outputs are sufficiently accurate and coherent to serve as a foundation for subsequent analyses and benchmarking.

\subsection{Fine-Tuning the Best-Performing LLM}
Furthermore, we finetuned \texttt{Qwen2.5-72B-Instruct}
on the \textit{train} subset of our dataset (type = \texttt{original}), then ran the evaluation again using our \textit{test} set, to empirically validate that the data supports improved model performance.

The model was trained on an NVIDIA Tesla V100 SXM2 $32$ GB, using a sequence block size of $1024$ and a maximum input length of $2048$ tokens, with mixed precision (bfloat16) to improve computational efficiency. Optimisation employed the AdamW optimiser in $8$-bit precision with a linear learning rate scheduler and an initial learning rate of $3×10^{-5}$. Training was conducted for $20$ epochs. Importantly, this training represents only a sample fine-tuning conducted without altering the model architecture or introducing any specialised components, intended solely to demonstrate the quality and utility of our dataset in improving model performance.

The results indicate that even this minimal fine-tuning on our \dataName dataset produces a substantial improvement in model performance, with an overall increase of 7–13\% across the evaluated metrics. This suggests that the dataset is of high quality, providing informative and relevant training signals. Moreover, the effectiveness of this simple fine-tuning mirrors the way humans improve performance through exposure to targeted practice, as recommended in prior work \cite{gunasekar2023textbooks, li2025scilitllmadaptllmsscientific, abdin2025phi, zhao2023beyond}: just as a human learner benefits from focused study on domain-specific material, the model demonstrates measurable gains when trained on specialised cryptography content. These findings underscore the importance of curated datasets and specialised training for enabling LLMs to capture cryptography-specific reasoning and knowledge.

\section{Discussion}\label{sec:discussion}

Given the observed limitations of current LLMs in performing cryptographic reasoning tasks, we summarise our main findings, potential future work, and limitations.

\subsection{Main Takeaways}
Our comprehensive evaluation of the SOTA LLMs on the \dataName dataset reveals the following key insights:

\begin{itemize}[leftmargin = *]
    \item \textbf{Overall performance and expertise gap}: LLMs demonstrate strong linguistic generalisation and stable semantic understanding, matching the performance of well-trained M.Sc. students across a broad range of cryptographic problems (\cref{fig:gold}). However, they are consistently outperformed by domain experts on questions requiring deep, specialised knowledge or advanced reasoning. Among all evaluated models, \texttt{Qwen2.5-72B-Instruct} achieves the most consistent performance advantage \cref{fig:main-res}. Although this edge may be partially influenced by dataset-source biases and alignment effects, our results are well aligned with the LLMs documented advantages in high-level reasoning domains (\cref{apx:llm-comparison}).

    \item \textbf{Strengths and weaknesses across task types:}
    Performance patterns indicate that LLMs excel at word-based, conceptual, and textbook-derived questions, but struggle with mathematical and formal reasoning tasks (\cref{fig:stratification}). While all models exhibit high consistency—producing stable outputs under paraphrasing and repeated queries (\cref{fig:backward-adversarial})—they fall short of comprehensive cryptographic reasoning and accurate numerical inference.

    \item \textbf{Consistency, factuality, and robustness}:
    Across all evaluations, models exhibit high consistency and paraphrase stability, reflecting robust semantic representation and reliable linguistic understanding. Furthermore, susceptibility to hallucinations is generally low, even under adversarial testing and backward reasoning conditions (\cref{fig:backward-adversarial}). Nonetheless, differences in factual grounding persist—particularly under more complex reasoning setups (\cref{fig:main-res})—indicating that LLMs still struggle to maintain accuracy when tasks involve implicit assumptions or cross-domain synthesis.

    \item \textbf{Improving reliability through referencing:} To mitigate errors and enhance factual grounding, we recommend the inclusion of explicit source references (e.g., DOIs) in prompts. This strategy substantially improves contextualization by anchoring responses to verifiable sources. Our experiments confirm that reference-augmented prompting increases factual accuracy (\cref{fig:sourced}), though not all models effectively retrieve or interpret such metadata.

    \item \textbf{Confidence calibration and user dependence:} Nearly all models display pronounced overconfidence, frequently assigning high certainty scores to incorrect answers. As such, confidence should not be interpreted in isolation but rather evaluated alongside accuracy and consistency. Given that most cryptographers now incorporate LLMs into their daily workflows, their use should be approached with caution. The utility of LLMs depends heavily on the user’s domain expertise—expert users tend to critically assess and contextualize outputs, while less experienced users may overestimate model correctness (\cref{fig:human}), leading to misplaced trust in erroneous conclusions.

    \item \textbf{Dataset Quality:} We curated content from reliable sources. Human evaluators assessed the \texttt{DeepSeek-V3}-generated dataset and rated it as accurate, complete, and clear (\cref{fig:human}), supporting its suitability for benchmarking tasks despite the presence of source-model biases. Furthermore, the observed improvement in LLM performance following fine-tuning (7–13\% increase in evaluation metrics) provides additional empirical evidence that the dataset effectively conveys meaningful training signals and is appropriate for evaluating domain-specific model capabilities.

\end{itemize}

\subsection{Implications and Future Work}

While current LLMs demonstrate promising adaptability and linguistic robustness in cryptographic applications, they remain constrained by limited formal reasoning, imperfect factual verification, and poor calibration. 
We underscore the pressing need for the development and training of domain-specific AI assistants tailored to the cryptography domain along with user-aware interpretability mechanisms to enable safe, trustworthy integration of LLMs in security-critical and cryptographic workflows. Such models should be grounded in a multi-layered foundation of prerequisite knowledge, beginning with rigorous scientific literature and structured mathematical datasets to capture the underlying theoretical principles (cf. \cref{sec:relatedwork}). This foundational layer should be complemented by comprehensive coverage of general cryptographic knowledge, such as that provided by our \dataName dataset. 
Additionally, building on our expert evaluation protocol, we suggest that Reinforcement Learning from Human Feedback (RLHF) \cite{ziegler2019fine} can be a valuable next step in refining these specialised cryptographic LLMs. Given the high degree of domain expertise required to assess correctness, completeness, and clarity in cryptographic reasoning, leveraging expert feedback during training can significantly enhance model reliability and performance.\\
Furthermore, although we sought to construct a comprehensive dataset by integrating all reliably sourced cryptographic materials, future research should expand beyond text-only corpora. In particular, incorporating additional modalities, such as cryptography-specific source code, graphical representations and visualisations, and instructional videos (e.g., lectures, demonstrations, and presentations), may further enhance the ecological validity and task relevance of model training and evaluation in cryptographic contexts.

\subsection{Limitations}
One limitation of our current approach is the reliance on\linebreak \texttt{DeepSeek-V3} for generating QA pairs. 
This decision was primarily driven by the impracticality of large-scale manual annotation, given the limited number of experts available with sufficient background knowledge in cryptography.
As discussed before, the exclusive reliance on \texttt{DeepSeek-V3} could induce model-specific biases.
Incorporating multiple LLMs as generators could solve this issue and enhance diversity.
However, doing so would entail significantly higher computational demands, longer processing time, and an increased carbon footprint—factors that must be carefully weighed in future extensions of this work (cf. \cref{tab:llm_comparison}). 
Another limitation of our study is the relatively small size of the qualitative evaluation sample.
Similarly to the generation of QA pairs, this constraint arises from the inherent limitations in human evaluation capacity, particularly when assessments require domain expertise in cryptography. 
Expert review demands careful, time-intensive analysis of model-generated answers, making it impractical to scale to large datasets.

\section{Conclusion}\label{sec:conclusion}

We introduced \dataName, the first large-scale data set for evaluating LLMs in cryptography, containing over two million QA pairs along with rich metadata. 
In \cref{sec:evaluation}, we evaluated state-of-the-art LLMs perform on our new dataset \dataName.
The LLMs performed well on conceptual and textbook-style questions but struggled with advanced reasoning, mathematical accuracy, and confidence calibration, frequently expressing unwarranted certainty in incorrect answers. 
Overall, the \texttt{Qwen2.5-72B-Instruct} model consistently achieves the highest performance across our experiments. 

Nevertheless, user expertise remains a critical factor for safe deployment, as experts can contextualise outputs while novices may over-trust them.
It is, therefore, essential to continually improve LLMs for cryptographic tasks.
This can be done by training or fine-tuning LLMs with \dataName.
As a proof of concept, we fine-tuned the best-performing model \texttt{Qwen2.5-72B-Instruct} with \dataName, resulting in performance gains of 7–13\%.
We expect that a deeper integration of \dataName~into the LLM training process in the future will lead to more significant improvements and ultimately to the development of more reliable, expert-level AI assistants in cryptography.


% ...

\bibliographystyle{ACM-Reference-Format}
\bibliography{literature}

@misc{code,
    author={Anonymous Author(s)},
    title ={CryptoQA dataset and implementation},
    note = {Available at: \href{https://www.dropbox.com/scl/fo/a4tbva3kz0m5wjxkkzhem/AJPgLIfDw73FxVJgtVbr9gk?rlkey=d8oebj59t2novfg4fg6epkyyp&st=m1v2ep80&dl=0}{https://tinyurl.com/popets2006-3-8}.},
year={2025}
}

@BOOK{test,
   author = "Donald E. Knuth",
   title = "Seminumerical Algorithms",
   volume = 2,
   series = "The Art of Computer Programming",
   publisher = "Addison-Wesley",
   address = "Reading, MA",
   edition = "2nd",
   month = "10~" # jan,
   year = "1981",
}

@ArtifactSoftware{R,
    title = {R: A Language and Environment for Statistical Computing},
    author = {{R Core Team}},
    organization = {R Foundation for Statistical Computing},
    address = {Vienna, Austria},
    year = {2019},
    url = {https://www.R-project.org/},
}

@article{jiang2024mixtral,
  title={Mixtral of experts},
  author={Jiang, Albert Q and Sablayrolles, Alexandre and Roux, Antoine and Mensch, Arthur and Savary, Blanche and Bamford, Chris and Chaplot, Devendra Singh and Casas, Diego de las and Hanna, Emma Bou and Bressand, Florian and others},
  journal={arXiv preprint arXiv:2401.04088},
  year={2024}
}

@misc{qwen2025technicalreport,
      title={Qwen2.5 Technical Report}, 
      author={Qwen and : and An Yang and Baosong Yang and others},
      year={2025},
      eprint={2412.15115},
      archivePrefix={arXiv},
      primaryClass={cs.CL},
      url={https://arxiv.org/abs/2412.15115}, 
}

@misc{meta2024llama3.1,
  author       = {Meta AI},
  title        = {Introducing Llama 3.1: Our most capable models to date},
  howpublished = {\url{https://ai.meta.com/blog/meta-llama-3-1/}},
  year         = {2024},
  note         = {Accessed: 2025-10-01}
}

@article{liu2024deepseek,
  title={Deepseek-v3 technical report},
  author={Liu, Aixin and Feng, Bei  and others},
  journal={arXiv preprint arXiv:2412.19437},
  year={2024}
}

@article{hurst2024gpt,
  title={Gpt-4o system card},
  author={Hurst, Aaron and Lerer, Adam and others},
  journal={arXiv preprint arXiv:2410.21276},
  year={2024}
}

@misc{comanici2025gemini25pushingfrontier,
      title={Gemini 2.5: Pushing the Frontier with Advanced Reasoning, Multimodality, Long Context, and Next Generation Agentic Capabilities}, 
      author={Gheorghe Comanici and  authors},
      year={2025},
      eprint={2507.06261},
      archivePrefix={arXiv},
      primaryClass={cs.CL},
      url={https://arxiv.org/abs/2507.06261}, 
}

@misc{anthropic2024claude35sonnet,
  author       = {Anthropic},
  title        = {Claude 3.5 Sonnet Model Card Addendum},
  year         = {2024},
  month        = oct,
  url          = {https://www-cdn.anthropic.com/fed9cc193a14b84131812372d8d5857f8f304c52/Model_Card_Claude_3_Addendum.pdf},
  note         = {Accessed: 2025-10-01}
}

@misc{deepseekr12025,
      title={DeepSeek-R1: Incentivizing Reasoning Capability in LLMs via Reinforcement Learning}, 
      author={DeepSeek-AI and Daya Guo and others},
      year={2025},
      eprint={2501.12948},
      archivePrefix={arXiv},
      primaryClass={cs.CL},
      url={https://arxiv.org/abs/2501.12948}, 
}

@misc{openai2024openaio1card,
      title={OpenAI o1 System Card}, 
      author={OpenAI and : and Aaron Jaech and 200 others},
      year={2024},
      eprint={2412.16720},
      archivePrefix={arXiv},
      primaryClass={cs.AI},
      url={https://arxiv.org/abs/2412.16720}, 
}

@misc{openai2025gpt5,
  author       = {OpenAI},
  title        = {Introducing GPT-5},
  year         = {2025},
  month        = aug,
  day          = 7,
  url          = {https://openai.com/index/introducing-gpt-5/},
  note         = {Accessed: 2025-10-01}
}

@misc{xai2025grok4,
  author       = {xAI},
  title        = {Grok 4},
  year         = {2025},
  month        = jul,
  day          = 9,
  url          = {https://x.ai/news/grok-4},
  note         = {Accessed: 2025-10-01}
}

@article{kumar2023mycrunchgpt,
  title={Mycrunchgpt: A llm assisted framework for scientific machine learning},
  author={Kumar, Varun and Gleyzer, Leonard and Kahana, Adar and Shukla, Khemraj and Karniadakis, George Em},
  journal={Journal of Machine Learning for Modeling and Computing},
  volume={4},
  number={4},
  year={2023},
  publisher={Begel House Inc.}
}

@article{zhang2025exploring,
  title={Exploring the role of large language models in the scientific method: from hypothesis to discovery},
  author={Zhang, Yanbo and Khan, Sumeer A and Mahmud, Adnan and Yang, Huck and Lavin, Alexander and Levin, Michael and Frey, Jeremy and Dunnmon, Jared and Evans, James and Bundy, Alan and others},
  journal={npj Artificial Intelligence},
  volume={1},
  number={1},
  pages={14},
  year={2025},
  publisher={Nature Publishing Group UK London}
}

@inproceedings{wang2024factuality,
  title={Factuality of Large Language Models: A Survey},
  author={Wang, Yuxia and Wang, Minghan and Manzoor, Muhammad Arslan and Liu, Fei and Georgiev, Georgi and Das, Rocktim Jyoti and Nakov, Preslav},
  booktitle={EMNLP},
  year={2024}
}

@inproceedings{ahn2024large,
  title={Large Language Models for Mathematical Reasoning: Progresses and Challenges},
  author={Ahn, Janice and Verma, Rishu and Lou, Renze and Liu, Di and Zhang, Rui and Yin, Wenpeng},
  booktitle={Proceedings of the 18th Conference of the European Chapter of the Association for Computational Linguistics: Student Research Workshop},
  pages={225--237},
  year={2024}
}

@inproceedings{yu2025training,
  title={Training with “paraphrasing the original text” teaches llm to better retrieve in long-context tasks},
  author={Yu, Yijiong and Huang, Yongfeng and Qi, Zhixiao and Zhou, Zhe},
  booktitle={Proceedings of the AAAI Conference on Artificial Intelligence},
  volume={39},
  number={24},
  pages={25751--25759},
  year={2025}
}

@inproceedings{ahmed2023better,
  title={Better patching using llm prompting, via self-consistency},
  author={Ahmed, Toufique and Devanbu, Premkumar},
  booktitle={2023 38th IEEE/ACM International Conference on Automated Software Engineering (ASE)},
  pages={1742--1746},
  year={2023},
  organization={IEEE}
}

@misc{chen2025reverse,
      title={Reverse Thinking Makes LLMs Stronger Reasoners}, 
      author={Justin Chih-Yao Chen and Zifeng Wang and Hamid Palangi and Rujun Han and Sayna Ebrahimi and Long Le and Vincent Perot and Swaroop Mishra and Mohit Bansal and Chen-Yu Lee and Tomas Pfister},
      year={2025},
      eprint={2411.19865},
      archivePrefix={arXiv},
      primaryClass={cs.CL},
      url={https://arxiv.org/abs/2411.19865}, 
}

@inproceedings{jiang-etal-2024-forward,
    title = "Forward-Backward Reasoning in Large Language Models for Mathematical Verification",
    author = "Jiang, Weisen  and
      Shi, Han  and
      Yu, Longhui  and
      Liu, Zhengying  and
      Zhang, Yu  and
      Li, Zhenguo  and
      Kwok, James",
    editor = "Ku, Lun-Wei  and
      Martins, Andre  and
      Srikumar, Vivek",
    booktitle = "Findings of the Association for Computational Linguistics: ACL 2024",
    month = aug,
    year = "2024",
    address = "Bangkok, Thailand",
    publisher = "Association for Computational Linguistics",
    url = "https://aclanthology.org/2024.findings-acl.397/",
    doi = "10.18653/v1/2024.findings-acl.397",
    pages = "6647--6661"
}

@inproceedings{zhang-xiong-2025-backmath,
    title = "{B}ack{MATH}: Towards Backward Reasoning for Solving Math Problems Step by Step",
    author = "Zhang, Shaowei  and
      Xiong, Deyi",
    editor = "Rambow, Owen  and
      Wanner, Leo  and
      Apidianaki, Marianna  and
      Al-Khalifa, Hend  and
      Eugenio, Barbara Di  and
      Schockaert, Steven  and
      Darwish, Kareem  and
      Agarwal, Apoorv",
    booktitle = "Proceedings of the 31st International Conference on Computational Linguistics: Industry Track",
    month = jan,
    year = "2025",
    address = "Abu Dhabi, UAE",
    publisher = "Association for Computational Linguistics",
    url = "https://aclanthology.org/2025.coling-industry.40/",
    pages = "466--482"
}

@misc{berglund2024reversalcurse,
      title={The Reversal Curse: LLMs trained on "A is B" fail to learn "B is A"}, 
      author={Lukas Berglund and Meg Tong and Max Kaufmann and Mikita Balesni and Asa Cooper Stickland and Tomasz Korbak and Owain Evans},
      year={2024},
      eprint={2309.12288},
      archivePrefix={arXiv},
      primaryClass={cs.CL},
      url={https://arxiv.org/abs/2309.12288}, 
}

@misc{wang2023aligning,
      title={Aligning Large Language Models with Human: A Survey}, 
      author={Yufei Wang and Wanjun Zhong and Liangyou Li and Fei Mi and Xingshan Zeng and Wenyong Huang and Lifeng Shang and Xin Jiang and Qun Liu},
      year={2023},
      eprint={2307.12966},
      archivePrefix={arXiv},
      primaryClass={cs.CL},
      url={https://arxiv.org/abs/2307.12966}, 
}

@misc{iacrEprint,
  author       = {{International Association for Cryptologic Research (IACR)}},
  title        = {{IACR ePrint Archive}},
  howpublished = {\url{https://eprint.iacr.org/}},
  note         = {Accessed: 2025-10-02},
  year = {2025}
}

@article{blanchet2018proverif,
  title={ProVerif 2.00: automatic cryptographic protocol verifier, user manual and tutorial},
  author={Blanchet, Bruno and Smyth, Ben and Cheval, Vincent and Sylvestre, Marc},
  journal={Version from},
  volume={16},
  pages={05--16},
  year={2018}
}

@inproceedings{meier2013tamarin,
  title={The TAMARIN prover for the symbolic analysis of security protocols},
  author={Meier, Simon and Schmidt, Benedikt and Cremers, Cas and Basin, David},
  booktitle={International conference on computer aided verification},
  pages={696--701},
  year={2013},
  organization={Springer}
}

@inproceedings{dy21,
  author = {Bhargavan, Karthikeyan and Bichhawat, Abhishek and Do, Quoc Huy and Hosseyni, Pedram and K{\"u}sters, Ralf and Schmitz, Guido and W{\"u}rtele, Tim},
  booktitle = {{IEEE European Symposium on Security and Privacy (EuroS{\&}P 2021)}},
  doi = {10.1109/EuroSP51992.2021.00042},
  pages = {523-542},
  publisher = {IEEE},
  title = {{DY*: A Modular Symbolic Verification Framework for Executable Cryptographic Protocol Code}},
  url = {https://publ.sec.uni-stuttgart.de/bhargavanbichhavatdohosseynikuestersschmitzwuertele-eurosp-2021.pdf},
  year = 2021
}

@inproceedings{blanchet2007cryptoverif,
  title={CryptoVerif: Computationally sound mechanized prover for cryptographic protocols},
  author={Blanchet, Bruno},
  booktitle={Dagstuhl seminar “Formal Protocol Verification Applied},
  volume={117},
  pages={156},
  year={2007}
}

@book{chlipala2013certified,
  title={Certified programming with dependent types: a pragmatic introduction to the Coq proof assistant},
  author={Chlipala, Adam},
  year={2013},
  publisher={MIT Press}
}

@book{nipkow2002isabelle,
  title={Isabelle/HOL: a proof assistant for higher-order logic},
  author={Nipkow, Tobias and Wenzel, Markus and Paulson, Lawrence C},
  year={2002},
  publisher={Springer}
}

@article{denny2013sage,
  title={SAGE: Open Source Mathematics Software System (http://sagemath. org)},
  author={Denny, JK},
  journal={The College Mathematics Journal},
  volume={44},
  number={2},
  pages={149--155},
  year={2013},
  publisher={Taylor \& Francis}
}

@article{ren2025towards,
  title={Towards scientific intelligence: A survey of llm-based scientific agents},
  author={Ren, Shuo and Jian, Pu and Ren, Zhenjiang and Leng, Chunlin and Xie, Can and Zhang, Jiajun},
  journal={arXiv preprint arXiv:2503.24047},
  year={2025}
}

@inproceedings{zhang2024mathverse,
  title={Mathverse: Does your multi-modal llm truly see the diagrams in visual math problems?},
  author={Zhang, Renrui and Jiang, Dongzhi and Zhang, Yichi and Lin, Haokun and Guo, Ziyu and Qiu, Pengshuo and Zhou, Aojun and Lu, Pan and Chang, Kai-Wei and Qiao, Yu and others},
  booktitle={European Conference on Computer Vision},
  pages={169--186},
  year={2024},
  organization={Springer}
}

@article{li2025cipherbank,
  title={Cipherbank: Exploring the boundary of llm reasoning capabilities through cryptography challenges},
  author={Li, Yu and Pei, Qizhi and Sun, Mengyuan and Lin, Honglin and Ming, Chenlin and Gao, Xin and Wu, Jiang and He, Conghui and Wu, Lijun},
  journal={arXiv preprint arXiv:2504.19093},
  year={2025}
}

@article{rogers2023qa,
  title={Qa dataset explosion: A taxonomy of nlp resources for question answering and reading comprehension},
  author={Rogers, Anna and Gardner, Matt and Augenstein, Isabelle},
  journal={ACM Computing Surveys},
  volume={55},
  number={10},
  pages={1--45},
  year={2023},
  publisher={ACM New York, NY}
}

@article{noever2023llmsforciphers,
  author       = {David Noever},
  title        = {Large Language Models for Ciphers},
  journal      = {International Journal of Artificial Intelligence and Applications (IJAIA)},
  volume       = {14},
  number       = {3},
  month        = may,
  year         = {2023},
  doi          = {10.5121/ijaia.2023.14301},
  url          = {https://aircconline.com/ijaia/V14N3/14323ijaia01.pdf},
  note         = {Accessed: 2025-10-02}
}

@misc{maskey2025benchmarkinglargelanguagemodels,
      title={Benchmarking Large Language Models for Cryptanalysis and Side-Channel Vulnerabilities}, 
      author={Utsav Maskey and Chencheng Zhu and Usman Naseem},
      year={2025},
      eprint={2505.24621},
      archivePrefix={arXiv},
      primaryClass={cs.CL},
      url={https://arxiv.org/abs/2505.24621}, 
}

@inproceedings{Wickramasinghe_2025,
   title={SoK: Decoding the Enigma of Encrypted Network Traffic Classifiers},
   url={http://dx.doi.org/10.1109/SP61157.2025.00165},
   DOI={10.1109/sp61157.2025.00165},
   booktitle={2025 IEEE Symposium on Security and Privacy (SP)},
   publisher={IEEE},
   author={Wickramasinghe, Nimesha and Shaghaghi, Arash and Tsudik, Gene and Jha, Sanjay},
   year={2025},
   month=may, pages={1825–1843} }

@inproceedings{wang2017malware,
  title={Malware traffic classification using convolutional neural network for representation learning},
  author={Wang, Wei and Zhu, Ming and Zeng, Xuewen and Ye, Xiaozhou and Sheng, Yiqiang},
  booktitle={2017 International conference on information networking (ICOIN)},
  pages={712--717},
  year={2017},
  organization={IEEE}
}

@article{sharafaldin2018toward,
  title={Toward generating a new intrusion detection dataset and intrusion traffic characterization.},
  author={Sharafaldin, Iman and Lashkari, Arash Habibi and Ghorbani, Ali A and others},
  journal={ICISSp},
  volume={1},
  number={2018},
  pages={108--116},
  year={2018}
}

@inproceedings{dadkhah2022towards,
  title={Towards the development of a realistic multidimensional IoT profiling dataset},
  author={Dadkhah, Sajjad and Mahdikhani, Hassan and Danso, Priscilla Kyei and Zohourian, Alireza and Truong, Kevin Anh and Ghorbani, Ali A},
  booktitle={2022 19th Annual International Conference on Privacy, Security \& Trust (PST)},
  pages={1--11},
  year={2022},
  organization={IEEE}
}

@article{koroniotis2019towards,
  title={Towards the development of realistic botnet dataset in the internet of things for network forensic analytics: Bot-iot dataset},
  author={Koroniotis, Nickolaos and Moustafa, Nour and Sitnikova, Elena and Turnbull, Benjamin},
  journal={Future Generation Computer Systems},
  volume={100},
  pages={779--796},
  year={2019},
  publisher={Elsevier}
}

@inproceedings{draper2016characterization,
  title={Characterization of encrypted and vpn traffic using time-related},
  author={Draper-Gil, Gerard and Lashkari, Arash Habibi and Mamun, Mohammad Saiful Islam and Ghorbani, Ali A},
  booktitle={Proceedings of the 2nd international conference on information systems security and privacy (ICISSP)},
  pages={407--414},
  year={2016}
}

@inproceedings{lashkari2017characterization,
  title={Characterization of tor traffic using time based features},
  author={Lashkari, Arash Habibi and Gil, Gerard Draper and Mamun, Mohammad Saiful Islam and Ghorbani, Ali A},
  booktitle={International conference on information systems security and privacy},
  volume={2},
  pages={253--262},
  year={2017},
  organization={SciTePress}
}

@inproceedings{kushman2014learning,
  title={Learning to automatically solve algebra word problems},
  author={Kushman, Nate and Artzi, Yoav and Zettlemoyer, Luke and Barzilay, Regina},
  booktitle={Proceedings of the 52nd Annual Meeting of the Association for Computational Linguistics (Volume 1: Long Papers)},
  pages={271--281},
  year={2014}
}

@inproceedings{roy2016illinois,
  title={Illinois math solver: Math reasoning on the web},
  author={Roy, Subhro and Roth, Dan},
  booktitle={Proceedings of the 2016 Conference of the North American Chapter of the Association for Computational Linguistics: Demonstrations},
  pages={52--56},
  year={2016}
}

@misc{roy2016solvinggeneralarithmeticword,
      title={Solving General Arithmetic Word Problems}, 
      author={Subhro Roy and Dan Roth},
      year={2016},
      eprint={1608.01413},
      archivePrefix={arXiv},
      primaryClass={cs.CL},
      url={https://arxiv.org/abs/1608.01413}, 
}

@article{ling2017program,
  title={Program induction by rationale generation: Learning to solve and explain algebraic word problems},
  author={Ling, Wang and Yogatama, Dani and Dyer, Chris and Blunsom, Phil},
  journal={arXiv preprint arXiv:1705.04146},
  year={2017}
}

@article{amini2019mathqa,
  title={Mathqa: Towards interpretable math word problem solving with operation-based formalisms},
  author={Amini, Aida and Gabriel, Saadia and Lin, Peter and Koncel-Kedziorski, Rik and Choi, Yejin and Hajishirzi, Hannaneh},
  journal={arXiv preprint arXiv:1905.13319},
  year={2019}
}

@article{cobbe2021training,
  title={Training verifiers to solve math word problems},
  author={Cobbe, Karl and Kosaraju, Vineet and Bavarian, Mohammad and Chen, Mark and Jun, Heewoo and Kaiser, Lukasz and Plappert, Matthias and Tworek, Jerry and Hilton, Jacob and Nakano, Reiichiro and others},
  journal={arXiv preprint arXiv:2110.14168},
  year={2021}
}

@misc{li2025scilitllmadaptllmsscientific,
      title={SciLitLLM: How to Adapt LLMs for Scientific Literature Understanding}, 
      author={Sihang Li and Jin Huang and Jiaxi Zhuang and Yaorui Shi and Xiaochen Cai and Mingjun Xu and Xiang Wang and Linfeng Zhang and Guolin Ke and Hengxing Cai},
      year={2025},
      eprint={2408.15545},
      archivePrefix={arXiv},
      primaryClass={cs.LG},
      url={https://arxiv.org/abs/2408.15545}, 
}

@article{abdin2025phi,
  title={Phi-4-reasoning technical report},
  author={Abdin, Marah and Agarwal, Sahaj and Awadallah, Ahmed and Balachandran, Vidhisha and Behl, Harkirat and Chen, Lingjiao and de Rosa, Gustavo and Gunasekar, Suriya and Javaheripi, Mojan and Joshi, Neel and others},
  journal={arXiv preprint arXiv:2504.21318},
  year={2025}
}

@article{gunasekar2023textbooks,
  title={Textbooks are all you need},
  author={Gunasekar, Suriya and Zhang, Yi and Aneja, Jyoti and Mendes, Caio C{\'e}sar Teodoro and Del Giorno, Allie and Gopi, Sivakanth and Javaheripi, Mojan and Kauffmann, Piero and de Rosa, Gustavo and Saarikivi, Olli and others},
  journal={arXiv preprint arXiv:2306.11644},
  year={2023}
}

@article{zhao2023beyond,
  title={Beyond one-model-fits-all: A survey of domain specialization for large language models},
  author={ZHAO, XUJIANG and LU, JIAYING and DENG, CHENGYUAN and ZHENG, C and WANG, JUNXIANG and CHOWDHURY, TANMOY and YUN, L and CUI, HEJIE and XUCHAO, ZHANG and Zhao, Tianjiao and others},
  journal={arXiv preprint arXiv},
  volume={2305},
  year={2023}
}

@inproceedings{alismail2025survey,
  title={A Survey of LLM-Based Methods for Synthetic Data Generation and the Rise of Agentic Workflows},
  author={Alismail, Ahmad and Lanquillon, Carsten},
  booktitle={International Conference on Human-Computer Interaction},
  pages={119--135},
  year={2025},
  organization={Springer}
}

@article{garg2022can,
  title={What can transformers learn in-context? a case study of simple function classes},
  author={Garg, Shivam and Tsipras, Dimitris and Liang, Percy S and Valiant, Gregory},
  journal={Advances in neural information processing systems},
  volume={35},
  pages={30583--30598},
  year={2022}
}

@article{brown2020language,
  title={Language models are few-shot learners},
  author={Brown, Tom and Mann, Benjamin and Ryder, Nick and Subbiah, Melanie and Kaplan, Jared D and Dhariwal, Prafulla and Neelakantan, Arvind and Shyam, Pranav and Sastry, Girish and Askell, Amanda and others},
  journal={Advances in neural information processing systems},
  volume={33},
  pages={1877--1901},
  year={2020}
}

@article{lewis2020retrieval,
  title={Retrieval-augmented generation for knowledge-intensive nlp tasks},
  author={Lewis, Patrick and Perez, Ethan and Piktus, Aleksandra and Petroni, Fabio and Karpukhin, Vladimir and Goyal, Naman and K{\"u}ttler, Heinrich and Lewis, Mike and Yih, Wen-tau and Rockt{\"a}schel, Tim and others},
  journal={Advances in neural information processing systems},
  volume={33},
  pages={9459--9474},
  year={2020}
}

@inproceedings{papineni2002machine,
  title={Machine Translation Evaluation: N-grams to the Rescue.},
  author={Papineni, Kishore},
  booktitle={LREC},
  year={2002}
}

@inproceedings{lin2004rouge,
  title={Rouge: A package for automatic evaluation of summaries},
  author={Lin, Chin-Yew},
  booktitle={Text summarization branches out},
  pages={74--81},
  year={2004}
}

@article{jelinek1977perplexity,
  title={Perplexity—a measure of the difficulty of speech recognition tasks},
  author={Jelinek, Fred and Mercer, Robert L and Bahl, Lalit R and Baker, James K},
  journal={The Journal of the Acoustical Society of America},
  volume={62},
  number={S1},
  pages={S63--S63},
  year={1977},
  publisher={Acoustical Society of America}
}

@inproceedings{banerjee2005meteor,
  title={METEOR: An automatic metric for MT evaluation with improved correlation with human judgments},
  author={Banerjee, Satanjeev and Lavie, Alon},
  booktitle={Proceedings of the acl workshop on intrinsic and extrinsic evaluation measures for machine translation and/or summarization},
  pages={65--72},
  year={2005}
}

@inproceedings{devlin2019bert,
  title={Bert: Pre-training of deep bidirectional transformers for language understanding},
  author={Devlin, Jacob and Chang, Ming-Wei and Lee, Kenton and Toutanova, Kristina},
  booktitle={Proceedings of the 2019 conference of the North American chapter of the association for computational linguistics: human language technologies, volume 1 (long and short papers)},
  pages={4171--4186},
  year={2019}
}

@misc{openrouter,
  author       = {OpenRouter, Inc.},
  title        = {OpenRouter — The Unified Interface for LLMs},
  howpublished = {\url{https://openrouter.ai/}},
  year         = {2025},
  note         = {Accessed: 2025-10-06}
}

@article{dietz2025llm,
  title={Llm-evaluation tropes: Perspectives on the validity of llm-evaluations},
  author={Dietz, Laura and Zendel, Oleg and Bailey, Peter and Clarke, Charles and Cotterill, Ellese and Dalton, Jeff and Hasibi, Faegheh and Sanderson, Mark and Craswell, Nick},
  journal={arXiv preprint arXiv:2504.19076},
  year={2025}
}

@article{liu2023secqa,
  title={Secqa: A concise question-answering dataset for evaluating large language models in computer security},
  author={Liu, Zefang},
  journal={arXiv preprint arXiv:2312.15838},
  year={2023}
}

@misc{jing2025secbench,
      title={SecBench: A Comprehensive Multi-Dimensional Benchmarking Dataset for LLMs in Cybersecurity}, 
      author={Pengfei Jing and Mengyun Tang and Xiaorong Shi and Xing Zheng and Sen Nie and Shi Wu and Yong Yang and Xiapu Luo},
      year={2025},
      eprint={2412.20787},
      archivePrefix={arXiv},
      primaryClass={cs.CR},
      url={https://arxiv.org/abs/2412.20787}, 
}

@inproceedings{farzand2024out,
  title={Out-of-device privacy unveiled: Designing and validating the out-of-device privacy scale (odps)},
  author={Farzand, Habiba and Marky, Karola and Khamis, Mohamed},
  booktitle={Proceedings of the 2024 CHI Conference on Human Factors in Computing Systems},
  pages={1--15},
  year={2024}
}

@article{hinkin1998brief,
  title={A brief tutorial on the development of measures for use in survey questionnaires},
  author={Hinkin, Timothy R},
  journal={Organizational research methods},
  volume={1},
  number={1},
  pages={104--121},
  year={1998},
  publisher={Sage Publications Sage CA: Thousand Oaks, CA}
}

@article{boateng2018best,
  title={Best practices for developing and validating scales for health, social, and behavioral research: a primer},
  author={Boateng, Godfred O and Neilands, Torsten B and Frongillo, Edward A and Melgar-Qui{\~n}onez, Hugo R and Young, Sera L},
  journal={Frontiers in public health},
  volume={6},
  pages={149},
  year={2018},
  publisher={Frontiers Media SA}
}

@article{madrueno25advancing,
title = {Advancing text adversarial example generation using large language models},
journal = {Knowledge-Based Systems},
volume = {329},
pages = {114361},
year = {2025},
issn = {0950-7051},
doi = {https://doi.org/10.1016/j.knosys.2025.114361},
url = {https://www.sciencedirect.com/science/article/pii/S0950705125014005},
author = {Natalia Madrueño and Alberto Fernández-Isabel and Rubén R. Fernández and Isaac {Martín de Diego}}
}

@misc{raina2024llm,
      title={Is LLM-as-a-Judge Robust? Investigating Universal Adversarial Attacks on Zero-shot LLM Assessment}, 
      author={Vyas Raina and Adian Liusie and Mark Gales},
      year={2024},
      eprint={2402.14016},
      archivePrefix={arXiv},
      primaryClass={cs.CL},
      url={https://arxiv.org/abs/2402.14016}, 
}

@misc{wang2023generating,
      title={Generating Valid and Natural Adversarial Examples with Large Language Models}, 
      author={Zimu Wang and Wei Wang and Qi Chen and Qiufeng Wang and Anh Nguyen},
      year={2023},
      eprint={2311.11861},
      archivePrefix={arXiv},
      primaryClass={cs.CL},
      url={https://arxiv.org/abs/2311.11861}, 
}

@article{fuad2024llm,
  title={Llm-ref: Enhancing reference handling in technical writing with large language models},
  author={Fuad, Kazi Ahmed Asif and Chen, Lizhong},
  journal={arXiv preprint arXiv:2411.00294},
  year={2024}
}

@article{patel2021nlp,
  title={Are NLP models really able to solve simple math word problems?},
  author={Patel, Arkil and Bhattamishra, Satwik and Goyal, Navin},
  journal={arXiv preprint arXiv:2103.07191},
  year={2021}
}

@article{wu2025estimating,
  title={Estimating LLM Consistency: A User Baseline vs Surrogate Metrics},
  author={Wu, Xiaoyuan and Lin, Weiran and Akgul, Omer and Bauer, Lujo},
  journal={arXiv preprint arXiv:2505.23799},
  year={2025}
}

@inproceedings{liu2025uncertainty,
  title={Uncertainty quantification and confidence calibration in large language models: A survey},
  author={Liu, Xiaoou and Chen, Tiejin and Da, Longchao and Chen, Chacha and Lin, Zhen and Wei, Hua},
  booktitle={Proceedings of the 31st ACM SIGKDD Conference on Knowledge Discovery and Data Mining V. 2},
  pages={6107--6117},
  year={2025}
}

@inproceedings{byun2024reference,
  title={This reference does not exist: an exploration of LLM citation accuracy and relevance},
  author={Byun, Courtni and Vasicek, Piper and Seppi, Kevin},
  booktitle={Proceedings of the Third Workshop on Bridging Human--Computer Interaction and Natural Language Processing},
  pages={28--39},
  year={2024}
}

@article{zhang2025reference,
  title={Reference decisions enhance LLM performance, amplified by source disclosure},
  author={Zhang, Yongxiang and Liu, Zhaobin and Bai, Shaosen and Xu, Ting and Lau, Raymond YK},
  journal={Digital Health},
  volume={11},
  pages={20552076251342078},
  year={2025},
  publisher={Sage Publications Sage UK: London, England}
}

@article{deb2023fill,
  title={Fill in the blank: Exploring and enhancing LLM capabilities for backward reasoning in math word problems},
  author={Deb, Aniruddha and Oza, Neeva and Singla, Sarthak and Khandelwal, Dinesh and Garg, Dinesh and Singla, Parag},
  journal={arXiv preprint arXiv:2310.01991},
  year={2023}
}

@article{berglund2023reversal,
  title={The Reversal Curse: LLMs trained on" A is B" fail to learn" B is A"},
  author={Berglund, Lukas and Tong, Meg and Kaufmann, Max and Balesni, Mikita and Stickland, Asa Cooper and Korbak, Tomasz and Evans, Owain},
  journal={arXiv preprint arXiv:2309.12288},
  year={2023}
}

@article{jiang2025survey,
  title={Survey of adversarial robustness in multimodal large language models},
  author={Jiang, Chengze and Wang, Zhuangzhuang and Dong, Minjing and Gui, Jie},
  journal={arXiv preprint arXiv:2503.13962},
  year={2025}
}

@article{panickssery2024llm,
  title={Llm evaluators recognize and favor their own generations},
  author={Panickssery, Arjun and Bowman, Samuel and Feng, Shi},
  journal={Advances in Neural Information Processing Systems},
  volume={37},
  pages={68772--68802},
  year={2024}
}

@article{liang2024mapping,
  title={Mapping the increasing use of LLMs in scientific papers},
  author={Liang, Weixin and Zhang, Yaohui and Wu, Zhengxuan and Lepp, Haley and Ji, Wenlong and Zhao, Xuandong and Cao, Hancheng and Liu, Sheng and He, Siyu and Huang, Zhi and others},
  journal={arXiv preprint arXiv:2404.01268},
  year={2024}
}

@article{liang2024monitoring,
  title={Monitoring ai-modified content at scale: A case study on the impact of chatgpt on ai conference peer reviews},
  author={Liang, Weixin and Izzo, Zachary and Zhang, Yaohui and Lepp, Haley and Cao, Hancheng and Zhao, Xuandong and Chen, Lingjiao and Ye, Haotian and Liu, Sheng and Huang, Zhi and others},
  journal={arXiv preprint arXiv:2403.07183},
  year={2024}
}

@article{wang25slm,
author = {Wang, Fali and Zhang, Zhiwei and Zhang, Xianren and Wu, Zongyu and Mo, TzuHao and Lu, Qiuhao and Wang, Wanjing and Li, Rui and Xu, Junjie and Tang, Xianfeng and He, Qi and Ma, Yao and Huang, Ming and Wang, Suhang},
title = {A Comprehensive Survey of Small Language Models in the Era of Large Language Models: Techniques, Enhancements, Applications, Collaboration with LLMs, and Trustworthiness},
year = {2025},
publisher = {Association for Computing Machinery},
address = {New York, NY, USA},
issn = {2157-6904},
url = {https://doi.org/10.1145/3768165},
doi = {10.1145/3768165},
note = {Just Accepted},
journal = {ACM Trans. Intell. Syst. Technol.},
month = sep
}

@article{ziegler2019fine,
  title={Fine-tuning language models from human preferences},
  author={Ziegler, Daniel M and Stiennon, Nisan and Wu, Jeffrey and Brown, Tom B and Radford, Alec and Amodei, Dario and Christiano, Paul and Irving, Geoffrey},
  journal={arXiv preprint arXiv:1909.08593},
  year={2019}
}

@misc{xu2025llmcyber,
      title={Large Language Models for Cyber Security: A Systematic Literature Review}, 
      author={Hanxiang Xu and Shenao Wang and Ningke Li and Kailong Wang and Yanjie Zhao and Kai Chen and Ting Yu and Yang Liu and Haoyu Wang},
      year={2025},
      eprint={2405.04760},
      archivePrefix={arXiv},
      primaryClass={cs.CR},
      url={https://arxiv.org/abs/2405.04760}, 
}

@inproceedings{bird-loper-2004-nltk,
    title = "{NLTK}: The Natural Language Toolkit",
    author = "Bird, Steven  and
      Loper, Edward",
    booktitle = "Proceedings of the {ACL} Interactive Poster and Demonstration Sessions",
    month = jul,
    year = "2004",
    address = "Barcelona, Spain",
    publisher = "Association for Computational Linguistics",
    url = "https://aclanthology.org/P04-3031/",
    pages = "214--217"
}

@misc{wang2025aicrypto,
      title={AICrypto: A Comprehensive Benchmark for Evaluating Cryptography Capabilities of Large Language Models}, 
      author={Yu Wang and Yijian Liu and Liheng Ji and Han Luo and Wenjie Li and Xiaofei Zhou and Chiyun Feng and Puji Wang and Yuhan Cao and Geyuan Zhang and Xiaojian Li and Rongwu Xu and Yilei Chen and Tianxing He},
      year={2025},
      eprint={2507.09580},
      archivePrefix={arXiv},
      primaryClass={cs.CR},
      url={https://arxiv.org/abs/2507.09580}, 
}

@misc{shao2025nyuctf,
      title={NYU CTF Bench: A Scalable Open-Source Benchmark Dataset for Evaluating LLMs in Offensive Security}, 
      author={Minghao Shao and Sofija Jancheska and Meet Udeshi and Brendan Dolan-Gavitt and Haoran Xi and Kimberly Milner and Boyuan Chen and Max Yin and Siddharth Garg and Prashanth Krishnamurthy and Farshad Khorrami and Ramesh Karri and Muhammad Shafique},
      year={2025},
      eprint={2406.05590},
      archivePrefix={arXiv},
      primaryClass={cs.CR},
      url={https://arxiv.org/abs/2406.05590}, 
}

@misc{zhu2025cvebench,
      title={CVE-Bench: A Benchmark for AI Agents' Ability to Exploit Real-World Web Application Vulnerabilities}, 
      author={Yuxuan Zhu and Antony Kellermann and Dylan Bowman and Philip Li and Akul Gupta and Adarsh Danda and Richard Fang and Conner Jensen and Eric Ihli and Jason Benn and Jet Geronimo and Avi Dhir and Sudhit Rao and Kaicheng Yu and Twm Stone and Daniel Kang},
      year={2025},
      eprint={2503.17332},
      archivePrefix={arXiv},
      primaryClass={cs.CR},
      url={https://arxiv.org/abs/2503.17332}, 
}

@misc{deng2024pentest,
      title={PentestGPT: An LLM-empowered Automatic Penetration Testing Tool}, 
      author={Gelei Deng and Yi Liu and Víctor Mayoral-Vilches and Peng Liu and Yuekang Li and Yuan Xu and Tianwei Zhang and Yang Liu and Martin Pinzger and Stefan Rass},
      year={2024},
      eprint={2308.06782},
      archivePrefix={arXiv},
      primaryClass={cs.SE},
      url={https://arxiv.org/abs/2308.06782}, 
}

@misc{wang2025cybergym,
      title={CyberGym: Evaluating AI Agents' Real-World Cybersecurity Capabilities at Scale}, 
      author={Zhun Wang and Tianneng Shi and Jingxuan He and Matthew Cai and Jialin Zhang and Dawn Song},
      year={2025},
      eprint={2506.02548},
      archivePrefix={arXiv},
      primaryClass={cs.CR},
      url={https://arxiv.org/abs/2506.02548}, 
}

@book{rosenzweig2013alarming,
  title={The alarming trend of cybersecurity breaches and failures in the US government},
  author={Rosenzweig, Paul},
  year={2013},
  publisher={Heritage Foundation}
}

@article{chung2017critical,
  title={Critical infrastructure, cybersecurity, and market failure},
  author={Chung, John J},
  journal={Or. L. Rev.},
  volume={96},
  pages={441},
  year={2017},
  publisher={HeinOnline}
}

@inproceedings{coffey2017ameliorating,
  title={Ameliorating sources of human error in cybersecurity: technological and human-centered approaches},
  author={Coffey, John W and others},
  booktitle={The 8th International Multi-Conference on Complexity, Informatics, and Cybernetics, Pensacola},
  pages={85--88},
  year={2017}
}

@article{el2021overview,
  title={Overview of the impact of human error on cybersecurity based on ISO/IEC 27001 information security management},
  author={El-Bably, Amar Y},
  journal={Journal of Information Security and Cybercrimes Research},
  volume={4},
  number={1},
  pages={95--102},
  year={2021}
}

@article{mohammed2023ai,
  title={AI and Machine Learning in Cybersecurity: Strategies, Threats, and Exploits},
  author={Mohammed, Anwar},
  journal={Innovative Computer Sciences Journal},
  volume={9},
  number={1},
  year={2023}
}

@article{al2023chatgpt,
  title={Chatgpt for cybersecurity: practical applications, challenges, and future directions},
  author={Al-Hawawreh, Muna and Aljuhani, Ahamed and Jararweh, Yaser},
  journal={Cluster Computing},
  volume={26},
  number={6},
  pages={3421--3436},
  year={2023},
  publisher={Springer}
}

@article{kadavath2022language,
  title={Language models (mostly) know what they know},
  author={Kadavath, Saurav and Conerly, Tom and Askell, Amanda and Henighan, Tom and Drain, Dawn and Perez, Ethan and Schiefer, Nicholas and Hatfield-Dodds, Zac and DasSarma, Nova and Tran-Johnson, Eli and others},
  journal={arXiv preprint arXiv:2207.05221},
  year={2022}
}

@article{jiang2021can,
  title={How can we know when language models know? on the calibration of language models for question answering},
  author={Jiang, Zhengbao and Araki, Jun and Ding, Haibo and Neubig, Graham},
  journal={Transactions of the Association for Computational Linguistics},
  volume={9},
  pages={962--977},
  year={2021},
  publisher={MIT Press One Rogers Street, Cambridge, MA 02142-1209, USA journals-info~…}
}

@misc{open_llm_leaderboard2025,
  title        = {Open LLM Leaderboard – a Hugging Face Space},
  author       = {{open-llm-leaderboard}},
  howpublished = {\url{https://huggingface.co/spaces/open-llm-leaderboard/open_llm_leaderboard\#/}},
  year         = {2025},
  note         = {Accessed: 2025-11-14}
}

@article{hendryckstest2021MMLU,
      title={Measuring Massive Multitask Language Understanding},
      author={Dan Hendrycks and Collin Burns and Steven Basart and Andy Zou and Mantas Mazeika and Dawn Song and Jacob Steinhardt},
      journal={Proceedings of the International Conference on Learning Representations (ICLR)},
      year={2021}
    }

@article{cobbe2021gsm8k,
  title={Training Verifiers to Solve Math Word Problems},
  author={Cobbe, Karl and Kosaraju, Vineet and Bavarian, Mohammad and Chen, Mark and Jun, Heewoo and Kaiser, Lukasz and Plappert, Matthias and Tworek, Jerry and Hilton, Jacob and Nakano, Reiichiro and Hesse, Christopher and Schulman, John},
  journal={arXiv preprint arXiv:2110.14168},
  year={2021}
}

@misc{chen2021evaluating,
      title={Evaluating Large Language Models Trained on Code},
      author={Mark Chen and Jerry Tworek and others},
      year={2021},
      eprint={2107.03374},
      archivePrefix={arXiv},
      primaryClass={cs.LG}
}

@article{lightman2023math500,
      title={Let's Verify Step by Step}, 
      author={Lightman, Hunter and Kosaraju, Vineet and Burda, Yura and Edwards, Harri and Baker, Bowen and Lee, Teddy and Leike, Jan and Schulman, John and Sutskever, Ilya and Cobbe, Karl},
      journal={arXiv preprint arXiv:2305.20050},
      year={2023}
}

@misc{rein2023gpqagraduatelevelgoogleproofqa,
      title={GPQA: A Graduate-Level Google-Proof Q\&A Benchmark}, 
      author={David Rein and Betty Li Hou and Asa Cooper Stickland and Jackson Petty and Richard Yuanzhe Pang and Julien Dirani and Julian Michael and Samuel R. Bowman},
      year={2023},
      eprint={2311.12022},
      archivePrefix={arXiv},
      primaryClass={cs.AI},
      url={https://arxiv.org/abs/2311.12022}, 
}

@misc{balunovic2025aime,
  title = {MathArena: Evaluating LLMs on Uncontaminated Math Competitions},
  author = {Mislav Balunović and Jasper Dekoninck and Ivo Petrov and Nikola Jovanović and Martin Vechev},
  copyright = {MIT},
  url = {https://matharena.ai/},
  publisher = {SRI Lab, ETH Zurich},
  month = feb,
  year = {2025},
}

@misc{arora2025healthbench,
      title={HealthBench: Evaluating Large Language Models Towards Improved Human Health}, 
      author={Rahul K. Arora and Jason Wei and Rebecca Soskin Hicks and Preston Bowman and Joaquin Quiñonero-Candela and Foivos Tsimpourlas and Michael Sharman and Meghan Shah and Andrea Vallone and Alex Beutel and Johannes Heidecke and Karan Singhal},
      year={2025},
      eprint={2505.08775},
      archivePrefix={arXiv},
      primaryClass={cs.CL},
      url={https://arxiv.org/abs/2505.08775}, 
}

@inproceedings{ding2025citations,
  title={Citations and trust in llm generated responses},
  author={Ding, Yifan and Facciani, Matthew and Joyce, Ellen and Poudel, Amrit and Bhattacharya, Sanmitra and Veeramani, Balaji and Aguinaga, Sal and Weninger, Tim},
  booktitle={Proceedings of the AAAI Conference on Artificial Intelligence},
  volume={39},
  number={22},
  pages={23787--23795},
  year={2025}
}

@article{li2025constructing,
  title={Constructing formal models of cryptographic protocols from Alice\&Bob style specifications via LLM},
  author={Li, Qiang and Han, Jihong and Yuan, Lin and Li, Xiangcheng and Wang, Xiaoyu},
  journal={Scientific Reports},
  volume={15},
  number={1},
  pages={11866},
  year={2025},
  publisher={Nature Publishing Group UK London}
}

@inproceedings{gligoroski2025llm,
  title={An LLM Framework For Cryptography Over Chat Channels},
  author={Gligoroski, Danilo and Raikwar, Mayank},
  booktitle={Proceedings of the Workshop on Privacy in Large Language Models (LLM) and Natural Language Processing (NLP) 2025},
  pages={32--43},
  year={2025}
}

\appendix
\crefalias{section}{appendix}
\section*{Appendix}
In this appendix, we adopt a color-coded convention to visually distinguish the different components of our examples: red-shaded boxes indicate prompts, green-shaded boxes denote data samples, and yellow-shaded boxes correspond to model or human responses.

\section{Further Dataset Statistics}\label{apx:stats}
\begin{figure}[h]
    \centering    \includegraphics[width=1\linewidth]{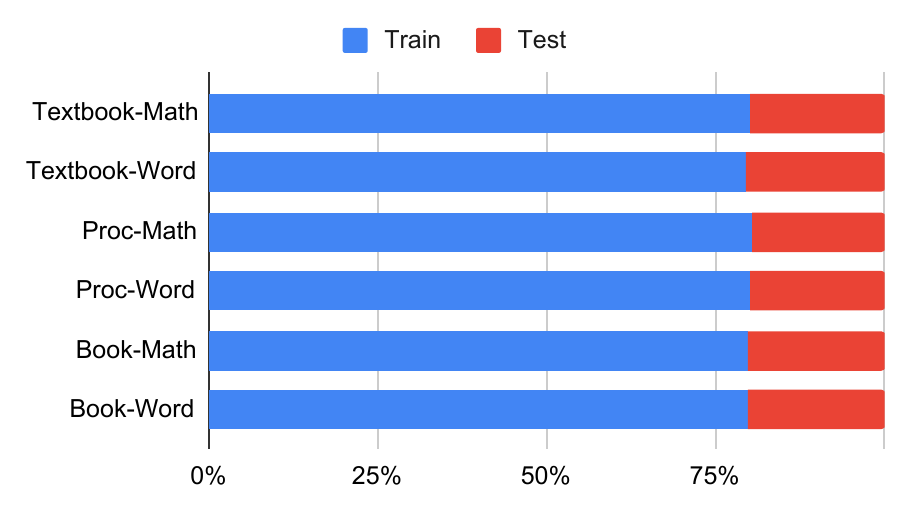}
    \caption{The \dataName's train and test subset split (80-20), respecting the metadata distributions}
    \label{fig:stats-percent}
\end{figure}

\begin{figure}[h]
    \centering
    \includegraphics[width=1\linewidth]{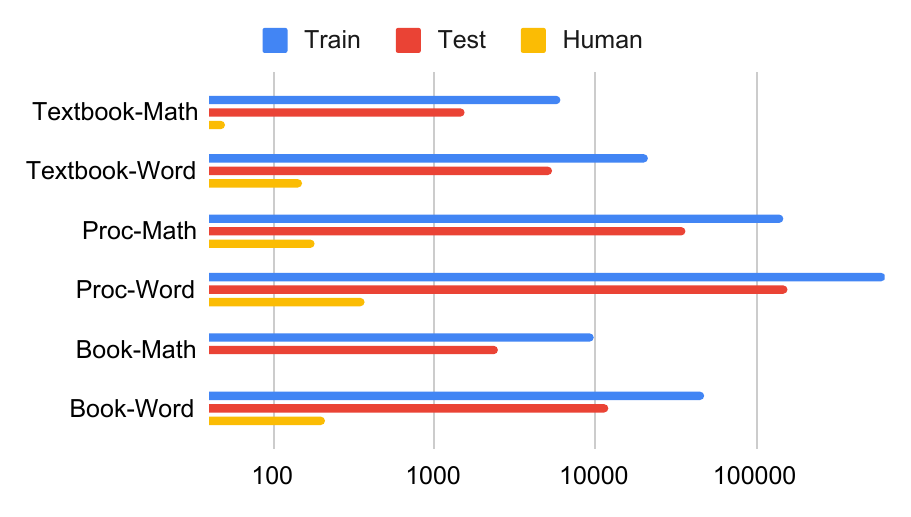}
    \caption{The number of samples in the \dataName's train, test, and human subsets, respecting the metadata distributions (on a log scale for visual clarity)}
    \label{fig:stats-log}
\end{figure}

\begin{figure*}
    \centering
    \includegraphics[width=1\linewidth]{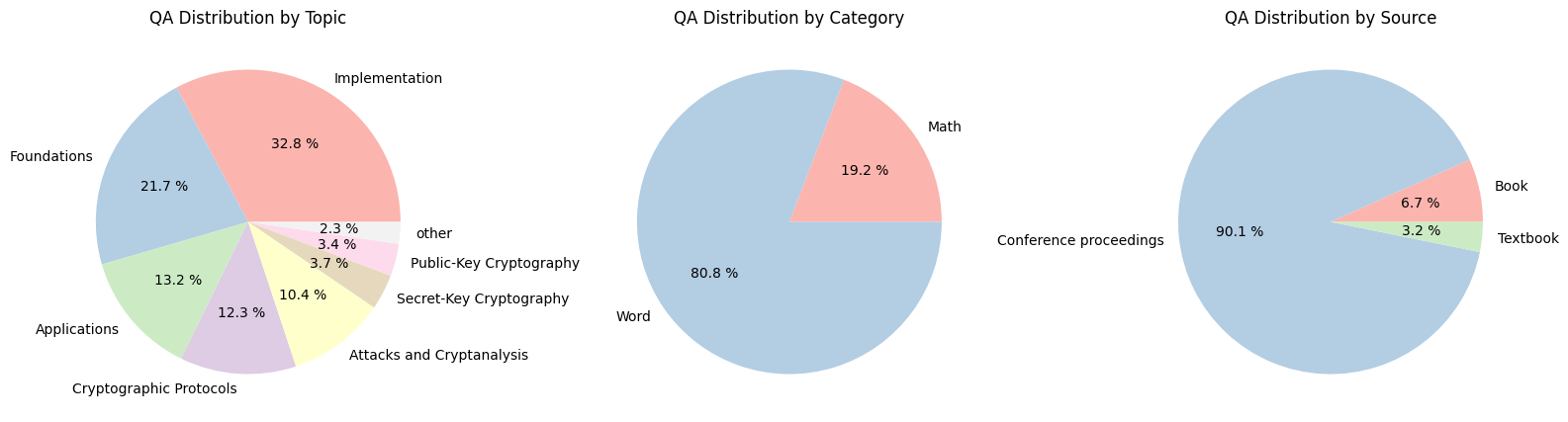}
    \caption{\dataName's statistics per metadata attribute}
    \label{fig:stats}
\end{figure*}

In addition to the statistics presented in Section \ref{sec:dataset}, we show in Figure \ref{fig:stats} the dataset distribution for each metadata attribute. We also show in Figures \ref{fig:stats-percent} and \ref{fig:stats-log} how the dataset was split intro training and test or subsampled for our human evaluation, respectively.

\section{QA Samples of the CryptoQA Dataset}\label{apx:examples}

In this section, we provide further examples of our dataset. We provide two random QA pairs for each metadata class as follows:

\subsection{Topics:}
\begin{itemize}[leftmargin = *]
    \item \textbf{Applications:}
    \begin{enumerate}[leftmargin = *]
        \item
        \begin{shadedquotationn}
        \textbf{Q:} \textit{What is the purpose of OAEP padding in RSA?}
        \\
        \textbf{A:} \textit{OAEP ensures probabilistic encryption (same plaintext yields different ciphertexts) and prevents attacks like chosen plaintext and malleability by mixing the message with randomness.}
        \end{shadedquotationn}
        
        \item 
        \begin{shadedquotationn}
        \textbf{Q:} \textit{What distinguishes dot matrix 2D barcodes (e.g., QR codes) from stacked barcodes in VC applications?}
        \\
        \textbf{A:} \textit{Dot matrix barcodes (like QR codes) use uniform dot arrays ideal for binary VC shares, while stacked barcodes inherit 1D limitations and are less space-efficient.}
        \end{shadedquotationn}
    \end{enumerate}

    \item \textbf{Cryptographic Protocols:}
    \begin{enumerate}[leftmargin = *]

        \item 
        \begin{shadedquotationn}
        \textbf{Q:} \textit{What is the key-switching step in FHE-DiNN, and why is it used?}
        \\
        \textbf{A:} \textit{Key-switching converts high-dimension LWE ciphertexts (e.g., \(N=1024\)) to lower dimensions (e.g., \(n=450\)) to reduce bootstrapping costs.}
        \end{shadedquotationn}

        \item 
        \begin{shadedquotationn}
        \textbf{Q:} \textit{What is the role of the identity set **ID** in the CL-Sign algorithm?}
        \\
        \textbf{A:} \textit{**ID** binds all signers’ identities to the multi-signature, ensuring accountability and preventing rogue-key attacks.}
        \end{shadedquotationn}

    \end{enumerate}

    \item \textbf{Foundations:}
    \begin{enumerate}[leftmargin = *]
        \item 
        \begin{shadedquotationn}
        \textbf{Q:}  \textit{What is the role of the Lagrangian parameter ($\mu$) in kriging?}
        \\
        \textbf{A:} \textit{$\mu$ is used in the kriging equations to ensure the sum of weights ($\lambda_i$) equals 1 for unbiased estimation.}
        \end{shadedquotationn}

        \item 
        \begin{shadedquotationn}
        \textbf{Q:} \textit{What is the main difference between symmetric and asymmetric key cryptography?}
        \\
        \textbf{A:} \textit{Symmetric cryptography uses the same key for encryption/decryption, while asymmetric cryptography (e.g., ElGamal) requires distinct keys, enabling secure key exchange.}
        \end{shadedquotationn}
    \end{enumerate}

\end{itemize}

\subsection{Categories:}

\begin{itemize}[leftmargin = *]
    \item \textbf{Math:}
    \begin{enumerate}[leftmargin = *]
        \item 
        \begin{shadedquotationn}
        \textbf{Q:} \textit{What is the spectrum entropy formula?}
        \\
        \textbf{A:} \textit{Spectrum entropy is:\[ E_S = -\sum_{i=1}^k \frac{\lambda_i}{\sum_{i=1}^k \lambda_i} \log \frac{\lambda_i}{\sum_{i=1}^k \lambda_i} \] where \( \lambda_i \) are eigenvalues of the network.}
        \end{shadedquotationn}
        \item 
        \begin{shadedquotationn}
        \textbf{Q:} \textit{What condition ensures full transmission (\(\xi = 1\)) in the Fabry-Perot interferometer?}
        \\
        \textbf{A:} \textit{The condition \(2kD = \pi\) (modulo \(2\pi\)) ensures full transmission, making \(|\xi| = 1\) and \(\eta = 0\).}
        \end{shadedquotationn}

    \end{enumerate}

    \item \textbf{Word:}
    \begin{enumerate}[leftmargin = *]
        \item 
        \begin{shadedquotationn}
        \textbf{Q:} \textit{What is the purpose of the 'soundness' property in transferable e-cash?}
        \\
        \textbf{A:} \textit{It guarantees that if an honest user accepts a coin during withdrawal or transfer, others (including the bank) will also accept it.}
        \end{shadedquotationn}

        \item 
        \begin{shadedquotationn}
        \textbf{Q:} \textit{How does TinyTable compare to Yao’s garbled circuits in malicious security?}
        \\
        \textbf{A:} \textit{TinyTable has lower communication but higher storage, while garbled circuits have lower rounds but higher computation.}
        \end{shadedquotationn}
    \end{enumerate}
\end{itemize}

\subsection{Sources:}

\begin{itemize}[leftmargin = *]
    \item \textbf{Books:}
    \begin{enumerate}[leftmargin = *]
        \item 
        \begin{shadedquotationn}
        \textbf{Q:} \textit{What is the Rényi entropy for a Laplace-distributed random variable \( X \sim \mathcal{L}^p(\mu, \Sigma) \)?}
        \\
        \textbf{A:} \textit{\( \mathrm{R}_\alpha(X) = p \frac{\log \alpha}{\alpha-1} + L(\Sigma) \), where \( L(\Sigma) \) is a term depending on \( \Sigma \).}
        \end{shadedquotationn}
        
        \item 
        \begin{shadedquotationn}
        \textbf{Q:} \textit{What is the fundamental principle behind neural cryptography?}
        \\
        \textbf{A:} \textit{Neural cryptography uses the synchronization property of neural networks to generate identical secret keys between two parties (Alice and Bob) while making it computationally infeasible for an eavesdropper (Eve) to derive the key.}
        \end{shadedquotationn}
        
    \end{enumerate}

    \item \textbf{Textbooks:}
    \begin{enumerate}[leftmargin = *]
        \item 
        \begin{shadedquotationn}
        \textbf{Q:} \textit{How is Euclid’s algorithm applied to polynomials?}
        \\
        \textbf{A:} \textit{Euclid’s algorithm for polynomials involves repeated polynomial division to find the greatest common divisor (GCD) of two polynomials. It can also be used to solve polynomial Diophantine equations of the form \( S(z)P(z)/P_i(z) + T(z)P_i(z) = 1 \).}
        \end{shadedquotationn}
        
        \item
        \begin{shadedquotationn}
        \textbf{Q:} \textit{What is the role of the General Data Protection Regulation (GDPR) in location privacy?}
        \\
        \textbf{A:} \textit{GDPR classifies location data as personally identifiable information (Article 4(1)) and highlights risks like movement analysis (Recital 75), mandating safeguards for its collection and processing.}
        \end{shadedquotationn}

    \end{enumerate}

    \item \textbf{Conference Proceedings:}
    \begin{enumerate}[leftmargin = *]
        \item 
        \begin{shadedquotationn}
        \textbf{Q:} \textit{What CVE was assigned to one of IFuzzer's findings?}
        \\
        \textbf{A:} \textit{CVE-2015-4507, a moderate-severity bug in SpiderMonkey's SavedStacks class.}
        \end{shadedquotationn}

        \item 
        \begin{shadedquotationn}
        \textbf{Q:} \textit{What is the probability of the differential transition \(1 \rightarrow 8\) through the TRIFLE-BC S-box?}
        \\
        \textbf{A:} \textit{The probability is \(2^{-3}\), as seen in the DDT (Table 2).}
        \end{shadedquotationn}

    \end{enumerate}
\end{itemize}

\section{LLM comparison} \label{apx:llm-comparison}

In \cref{tab:llm_comparison}, we compare our LLM candidates in terms of context length (i.e. the maximum number of tokens that the LLM can take into account at once; e.g. a book chapter), the number of parameters (i.e. the internal weights and biases the model learns during training, active parameters are the subset of those weights that are actually used during a single forward pass during inference), the input and output price (according to the OpenAI/OpenRouter API \cite{openrouter}), quantization (i.e. the process of reducing the precision of the numbers (weights and activations) used in the model, in order to make it smaller and faster to run), reasoning (i.e. the model’s ability to follow logical steps, connect facts, and solve multi-step problems), and caching (i.e. the storage of intermediate values, e.g. key–value pairs from the self-attention mechanism for later re-usage).

These are the SOTA LLM models at the time of writing this paper. However, our dataset can be used with future LLMs. The LLM trends can be found in \hyperlink{https://openrouter.ai/rankings}{OpenRouter Rankings}.

\begin{table*}
    \centering
    \small
    \caption{Comparison of the LLM candidates}
    \begin{tabular}{l|ccccccccc}
        \textbf{LLM candidates} & \textbf{Context Length} & \textbf{\#Parameters} & \textbf{Input Price}  & \textbf{Output Price}  & \textbf{Quantization} & \textbf{Reasoning} & \textbf{Caching}\\
        \hline
        \textbf{Open-source LLMs:}\\
        mistralai/mixtral-8x22b-instruct & 66K & 12.9B active/141B & \$0.90/M tokens & \$0.90/M tokens & unknown & x & x  \\
        qwen/qwen-2.5-72b-instruct & 33K & 72B & \$0.07/M tokens & \$0.26/M tokens & bf16 & x & x  \\
        deepseek/deepseek-chat (V3) & 164K & 37B active/671B & \$0.25/M tokens & \$1/M tokens & unknown  & x & x \\
        deepseek/deepseek-r1 & 164K & 37B active/671B & \$0.40/M tokens & \$2/M tokens & fp8 & $\checkmark$ &  x \\
        \hline
        \textbf{Closed-source LLMs:}\\
        meta-llama/llama-3.1-70b-instruct & 131K & 70B  &\$0.10/M tokens & \$0.28/M tokens & fp8 & x & x \\
        meta-llama/llama-3.3-70b-instruct & 131K & 70B & \$0.04/M tokens & \$0.12/M tokens & bf16 & x & x  \\     
        openai/gpt-4o & 128K & 70B & \$2.50/M tokens & \$10/M tokens & unknown & x &  $\checkmark$ \\
        openai/gpt-4o-mini & 128K & 8B & \$0.15/M tokens & \$0.60/M tokens & unknown & x & $\checkmark$  \\
        openai/o1 & 200K & unknown & \$15/M tokens & \$60/M tokens & unknown & x &  $\checkmark$ \\
        openai/o1-mini & 128K & unknown & \$1.10/M tokens & \$4.40/M tokens & unknown & $\checkmark$ &  $\checkmark$   \\
        openai/gpt-5 & 400K & 635B & \$0.625/M tokens & \$5/M tokens & unknown & $\checkmark$ &  $\checkmark$ \\
        x-ai/grok-4  & 256K & 1.7B & \$3/M tokens & \$15/M tokens &  unknown & $\checkmark$ &  $\checkmark$ \\
        google/gemini-2.5-pro & 1.05M & unknown & \$1.25/M tokens & \$10/M tokens & unknown & $\checkmark$ &  $\checkmark$ \\
        google/gemini-2.5-flash & 1.05M & unknown & \$0.30/M tokens & \$2.50/M tokens & unknown & $\checkmark$ &  $\checkmark$  \\
        anthropic/claude-3.5-sonnet & 200K & unknown & \$3/M tokens & \$15/M tokens & unknown & x & $\checkmark$  \\
    \end{tabular}
    \label{tab:llm_comparison}
\end{table*}

We provide a short description for each LLM, along with their performance on generic capabilities sourced from official model cards, papers, and reputable third-party evaluations (like the Hugging Face Open LLM Leaderboard \cite{open_llm_leaderboard2025}). We report scores on three main datasets: MMLU (Massive Multitask Language Understanding) \cite{hendryckstest2021MMLU}, GSM8K (Math) \cite{cobbe2021gsm8k}, and HumanEval (Code) \cite{chen2021evaluating}.
Note that performance can vary based on the exact prompting and evaluation setup. Hence, direct comparisons should be interpreted cautiously.

\begin{itemize}[leftmargin = *]
    \item \textbf{Mixtral‑8x22B Instruct \cite{jiang2024mixtral}:} Mixtral-8x22B Instruct (Mistral AI) is a sparse Mixture-of-Experts (MoE) model. It has a total of 141B parameters (22B per expert), but only activates 12.9B parameters per token (using 2 out of 8 experts). This makes it as efficient as a 13B model during inference while possessing the knowledge and capacity of a much larger model. It was pre-trained on a massive, multilingual corpus. The "Instruct" version is fine-tuned using Supervised Fine-Tuning (SFT) and Direct Preference Optimization (DPO). It offers an exceptional performance-to-cost ratio; very fast inference; strong in code generation, reasoning, and multilingual tasks. It scored 77.6\% on MMLU, 86.7\% on GSM8K, and 74.4\% on HumanEval.

    \item \textbf{Qwen2.5‑72B Instruct \cite{qwen2025technicalreport}:} Qwen2.5-72B-Instruct (Alibaba) is a dense, transformer-based autoregressive model with 72 billion parameters. It is pre-trained on a vast and diverse dataset with a strong focus on multilingual content (high Chinese/English balance). Instruction-tuning employs SFT and RLHF. It is currently the state-of-the-art (SOTA) performant among open-source models, particularly strong in reasoning, coding, and mathematical tasks. Its bilingual training makes it exceptionally capable in Chinese. It scores 86.5\% on MMLU, 93.5\% on GSM8K, 88.4\% on HumanEval, and 90.6\% on C-Eval (Chinese).

    \item \textbf{DeepSeek‑V3 \cite{liu2024deepseek}:} DeepSeek-V3 (DeepSeek) is a massive MoE model with 671B total parameters, but only 37B activated per token. This is a significantly larger and more complex architecture than Mixtral. It is trained on a 2 trillion token multilingual corpus. It scores 85.6\% on MMLU, 95.2\% on GSM8K, and 80.5\% on HUmanEval.

    \item \textbf{DeepSeek‑R1 \cite{deepseekr12025}}: DeepSeek-R1 (DeepSeek) is a Reasoning Specialist Model trained specifically for complex, multi-step reasoning. It is fine-tuned from DeepSeek-V3 using Reinforcement Learning (RL) on trajectories of correct reasoning (process-based reward models). It excels at mathematical problem-solving, scientific reasoning, and complex code debugging. It shows "chain-of-thought" (CoT) reasoning by default. It scored 78.2\% on MATH-500 \cite{lightman2023math500} (significantly higher than its base model) and 56.1\% on GPQA Diamond (expert-level QA) \cite{rein2023gpqagraduatelevelgoogleproofqa}. Although it represents a major shift in the open-source world towards specialized agents for specific cognitive tasks, mirroring the "reasoning model" trend, it is slower and more computationally expensive per token than standard models due to its long internal "thought" process.

    \item \textbf{Llama 3.1‑70B Instruct and Llama 3.3‑70B Instruct \cite{meta2024llama3.1}:} Meta’s Llama-3.1-70B-Instruct and Llama-3.3-70B-Instruct are dense transformer-based models. Both are instruction-tuned versions of the 70B parameter Llama 3.1 base model. The 3.3-70B is a more recent, highly refined iteration with significant improvements in reasoning and instruction-following, demonstrating the importance of post-pretraining alignment. They scored 82.5\% (3.3) and 80.6\% (3.1) on MMLU, 92.0\% (3.3) and 86.5\% (3.1) on GSM8K, and 84.1\% (3.3) and 81.7\% (3.1) on HumanEval.

    \item \textbf{GPT‑4o and GPT‑4o‑mini \cite{hurst2024gpt}:} OpenAI’s GPT-4o ("omni") and 4o-mini are dense multimodal models (text, audio, vision) from the ground up. GPT-4o is optimized for real-time, multimodal reasoning with low latency. GPT-4o-mini is a highly efficient distilled version designed to be a cost-effective default model. They scored 88.7\% (4o) and 82.1\% (mini) on MMLU, 95.1\% (4o) and 90.7\% (mini) on GSM8K, and 90.5\% (4o) and 76.6\% (mini) on HumanEval. They represent a move towards unified, low-latency multimodal architectures, breaking away from separate component models for different modalities.

    \item \textbf{o1 and o1-mini \cite{openai2024openaio1card}:} OpenAI’s o1 and o1-mini are Reasoning-optimized Models. They use a search-augmented reasoning process, spending significantly more compute to 'think' before producing an answer. They move beyond next-token prediction towards a deliberative process. The model is trained to use a chain-of-thought that is scored and guided by a reward model, leading to dramatically higher reasoning accuracy. O1 reported to exceed 95\% on MATH. It sets a new SOTA on benchmarks like GPQA (78\%) \cite{rein2023gpqagraduatelevelgoogleproofqa} and AIME (Math - 83.3\%) \cite{balunovic2025aime}. However, it is extremely slow (latency in the minutes) and expensive compared to standard models. The 'thought' process is often internal and not fully visible. O1 is a paradigm shift in LLM design, prioritizing 'right-for-a-reason' answers over 'fast' ones. It validates the approach of using process-based supervision for complex reasoning.

    \item \textbf{GPT-5 \cite{openai2025gpt5}:} GPT-5 is described as OpenAI’s next-generation generalist model, achieving significant improvements in reasoning, multimodal integration and long-context comprehension. It scores 94.6\% on AIME, 74.9\% on SWE-Bench Verified (coding), 84.2\% on MMMU (multimodal), and 46.2\% on HealthBench Hard \cite{arora2025healthbench}. Overall, GPT-5 excels in scientific and mathematical reasoning, coding, healthcare analytics, and multimodal understanding, offering higher factual reliability and steerability than previous OpenAI models. However, it remains closed-source, with limited transparency on training data, architecture, and alignment methodology.

    \item \textbf{Grok-4 \cite{xai2025grok4}:} X-AI’s Grok-4 is a dense transformer model, significantly larger and more capable than its predecessor Grok-1. It is known for its integration with the X (Twitter) platform for real-time knowledge and a distinct, humorous personality in its responses. It scored 84\% on MMLU, 90\% on GSM8K, and 74\% on HumanEval.

    \item \textbf{Gemini-2.5-pro and gemini-2.5-flash \cite{comanici2025gemini25pushingfrontier}:} Goggle DeepMind’s Gemini 2.5 Pro and  Gemini 2.5 Flash are proprietary, multimodal (text, image, video, audio) models. Gemini 2.5 Pro uses a novel MoE architecture, while Flash is a distilled, highly efficient version for speed. The original Pro version scored 91.1\% on MMLU, 95.4\% on GSM8K, and 91.1\% on HumanEval. It pushes the boundaries of context length and efficient MoE design for multimodal models, enabling entirely new applications involving long documents or video analysis.

    \item \textbf{Claude-3.5-sonnet \cite{anthropic2024claude35sonnet}: } Anthropic’s Claude 3.5 Sonnet is a proprietary dense transformer. It focuses on robust safety, reduced refusal rates on benign requests, and strong 'steerability'. It is known for its high 'honesty' and low rates of hallucination. It scored 88.3\% on MMLU, 96.2\% on GSM8K, and 91.5\% on HumanEval. It highlights the importance of model alignment and safety as a core, differentiable feature, proving that high capability and strong safety measures are not mutually exclusive.

\end{itemize}

\section{Paraphrasing metrics} \label{apx: paraphrasing}

The paraphrasing outputs exhibit moderate quality across evaluation metrics: 
Lexical overlap scores are consistent with partial preservation of reference content, with ROUGE-1 = 0.560, ROUGE-2 = 0.340, ROUGE-L = 0.514, and ROUGE-Lsum = 0.516, indicating reasonable structural and content alignment. 
The BLEU score (0.264) is relatively low, which is expected in paraphrasing tasks where surface similarity is not the primary goal. 
In contrast, METEOR (0.567) reflects stronger semantic adequacy due to its sensitivity to synonymy and morphological variation. Collectively, the results suggest that the model maintains semantic fidelity while introducing lexical diversity, as shown in the examples below:

\begin{enumerate}[leftmargin = *]
    \item 
    \begin{shadedanswer}
        \textbf{Q-orig:} What method does the JRTM employ to progressively adjust the risk probabilities $r_{\varepsilon(i)}, r_{\phi(i)}, r_{\rho(i)}$ over time?\\
        \textbf{Q-para:} How does JRTM adjust risk probabilities (\( r_{\varepsilon(i)}, r_{\phi(i)}, r_{\rho(i)} \)) over time?
    \end{shadedanswer}

    \item 
    \begin{shadedanswer}
        \textbf{Q-orig:} Using the constraints \(a+c+d=l\) and \(a'+c'+d'=h\), express how the two basis matrices \(M_{0}\) and \(M_{1}\) in a \((2,2)\)-visual cryptography scheme are related to one another.\\
        \textbf{Q-para:} Derive the relationship between the basis matrices \(M_0\) and \(M_1\) for a (2,2)-VCS, given \(a + c + d = l\) and \(a' + c' + d' = h\).
    \end{shadedanswer}
\end{enumerate}

\section{Confidence Metrics}\label{apx:confidence}
When a language model produces text sequentially, it predicts one token at a time by assigning a probability distribution over the entire vocabulary at each generation step. The log-probability associated with the chosen token can be interpreted as a measure of the model’s internal confidence that this token is the most appropriate continuation given the preceding context. By summing or averaging these log-probabilities across all tokens in a generated response, we can obtain an estimate of the model’s overall confidence or self-assessed likelihood for the complete output. This aggregated metric provides insight into which answers the model “believes” are more probable, enabling analyses of reliability, uncertainty, or alignment with the input prompt.

For instance, in the following example, the model assigns higher log-probabilities (i.e., greater confidence) to common functional tokens such as 'the', and lower values to content words like 'difficulty', which admit more alternatives. The sum of all token log-probabilities represents the model’s overall confidence in the generated answer sequence.

\begin{shadedquotation}
        \textbf{Confidence example:} \\
        \textbf{Q:} \textit{What is the primary security guarantee of the RSA algorithm?}\\
        \textbf{A:} 
            \begin{tabular}{c|c|c}
              \textbf{Token}	& \textbf{Probability}	& \textbf{Log-probability}\\
            'The'	& 0.35	& -1.05\\
            'primary' &	0.22 &	-1.51\\
            'security'	& 0.18 &	-1.71\\
           .	& .	& .\\
            'difficulty & 0.40 & -0.92
            \end{tabular}
    \end{shadedquotation}

\section{Other LLMs as Judges}\label{apx:other-judges}
To mitigate source-model bias in our evaluation, we employed other LLMs—specifically \texttt{GPT-5} and \texttt{Claude-3.5-Sonnet}—as judges, together with \texttt{DeepSeek-V3}. Because these models are likely to appear as underlying assistants in many of the papers included in our corpus, incorporating them as evaluators reduces the risk that our results disproportionately reflect the behaviour or stylistic tendencies of a single source model (e.g. \texttt{DeepSeek-V3}).

Moreover, leveraging multiple high-performing models enables cross-model validation of judgments. The evaluation outcomes produced by \texttt{GPT-5} and \texttt{Claude-3.5-Sonnet} remained consistent with those obtained from our primary setup. This consistency suggests that our conclusions are robust and not an artifact of any individual model’s inductive biases or evaluation idiosyncrasies.

\begin{figure}[h]
    \centering
    \includegraphics[width=0.98\linewidth]{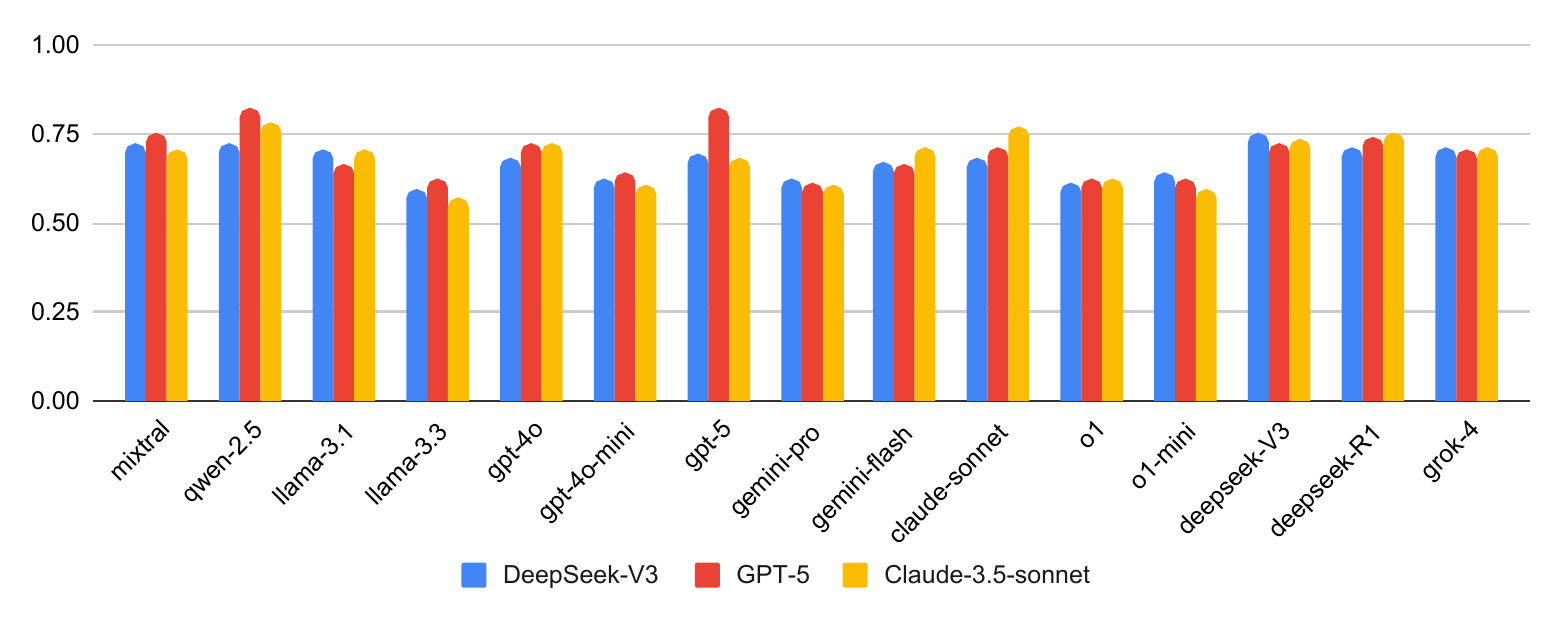}
    \caption{Ratings of different LLMs as judges}
    \label{fig:judges}
\end{figure}

As shown in \cref{fig:judges}, across all LLMs evaluated, LLM-as-judge assessments exhibit highly consistent rating patterns. At the same time, each LLM systematically assigns higher scores to its own generations, a well-documented form of self-preferential bias as shown in \cref{sec:results}.

\section{Human Evaluation} \label{apx:human_evaluation}
Here, we provide two sample questions from our user survey along with two random answers from the participants:
\begin{enumerate}[leftmargin = *]
    \item \begin{shadedanswer}
        \textbf{Q1:} What is the significance of exponential growth in cryptographic key spaces?\\
        \textbf{Participant-answer (BSc):}
        It avoids brute-force attacks, attackers cannot try all possible keys.
        \\
        \textbf{LLM-answer:} Exponential key-space growth ($2^n$) means each added bit doubles the number of possible keys, so brute-force attack cost explodes rapidly. Thus, small increases in key length yield huge security gains and keep exhaustive search infeasible despite hardware advances and parallelism. Conversely, any attack that cuts this growth (e.g., subexponential or quadratic speedups) drastically weakens security and forces larger keys.
        \\
        \textbf{Correctness:} 4, \textbf{Completeness:} 4, \textbf{Clarity:} 5
    \end{shadedanswer}

    \item \begin{shadedanswer}
        \textbf{Q2:} What is the Taylor series expansion of $\frac{e^x - 1}{x}$ up to order 10?\\
        \textbf{Participant-answer (PhD):} $1 + x/2 + (x^2)/2 + (x^3)/24 + ...$
        \\
        \textbf{LLM-answer:}
        The Taylor series expansion of $\frac{e^x - 1}{x}$ up to order 10 is: $\frac{e^x - 1}{x} = 1 + \frac{x}{2} + \frac{x^2}{6} + \frac{x^3}{24} + \frac{x^4}{120} + \frac{x^5}{720} + \frac{x^6}{5040} + \frac{x^7}{40320} + \frac{x^8}{362880} + \frac{x^9}{3628800} + \mathcal{O}(x^{10})$
        \\
        \textbf{Correctness:} 5, \textbf{Completeness:} 4, \textbf{Clarity:} 5
        
    \end{shadedanswer}
\end{enumerate}

\section{Sample Responses}\label{apx:sample_responses}
In this section, we provide examples of responses generated by the evaluated LLMs, accompanied by evaluation metrics that assess their performance. We highlight some interesting cases in \textcolor{red}{red}.

\begin{enumerate}[leftmargin = *]
    \item Responses of all LLMs on the same question with type 'word':
    \begin{shadedanswer}
        \textbf{Q:} What is the significance of exponential growth in cryptographic key spaces?\\
        \textbf{A:} Exponential growth (e.g., $ 2^n $ for $n$-bit keys) ensures that brute-force attacks are impractical, as the number of possible keys becomes astronomically large with modest increases in $n$.\\
        \textbf{Metadata:} original, word, foundations, source:	10.1007/978-3-031-59922-4\_5, textbook: 978-3-031-59922-4
        \\
        \\
        \textbf{Mixtral-8x22b-instruct:} Exponential growth in cryptographic key spaces is significant because it ensures that the number of possible keys increases rapidly with key size, making it extremely difficult for attackers to find the correct key through brute force attacks. This property is crucial for maintaining the security of encryption systems, as the time and computational resources required for an attacker to try every possible key combination grow exponentially with key size.\\
        \textbf{Rouge1:} 0.34,
        \textbf{Rouge2:} 0.18, \textbf{ROUGE-L:} 0.22, \textbf{ROUGELSum:} 0.22,
        \textbf{BLEU:} 0.08,
        \textbf{Meteor:} 0.33,
        \textbf{BERT-precision:} 0.60,
        \textbf{BERT-recall:} 0.59,
        \textbf{BERT-F1:} 0.59,
        \textbf{LLM-as-judge:} 0.8\\
        \\
        \textbf{Qwen-2.5-72b-instruct:} Exponential growth in cryptographic key spaces is significant because it ensures that the number of possible keys increases dramatically with each additional bit, making brute-force attacks computationally infeasible. This rapid expansion is crucial for maintaining the security and integrity of encrypted data against unauthorized access.\\
        \textbf{Rouge1:} 0.55, \textbf{ROUGE-2:} 0.42,
        \textbf{ROUGE-L:} 0.53,
        \textbf{ROUGELSum:} 0.53,
        \textbf{BLEU:} 0.32,
        \textbf{Meteor:} 0.59,
        \textbf{BERT-precision:} 0.75,
        \textbf{BERT-recall:} 0.78,
        \textbf{BERT-F1:} 0.77,
        \textbf{LLM-as-judge:} 0.9\\
        \\
        \textbf{Deepseek-V3:} Because key spaces grow exponentially (e.g., $2^n$ for an $n$-bit key), even small increases in $n$ cause the number of possible keys to become overwhelmingly large, rendering brute-force attacks computationally infeasible.\\
        \textbf{Rouge1:} 0.57, 
        \textbf{Rouge2:} 0.40, \textbf{ROUGE-L:} 0.39, \textbf{ROUGELSum:} 0.39,
        \textbf{BLEU:} 0.43,
        \textbf{Meteor:} 0.76,
        \textbf{BERT-precision:} 0.79,
        \textbf{BERT-recall:} 0.83,
        \textbf{BERT-F1:} 0.81,
        \textbf{LLM-as-judge:} 0.9\\
        \\
        \textbf{Deepseek-r1:} Exponential growth in cryptographic key spaces is fundamental to security because it makes brute-force attacks computationally infeasible. Each additional bit in a key doubles the number of possible keys, causing the search space to expand at an astronomical rate. This ensures that even with vast computational resources, an attacker cannot guess the correct key in a reasonable amount of time.\\
        \textbf{Rouge1:} 0.39, 
        \textbf{Rouge2:} 0.19, \textbf{ROUGE-L:} 0.20, \textbf{ROUGELSum:} 0.20,
        \textbf{BLEU:} 0.12,
        \textbf{Meteor:} 0.25,
        \textbf{BERT-precision:} 0.79,
        \textbf{BERT-recall:} 0.83,
        \textbf{BERT-F1:} 0.81,
        \textbf{LLM-as-judge:} 0.8
        \\
        \\
        \textbf{Llama-3.1-70b-instruct:} Exponential growth in cryptographic key spaces means that a small increase in key size leads to an enormous increase in the number of possible keys, making it computationally infeasible for an attacker to try all possible keys through brute force. This property provides strong security guarantees, as even the fastest computers would take an impractically long time to exhaustively search the key space. As a result, cryptographic systems can rely on keys of moderate size to achieve robust security.\\
        \textbf{Rouge1:} 0.25, 
        \textbf{Rouge2:} 0.11, \textbf{ROUGE-L:} 0.16, \textbf{ROUGELSum:} 0.16,
        \textbf{BLEU:} 0.05,
        \textbf{Meteor:} 0.27,
        \textbf{BERT-precision:} 0.58,
        \textbf{BERT-recall:} 0.59,
        \textbf{BERT-F1:} 0.58,
        \textbf{LLM-as-judge:} 0.7\\
        \\
        \textbf{Llama-3.3-70b-instruct:} Exponential growth in cryptographic key spaces refers to the rapid increase in the number of possible keys as the key size increases, making it extremely difficult for attackers to brute-force guess the key. This property is significant because it provides a high level of security against brute-force attacks, as the number of possible keys becomes virtually ung guessable with even moderately sized keys. As a result, exponential growth in key spaces is a fundamental principle in cryptography, enabling the creation of secure encryption systems that can protect sensitive information.\\
        \textbf{Rouge1:} 0.29, 
        \textbf{Rouge2:} 0.15, \textbf{ROUGE-L:} 0.24, \textbf{ROUGELSum:} 0.24,
        \textbf{BLEU:} 0.09,
        \textbf{Meteor:} 0.28,
        \textbf{BERT-precision:} 0.58,
        \textbf{BERT-recall:} 0.59,
        \textbf{BERT-F1:} 0.58,
        \textbf{LLM-as-judge:} 0.8\\
        \\
        \textbf{GPT-4o :} Exponential growth in cryptographic key spaces is significant because it vastly increases the number of potential keys, making it computationally infeasible for adversaries to perform brute force attacks to guess the correct key. As the key length increases, the possible combinations grow exponentially, thereby strengthening the security and resilience of cryptographic systems against unauthorized decryption. This exponential growth ensures that even rapid advancements in computational power do not easily compromise the security provided by long key lengths.\\
        \textbf{Rouge1:} 0.29, 
        \textbf{Rouge2:} 0.13, \textbf{ROUGE-L:} 0.18, \textbf{ROUGELSum:} 0.18,
        \textbf{BLEU:} 0.0,
        \textbf{Meteor:} 0.26,
        \textbf{BERT-precision:} 0.56,
        \textbf{BERT-recall:} 0.57,
        \textbf{BERT-F1:} 0.57,
        \textbf{LLM-as-judge:} 0.7 \\
        \\
        \textbf{GPT-4o-mini:} Exponential growth in cryptographic key spaces means that each added bit doubles the number of possible keys, making brute-force attacks computationally infeasible very quickly. This steep increase in search complexity ensures that even modest increases in key length massively boost security against exhaustive key searches.\\
        \textbf{Rouge1:} 0.41, 
        \textbf{Rouge2:} 0.26, \textbf{ROUGE-L:} 0.28, \textbf{ROUGELSum:} 0.28,
        \textbf{BLEU:} 0.12,
        \textbf{Meteor:} 0.40,
        \textbf{BERT-precision:} 0.66,
        \textbf{BERT-recall:} 0.61,
        \textbf{BERT-F1:} 0.63,
        \textbf{LLM-as-judge:} 0.7 \\
        \\
        \textbf{O1:} Exponential growth in cryptographic key spaces means each extra bit doubles the total number of possible keys, making brute force attacks increasingly impractical. It forces attackers to invest exponentially more resources to break the key as its size increases. Consequently, properly chosen key lengths ensure long-term security.\\
        \textbf{Rouge1:} 0.37, 
        \textbf{Rouge2:} 0.15, \textbf{ROUGE-L:} 0.25, \textbf{ROUGELSum:} 0.25,
        \textbf{BLEU:} 0.06,
        \textbf{Meteor:} 0.32,
        \textbf{BERT-precision:} 0.65,
        \textbf{BERT-recall:} 0.62,
        \textbf{BERT-F1:} 0.63,
        \textbf{LLM-as-judge:} 0.7\\
        \\
        \textbf{O1-mini:} Exponential growth in cryptographic key spaces significantly increases the number of possible keys, making brute-force attacks computationally infeasible. As the key length doubles, the effort required to search the entire key space grows exponentially, thereby enhancing the security of cryptographic systems. This scalability ensures that even with advancing technology, the likelihood of successfully cracking the encryption remains exceedingly low.\\
        \textbf{Rouge1:} 0.35, 
        \textbf{Rouge2:} 0.20, \textbf{ROUGE-L:} 0.21, \textbf{ROUGELSum:} 0.21,
        \textbf{BLEU:} 0.08,
        \textbf{Meteor:} 0.35,
        \textbf{BERT-precision:} 0.62,
        \textbf{BERT-recall:} 0.61,
        \textbf{BERT-F1:} 0.61,
        \textbf{LLM-as-judge:} 0.7 \\
        \\
        \textbf{GPT-5:} Exponential key-space growth ($2^n$) means each added bit doubles the number of possible keys, so brute-force attack cost explodes rapidly. Thus, small increases in key length yield huge security gains and keep exhaustive search infeasible despite hardware advances and parallelism. Conversely, any attack that cuts this growth (e.g., subexponential or quadratic speedups) drastically weakens security and forces larger keys.\\
        \textbf{Rouge1:} 0.38, 
        \textbf{Rouge2:} 0.19, \textbf{ROUGE-L:} 0.25, \textbf{ROUGELSum:} 0.25,
        \textbf{BLEU:} 0.15,
        \textbf{Meteor:} 0.4,
        \textbf{BERT-precision:} 0.61,
        \textbf{BERT-recall:} 0.68,
        \textbf{BERT-F1:} 0.64,
        \textbf{LLM-as-judge:} 0.8\\
        \\
        \textbf{Grok-4:} Exponential growth in cryptographic key spaces refers to how the number of possible keys increases dramatically with each additional bit of key length, such as doubling for every bit in a binary system (e.g., $2^n$ possibilities for an n-bit key). This makes brute-force attacks computationally infeasible for sufficiently large keys, as even modern supercomputers would take an impractically long time—potentially billions of years—to test all possibilities. As a result, it underpins the security of modern encryption algorithms, ensuring that breaking them requires more than just raw computing power and instead demands sophisticated cryptanalysis or exploiting weaknesses.\\
        \textbf{Rouge1:} 0.36, 
        \textbf{Rouge2:} 0.16, \textbf{ROUGE-L:} 0.22, \textbf{ROUGELSum:} 0.22,
        \textbf{BLEU:} 0.10,
        \textbf{Meteor:} 0.39,
        \textbf{BERT-precision:} 0.57,
        \textbf{BERT-recall:} 0.71,
        \textbf{BERT-F1:} 0.63,
        \textbf{LLM-as-judge:} 0.8 \\
        \\
        \textbf{Gemini-2.5-pro:} Exponential growth in key spaces drastically increases the number of possible keys with each added bit of key length, making brute-force attacks computationally infeasible.  This ensures strong encryption, as attackers would require impractical amounts of time and resources to try all possible keys. A larger key space directly translates to a higher level of security for the encrypted information.\\
        \textbf{Rouge1:} 0.41, 
        \textbf{Rouge2:} 0.15, \textbf{ROUGE-L:} 0.21, \textbf{ROUGELSum:} 0.21,
        \textbf{BLEU:} 0.08,
        \textbf{Meteor:} 0.35,
        \textbf{BERT-precision:} 0.62,
        \textbf{BERT-recall:} 0.62,
        \textbf{BERT-F1:} 0.62,
        \textbf{LLM-as-judge:} 0.7 \\
        \\
        \textbf{Gemini-2.5-flash:} Exponential growth in cryptographic key spaces represents the vast number of possible keys available. This ensures the security of encryption by making brute-force attacks computationally infeasible, as the time required to try all possible keys increases dramatically. Consequently, larger key spaces effectively protect sensitive information from unauthorized access.\\
        \textbf{Rouge1:} 0.37, 
        \textbf{Rouge2:} 0.17, \textbf{ROUGE-L:} 0.30, \textbf{ROUGELSum:} 0.30,
        \textbf{BLEU:} 0.08,
        \textbf{Meteor:} 0.32,
        \textbf{BERT-precision:} 0.63,
        \textbf{BERT-recall:} 0.61,
        \textbf{BERT-F1:} 0.62,
        \textbf{LLM-as-judge:} 0.7\\
        \\
        \textbf{Claude-3.5-sonnet:} Exponential growth in cryptographic key spaces is crucial for security because it makes brute-force attacks computationally infeasible. As key length increases linearly, the number of possible keys grows exponentially ($2^n$ for n-bit keys), making it increasingly difficult for attackers to try all possibilities. This exponential relationship ensures that even small increases in key length result in massive increases in security against brute-force attacks.\\
        \textbf{Rouge1:} 0.42, 
        \textbf{Rouge2:} 0.29, \textbf{ROUGE-L:} 0.28, \textbf{ROUGELSum:} 0.28,
        \textbf{BLEU:} 0.09,
        \textbf{Meteor:} 0.43,
        \textbf{BERT-precision:} 0.66,
        \textbf{BERT-recall:} 0.68,
        \textbf{BERT-F1:} 0.67,
        \textbf{LLM-as-judge:} 0.7\\
    \end{shadedanswer}
    \item Responses of all LLMs on the same question with type 'math':
    \begin{shadedanswer}
        \textbf{Q:} Convert $(10000)_{10}$ to base-8.\\
        \textbf{A:} Repeated division by 8: 
        $10000 \div 8 = 1250, \quad 1250 \div 8 = 156, \quad 156 \div 8 = 19, \quad 19 \div 8 = 2, \quad 2 \div 8 = 0$.
        Reading remainders backward: $(23420)_8$.\\
        \textbf{Metadata:} original, math, foundations, source:	10.1007/978-3-658-40423-9\_3, textbook: 978-3-658-40423-9
        \\
        \\
        \textbf{Mixtral-8x22b-instruct:} \textcolor{red}{The conversion from base-10 to base-8 involves dividing the base-10 number by 8 repeatedly until the quotient is 0, recording the remainders from each division in reverse order. For $(10000)_{10}$, the divisions yield a quotient of 1250 (which leaves no remainder) after the first division, 156 after the second division (with no remainder), and 19 after the third division (with no remainder), leaving a final quotient of 2 and a remainder of 3. Therefore, $(10000)_{10} = (23000)_8$.}\\
        \textbf{Rouge1:} 0.20,
        \textbf{Rouge2:} 0.01, \textbf{ROUGE-L:} 0.13, \textbf{ROUGELSum:} 0.13,
        \textbf{BLEU:} 0.00,
        \textbf{Meteor:} 0.17,
        \textbf{BERT-precision:} 0.53,
        \textbf{BERT-recall:} 0.60,
        \textbf{BERT-F1:} 0.56,
        \textbf{LLM-as-judge:} 0.4\\
        \\
        \textbf{Qwen-2.5-72b-instruct:} To convert the decimal number $(10000)_{10}$ to base-8, we repeatedly divide by 8 and keep track of the remainders. The process yields the octal number  $(23420)_8$. \\
        \textbf{Rouge1:} 0.22, 
        \textbf{Rouge2:} 0.06, \textbf{ROUGE-L:} 0.19, \textbf{ROUGELSum:} 0.19,
        \textbf{BLEU:} 0.07,
        \textbf{Meteor:} 0.13,
        \textbf{BERT-precision:} 0.59,
        \textbf{BERT-recall:} 0.56,
        \textbf{BERT-F1:} 0.57,
        \textbf{LLM-as-judge:} 0.8\\
        \\
        \textbf{Deepseek-V3:} We convert $(10000)_10$ to base 8 by repeated division.
        $10000 \div 8 = 1250 \quad \text{remainder } 0, 1250 \div 8 = 156 \quad \text{remainder } 2 156 \div 8 = 19 \quad \text{remainder } 4 19 \div 8 = 2 \quad \text{remainder } 3 2 \div 8 = 0 \quad \text{remainder } 2$. Reading the remainders from bottom to top gives $(23420)_8$. Verification: $23420_8 = 2×8^4 + 3×8^3+4×8^2+2×8^1+0×8^0 = 2×4096+3×512+4×64+2×8 = 8192+1536+256+16=10000$\\
        \textbf{Rouge1:} 0.47, 
        \textbf{Rouge2:} 0.28, \textbf{ROUGE-L:} 0.42, \textbf{ROUGELSum:} 0.42,
        \textbf{BLEU:} 0.13,
        \textbf{Meteor:} 0.60,
        \textbf{BERT-precision:} 0.64,
        \textbf{BERT-recall:} 0.80,
        \textbf{BERT-F1:} 0.71,
        \textbf{LLM-as-judge:} 1.0\\
        \\
        \textbf{Deepseek-r1:} To convert the decimal number $10000_{10}$ to base-8, repeatedly divide by 8 and note the remainders:   $10000 ÷ 8 = 1250$ remainder 0,  $1250 ÷ 8 = 156$ remainder 2,  $156 ÷ 8 = 19$ remainder 4,  $19 ÷ 8 = 2$ remainder 3,  $2 ÷ 8 = 0$ remainder 2.  
        Reading the remainders from last to first gives $23420_8$.  Note: The answer is in base-8, as $(23420)_8$.\\
        \textbf{Rouge1:} 0.42, 
        \textbf{Rouge2:} 0.14, \textbf{ROUGE-L:} 0.42, \textbf{ROUGELSum:} 0.42,
        \textbf{BLEU:} 0.09,
        \textbf{Meteor:} 0.51,
        \textbf{BERT-precision:} 0.63,
        \textbf{BERT-recall:} 0.68,
        \textbf{BERT-F1:} 0.66,
        \textbf{LLM-as-judge:} 1.0\\
        \\
        \textbf{Llama-3.1-70b-instruct:}
        \#\# Step 1: To convert a decimal number to base-8 (octal), we need to divide the decimal number by 8 and note the remainder until the quotient becomes less than 8.\\ First, we divide 10000 by 8.\\ \#\# Step 2: We continue the division process with the quotient obtained in the previous step until the quotient is less than 8, noting the remainder at each step.\\ 10000 divided by 8 gives a quotient of 1250 with a remainder of 0. We continue this process: 1250 divided by 8 gives 156 with a remainder of 2, 156 divided by 8 gives 19 with a remainder of 4, 19 divided by 8 gives 2 with a remainder of 3, and finally, 2 divided by 8 gives 0 with a remainder of 2.\\ \#\# Step 3: The base-8 equivalent of the decimal number is obtained by writing the remainders in reverse order.\\ The remainders are 2, 3, 4, 2, 0, which when written in reverse order gives the octal equivalent.\\ The final answer is: $\boxed{23420}$.\\
        \textbf{Rouge1:} 0.20, 
        \textbf{Rouge2:} 0.00, \textbf{ROUGE-L:} 0.18, \textbf{ROUGELSum:} 0.18,
        \textbf{BLEU:} 0.00,
        \textbf{Meteor:} 0.25,
        \textbf{BERT-precision:} 0.48,
        \textbf{BERT-recall:} 0.59,
        \textbf{BERT-F1:} 0.53,
        \textbf{LLM-as-judge:} 0.9\\
        \\
        \textbf{Llama-3.3-70b-instruct:}
        To convert $(10000)_{10}$ to base-8, we divide 10000 by 8 repeatedly and record the remainders. The remainders in reverse order give the base-8 representation: 10000 divided by 8 is 1250 with a remainder of 0, 1250 divided by 8 is 156 with a remainder of 2, 156 divided by 8 is 19 with a remainder of 4, 19 divided by 8 is 2 with a remainder of 3, and 2 divided by 8 is 0 with a remainder of 2. Therefore, $(10000)_{10}$ in base-8 is $(23420)_{8}$.\\
        \textbf{Rouge1:} 0.32, 
        \textbf{Rouge2:} 0.03, \textbf{ROUGE-L:} 0.30, \textbf{ROUGELSum:} 0.30,
        \textbf{BLEU:} 0.03,
        \textbf{Meteor:} 0.26,
        \textbf{BERT-precision:} 0.59,
        \textbf{BERT-recall:} 0.63,
        \textbf{BERT-F1:} 0.61,
        \textbf{LLM-as-judge:} 0.9\\
        \\
        \textbf{GPT-4o :} To convert  $(10000)_{10}$ to base-8, perform successive division by 8 and record the remainders. Dividing:\\ 1.  $10000 \div 8 = 1250$ remainder 0 \\ 2.  $1250 \div 8 = 156$  remainder 2 \\ 3. $156 \div 8 = 19$ remainder 4 \\ 4.  $19 \div 8 = 2$ remainder 3 \\ 5. $2 \div 8 = 0$ remainder 2\\ The base-8 (octal) representation, reading \text{remainder }s from bottom to top, is $(23420)_8$.\\
        \textbf{Rouge1:} 0.54, 
        \textbf{Rouge2:} 0.39, \textbf{ROUGE-L:} 0.54, \textbf{ROUGELSum:} 0.54,
        \textbf{BLEU:} 0.23,
        \textbf{Meteor:} 0.64,
        \textbf{BERT-precision:} 0.67,
        \textbf{BERT-recall:} 0.77,
        \textbf{BERT-F1:} 0.72,
        \textbf{LLM-as-judge:} 0.9\\
        \\
        \textbf{GPT-4o-mini:} Dividing repeatedly by 8 gives  10000 ÷ 8 = 1250 r0,  \\ 1250 ÷ 8 = 156 r2,  \\ 156 ÷ 8 = 19 r4,  \\ 19 ÷ 8 = 2 r3,  \\ 2 ÷ 8 = 0 r2;  \\ reading the remainders from last to first yields  \\ $(10000)_{10}$ = $(23420)_8$.\\
        \textbf{Rouge1:} 0.58, 
        \textbf{Rouge2:} 0.18, \textbf{ROUGE-L:} 0.58, \textbf{ROUGELSum:} 0.58,
        \textbf{BLEU:} 0.00,
        \textbf{Meteor:} 0.38,
        \textbf{BERT-precision:} 0.68,
        \textbf{BERT-recall:} 0.63,
        \textbf{BERT-F1:} 0.65,
        \textbf{LLM-as-judge:} 0.9\\
        \\
        \textbf{O1:} 10000 in decimal is 23420 in base 8.\\
        \textbf{Rouge1:} 0.14, 
        \textbf{Rouge2:} 0.00, \textbf{ROUGE-L:} 0.14, \textbf{ROUGELSum:} 0.14,
        \textbf{BLEU:} 0.00,
        \textbf{Meteor:} 0.04,
        \textbf{BERT-precision:} 0.56,
        \textbf{BERT-recall:} 0.39,
        \textbf{BERT-F1:} 0.46,
        \textbf{LLM-as-judge:} 0.8\\
        \\
        \textbf{O1-mini:} The base-8 (octal) representation of the decimal number $10000_{10}$ is $23420_{8}$.\\
        \textbf{Rouge1:} 0.17, 
        \textbf{Rouge2:} 0.04, \textbf{ROUGE-L:} 0.17, \textbf{ROUGELSum:} 0.17,
        \textbf{BLEU:} 0.00,
        \textbf{Meteor:} 0.04,
        \textbf{BERT-precision:} 0.55,
        \textbf{BERT-recall:} 0.50,
        \textbf{BERT-F1:} 0.53,
        \textbf{LLM-as-judge:} 0.8\\
        \\
        \textbf{GPT-5:} To convert $10000_{10}$ to base 8, we repeatedly divide by 8 and record the remainders:

        $\begin{aligned}
        10000 \div 8 = 1250 \quad \text{remainder } 0 \\
        1250 \div 8 = 156 \quad \text{remainder } 2 \\
        156 \div 8 = 19 \quad \text{remainder } 4 \\
        19 \div 8 = 2 \quad \text{remainder } 3 \\
        2 \div 8 = 0 \quad \text{remainder } 2
        \end{aligned}$

        Reading the remainders from bottom to top gives:
        
        \[
        10000_{10} = 23420_8
        \]\\
        \textbf{Rouge1:} 0.57, 
        \textbf{Rouge2:} 0.40, \textbf{ROUGE-L:} 0.57, \textbf{ROUGELSum:} 0.57,
        \textbf{BLEU:} 0.12,
        \textbf{Meteor:} 0.51,
        \textbf{BERT-precision:} 0.63,
        \textbf{BERT-recall:} 0.76,
        \textbf{BERT-F1:} 0.69,
        \textbf{LLM-as-judge:} 1.0\\
        \\
        \textbf{Grok-4:} To convert  $(10000)_{10}$ to base-8, repeatedly divide the number by 8 and record the remainders: 10000 ÷ 8 = 1250 remainder 0, 1250 ÷ 8 = 156 remainder 2, 156 ÷ 8 = 19 remainder 4, 19 ÷ 8 = 2 remainder 3, 2 ÷ 8 = 0 remainder 2.  \\ Reading the remainders from bottom to top gives  $(23420)_8$.  \\ This can be verified as  $2 \times 8^4 + 3 \times 8^3 + 4 \times 8^2 + 2 \times 8^1 + 0 \times 8^0 = 10000.$\\
        \textbf{Rouge1:} 0.37, 
        \textbf{Rouge2:} 0.12, \textbf{ROUGE-L:} 0.37, \textbf{ROUGELSum:} 0.37,
        \textbf{BLEU:} 0.08,
        \textbf{Meteor:} 0.32,
        \textbf{BERT-precision:} 0.64,
        \textbf{BERT-recall:} 0.68,
        \textbf{BERT-F1:} 0.66,
        \textbf{LLM-as-judge:} 1.0\\
        \\
        \textbf{Gemini-2.5-pro:} \textcolor{red}{To convert 10000 to base-8, we repeatedly divide by 8 and record the remainders. Performing these divisions gives remainders of 4, 2, 3, 2, so $(10000)_{10} = (23204)_8$.}\\
        \textbf{Rouge1:} 0.25, 
        \textbf{Rouge2:} 0.03, \textbf{ROUGE-L:} 0.19, \textbf{ROUGELSum:} 0.19,
        \textbf{BLEU:} 0.08,
        \textbf{Meteor:} 0.26,
        \textbf{BERT-precision:} 0.63,
        \textbf{BERT-recall:} 0.56,
        \textbf{BERT-F1:} 0.59,
        \textbf{LLM-as-judge:} 0.6\\
        \\
        \textbf{Gemini-2.5-flash:} To convert the decimal number 10000 to base-8, repeatedly divide by 8 and record the remainders. The remainders, read from bottom to top, form the base-8 representation. Therefore, $(10000)_{10} = (23420)_8$.\\
        \textbf{Rouge1:} 0.23, 
        \textbf{Rouge2:} 0.06, \textbf{ROUGE-L:} 0.17, \textbf{ROUGELSum:} 0.17,
        \textbf{BLEU:} 0.08,
        \textbf{Meteor:} 0.17,
        \textbf{BERT-precision:} 0.57,
        \textbf{BERT-recall:} 0.57,
        \textbf{BERT-F1:} 0.57,
        \textbf{LLM-as-judge:} 0.8\\
        \\
        \textbf{Claude-3.5-sonnet:} Let's divide 10000 repeatedly by 8, keeping track of remainders from bottom to top: 10000 ÷ 8 = 1250 remainder 0, 1250 ÷ 8 = 156 remainder 2, 156 ÷ 8 = 19 remainder 4, 19 ÷ 8 = 2 remainder 3, 2 ÷ 8 = 0 remainder 2. Reading the remainders from bottom to top, $(10000)_{10} = (23420)_8$.\\
        \textbf{Rouge1:} 0.50, 
        \textbf{Rouge2:} 0.17, \textbf{ROUGE-L:} 0.50, \textbf{ROUGELSum:} 0.50,
        \textbf{BLEU:} 0.12,
        \textbf{Meteor:} 0.38,
        \textbf{BERT-precision:} 0.67,
        \textbf{BERT-recall:} 0.68,
        \textbf{BERT-F1:} 0.67,
        \textbf{LLM-as-judge:} 0.9\\
    \end{shadedanswer}

    \item The effect of including the source for context:
    \begin{shadedanswer}
        \textbf{Q (without DOI):} How is the U-NOT gate related to the quantum cloner?\\
        \textbf{Llama-3.1-70b-instruct:} \textcolor{red}{I am sorry but I don't know the answer to this question.}\\
        \textbf{Rouge1:} 0.04, 
        \textbf{Rouge2:} 0.00, \textbf{ROUGE-L:} 0.04, \textbf{ROUGELSum:} 0.04,
        \textbf{BLEU:} 0.00,
        \textbf{Meteor:} 0.02,
        \textbf{BERT-precision:} 0.34,
        \textbf{BERT-recall:} 0.26,
        \textbf{BERT-F1:} 0.29,
        \textbf{LLM-as-judge:} 0\\
        \\
        \textbf{Q (with DOI):} \textcolor{blue}{According to 10.1007/978-3-030-75436-5\_8,} how is the U-NOT gate related to the quantum cloner?\\
        \textbf{Llama-3.1-70b-instruct:} The U-NOT gate is realized by performing a unitary transformation on the cloner's third output qubit, yielding the optimal approximation of the orthogonal state with a fidelity of $\frac{2}{3}$.\\
        \textbf{Rouge1:} 0.73, 
        \textbf{Rouge2:} 0.51, \textbf{ROUGE-L:} 0.70, \textbf{ROUGELSum:} 0.70,
        \textbf{BLEU:} 0.38,
        \textbf{Meteor:} 0.67,
        \textbf{BERT-precision:}  0.87,
        \textbf{BERT-recall:}  0.84,
        \textbf{BERT-F1:}  0.86,
        \textbf{LLM-as-judge:} 9\\        
    \end{shadedanswer}
\end{enumerate}

\section{Comprehensive Results}\label{apx:comprehensive-results}
The following figures present more comprehensive results, incorporating all evaluation metrics used in our study.

\begin{figure*}[h]
    \centering
    \begin{subfigure}[b]{0.8\textwidth}
        \centering
        \includegraphics[width=\textwidth]{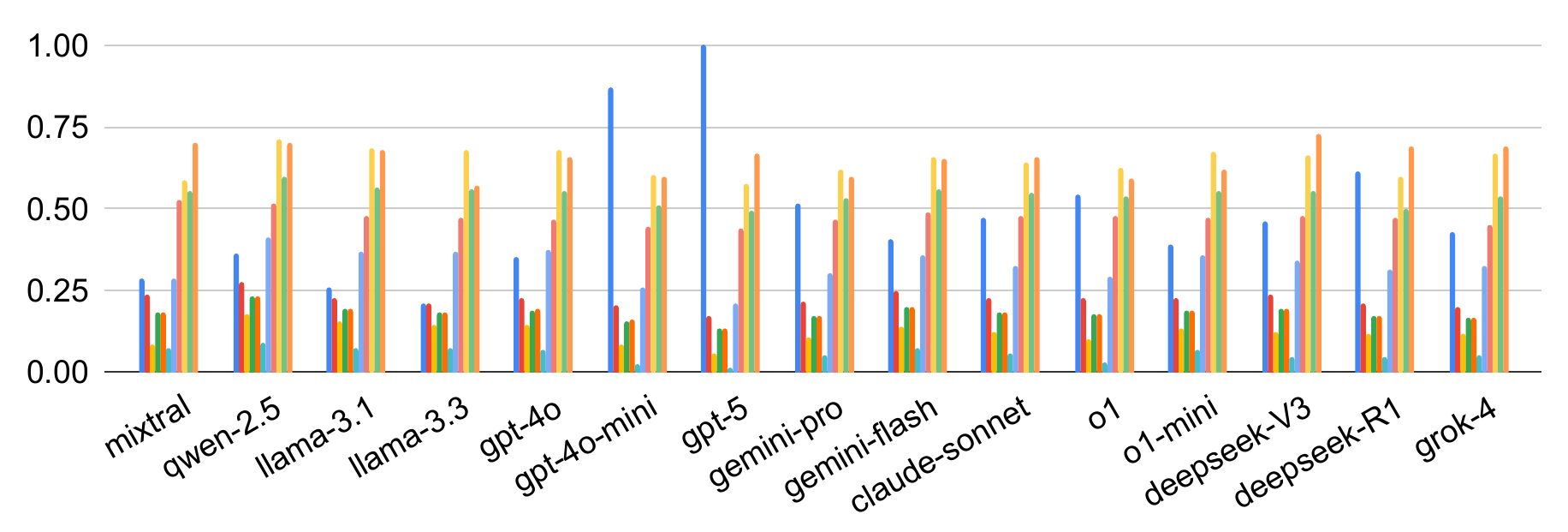}
        \caption{Original dataset}
        \label{fig-apx:original}
    \end{subfigure}
    \begin{subfigure}[b]{0.8\textwidth}
        \centering
        \includegraphics[width=\textwidth]{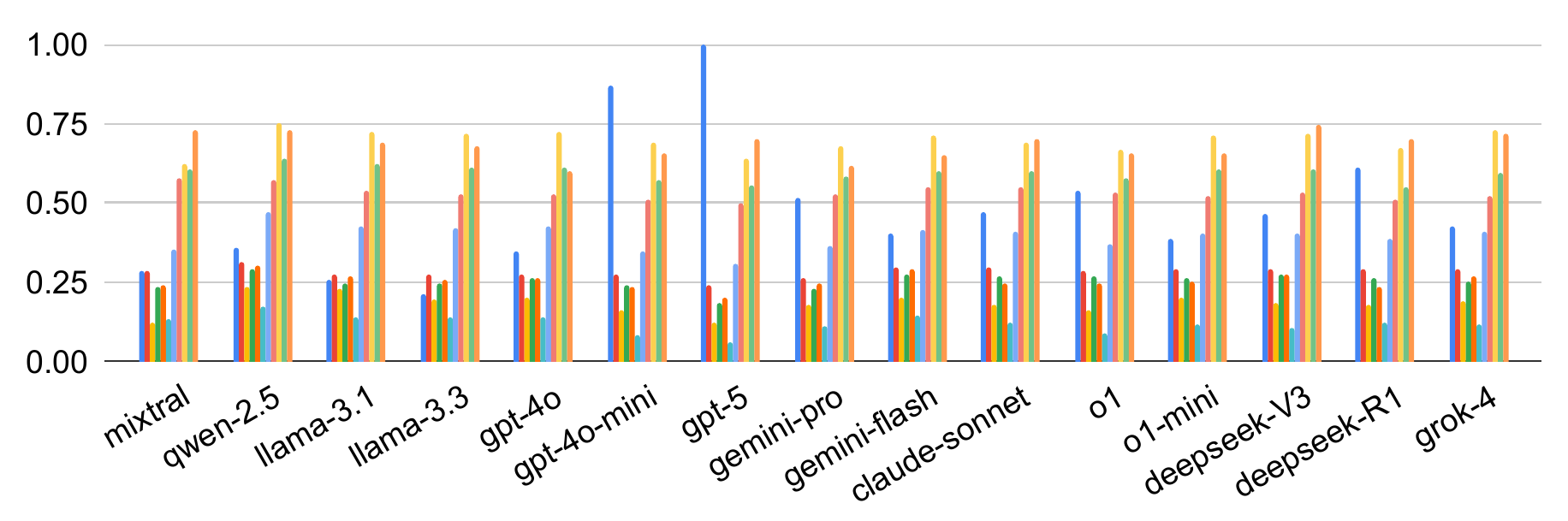}
        \caption{Paraphrased dataset}
        \label{fig-apx:paraphrased}
    \end{subfigure}

    \begin{subfigure}[b]{0.8\textwidth}
        \centering
        \includegraphics[width=\textwidth]{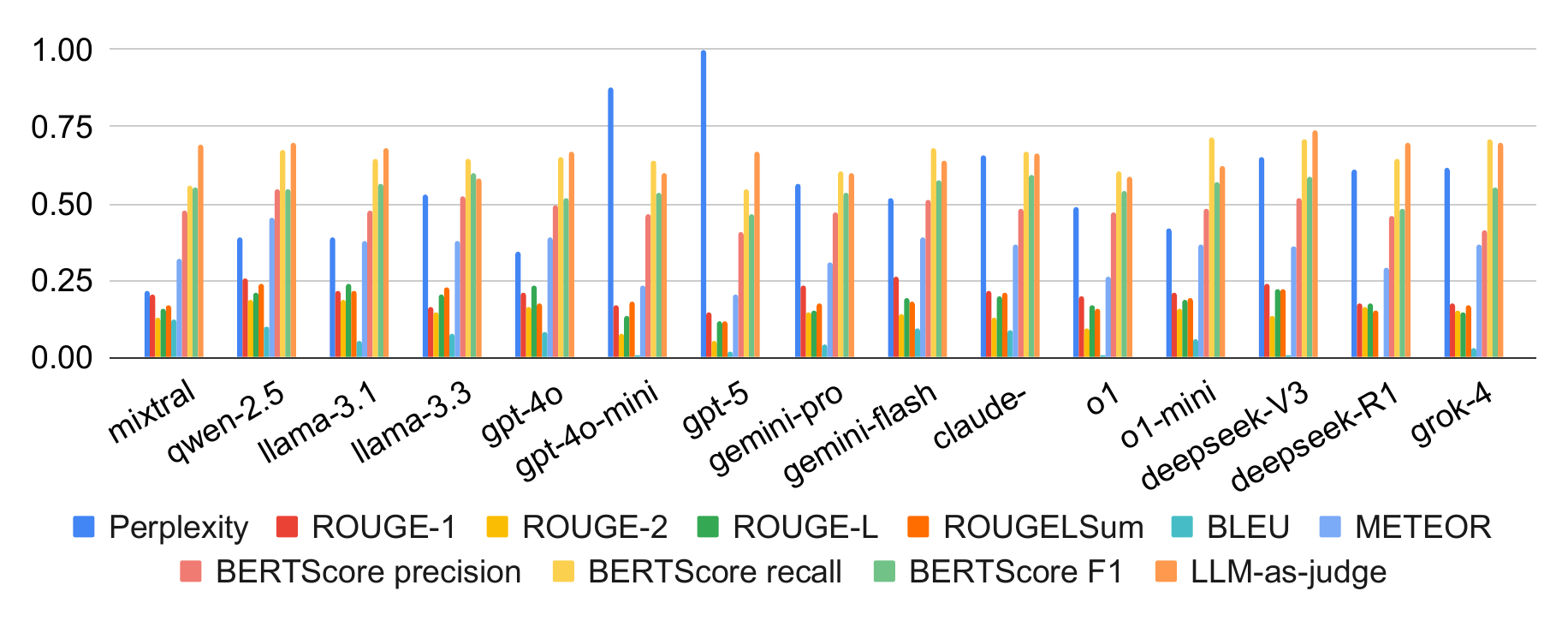}
        \caption{Sourced dataset}
        \label{fig-apx:sourced}
        
    \end{subfigure}
    \caption{ \Cref{fig:original,fig:paraphrased,fig:sourced} represent all evaluation metrics when prompting our candidate LLMS with the original dataset, the paraphrased dataset, and the source (as DOI) along the original dataset, respectively.}
    \label{fig-apx:main-res}
\end{figure*}

\begin{figure*}[h]
    \centering
    \includegraphics[width=0.8\textwidth]{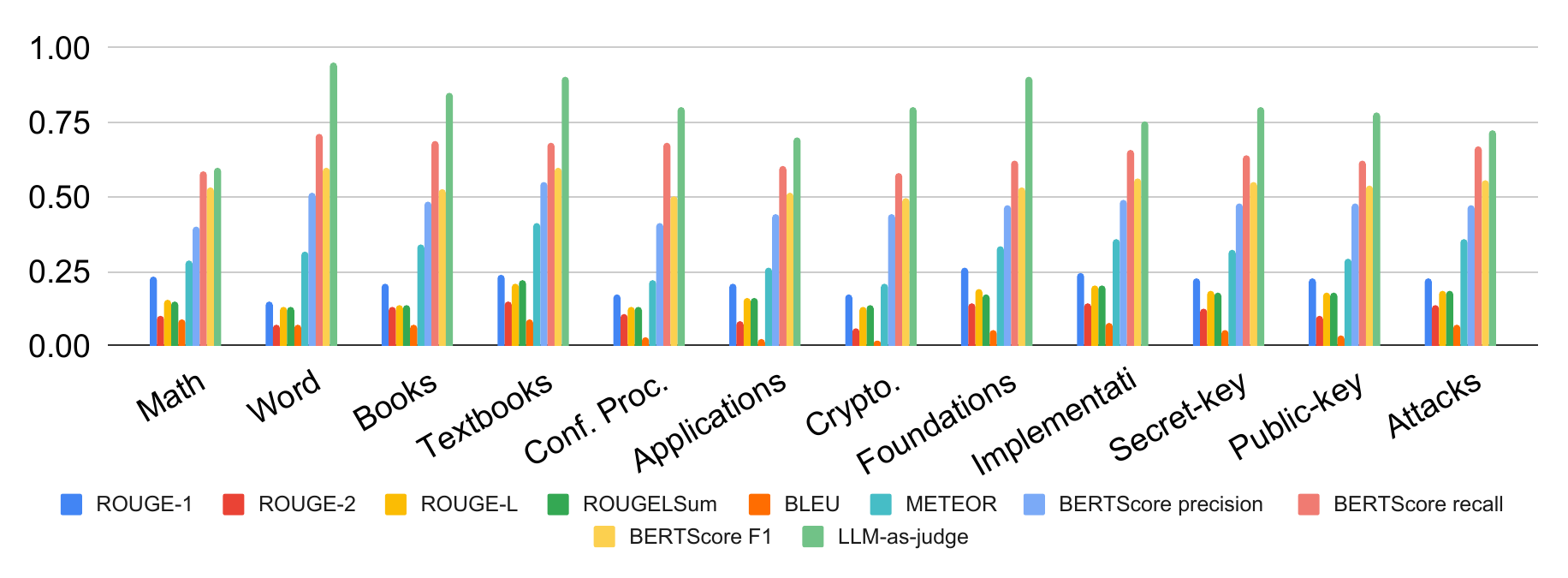}
    \caption{The averaged scores over all LLMs for the different metadata attributes.}
    \label{fig-apx:stratification}
    
\end{figure*}

\begin{figure*}[h]
        \centering
        \includegraphics[width=0.8\textwidth]{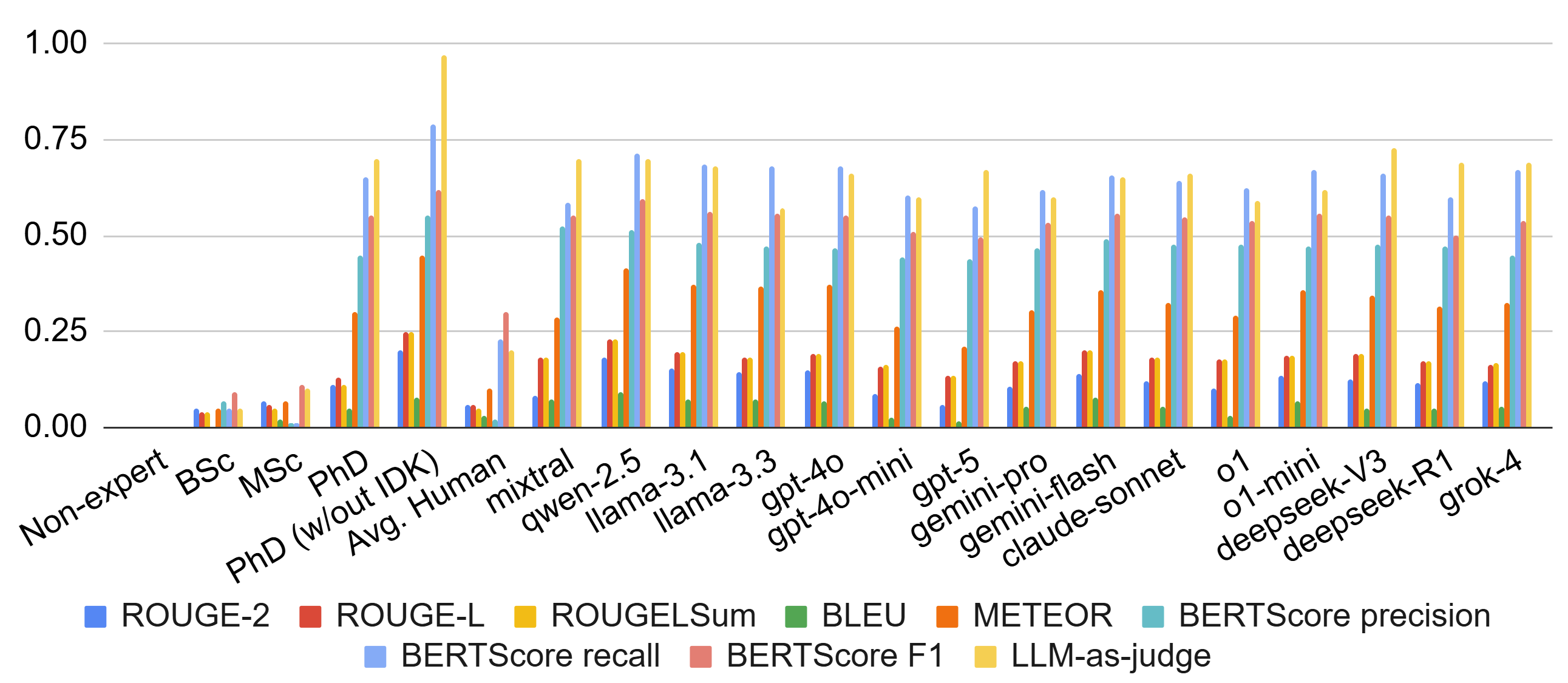} 
        \caption{Gold Standard: The LLMs' as well as the human (expert)'s performance on our qualitative subset. 
        }
        \label{fig-apx:gold}
               
    \end{figure*}

    \label{sec:appendix}

\end{document}